\documentclass[twocolumn,aps,floats,longbibliography]{revtex4-1}
\usepackage[latin9]{inputenc}
\usepackage{epsfig}
\usepackage{graphicx}
\usepackage[figuresright]{rotating}

\usepackage{amsfonts}
\usepackage{amsmath}
\usepackage{amssymb}
\usepackage{wasysym}
\PassOptionsToPackage{normalem}{ulem}
\usepackage{ulem}
\usepackage[unicode=true,pdfusetitle,
 bookmarks=true,bookmarksnumbered=false,bookmarksopen=false,
 breaklinks=false,pdfborder={0 0 1},backref=false,colorlinks=false]
 {hyperref}
\hypersetup{colorlinks,linkcolor=blue,citecolor=blue,urlcolor=blue}

\makeatletter

\usepackage{xcolor}
\usepackage{pdfcolmk}
\usepackage{wasysym}

\usepackage{txfonts}

\def\avg#1{\langle#1\rangle}

\def\be{\begin{equation}} \def\ee{\end{equation}}
\def\bea{\begin{eqnarray}} \def\eea{\end{eqnarray}}

\def\nn{\nonumber}

\makeatother

\begin{document}
\title{Multiflavor Mott insulators in quantum materials and ultracold atoms}
\author{Gang V. Chen}
\email{gangchen.physics@gmail.com}
 \affiliation{International Center for Quantum Materials, School of Physics, Peking University, Beijing 100871, China}
\affiliation{Department of Physics and HKU-UCAS Joint Institute
for Theoretical and Computational Physics at Hong Kong,
The University of Hong Kong, Hong Kong, China}
\affiliation{Collaborative Innovation Center of Quantum Matter, 100871, Beijing, China}

\author{Congjun Wu}
\email{wucongjun@westlake.edu.cn}
\affiliation{New Cornerstone Science Laboratory, Department of Physics, School of Science, Westlake University, Hangzhou 310024, Zhejiang, China}
\affiliation{Institute for Theoretical Sciences, Westlake University, Hangzhou 310024, Zhejiang, China}
\affiliation{Key Laboratory for Quantum Materials of Zhejiang Province, School of Science, Westlake University, Hangzhou 310024, Zhejiang, China}
\affiliation{Institute of Natural Sciences, Westlake Institute for Advanced Study, Hangzhou 310024, Zhejiang, China}

\date{\today}
\begin{abstract}
Mott insulators with large and active (or multiflavor) local Hilbert spaces
widely occur in quantum materials and ultracold atomic systems, and are 
dubbed ``multiflavor Mott insulators''. For these multiflavored Mott insulating materials, 
the spin-only description with the quadratic spin interactions is often
insufficient to capture the major physical processes. 
In the situation with active orbitals, the Kugel-Khomskii superexchange 
model was then proposed. We briefly review this historical model 
and discuss the modern developments beyond the original spin-orbital context.
These include and are not restricted to the $4d$/$5d$ transition metal
compounds with the spin-orbit-entangled ${J=3/2}$ quadruplets, 
the rare-earth magnets with two weakly-separated crystal field doublets, 
breathing magnets and/or the cluster and molecular magnets,
{\sl et al}. We explain the microscopic origin of the emergent 
Kugel-Khomskii physics in each realization with some emphasis 
on the ${J=3/2}$ quadruplets, and refer the candidate multiflavor Mott insulators 
as ``{${J=3/2}$ Mott insulators}''.  For the ultracold atoms, 
we review the multiflavor Mott insulator realization with the 
ultracold alkaline and alkaline-earth atoms on the optical lattices. 
Despite a large local Hilbert space from the atomic hyperfine spin states, 
the system could naturally realize a large symmetry group such as the Sp($N$) 
and SU($N$) symmetries. These ultracold atomic systems lie in the large-$N$ 
regime of these symmetry groups and are characterized by strong quantum 
fluctuations. The Kugel-Khomskii physics and the exotic quantum ground states
with the ``baryon-like" physics can appear in various limits.
We conclude with our vision and outlook on this subject. 
\end{abstract}

\maketitle

\section{Introduction}
\label{sec1}

There are many different ways to classify and view the physics 
related to correlated quantum many-body systems~\cite{Anderson1984}.
One classification scheme may be insufficient 
and thus misses other complementary views, while different
views could raise different types of physical questions and 
point to new directions of our field. 
Here we sketch some popular schemes and views in the field. 
One could classify the quantum many-body systems 
from the relevant and emergent phases and phase transitions. 
Alternatively, one could summarize various physical phenomena 
and classify the qualitative behaviors that may or may not be unique to 
particular phases, but these efforts can be quite useful experimentally
and practically.  
One can further identify the internal structures of the underlying
systems such as the intrinsic topological structures~\cite{ChenXie2013}
or emergent symmetries and the related experimental signatures.
One may also discuss various universal properties pertinent to
certain phases or focus on the physical realization of these phases
in quantum materials with the relevant physical degrees of freedom.
The last view may necessarily involve a significant amount of 
specific physics and specific features of the degrees of freedom
and the underlying quantum materials~\cite{TMO,khomskii2014,RevModPhys.87.1}.
The universal parts of the physics, however, are inevitably entangled
with the specific physics and manifest themselves in terms of the
specific degrees of freedom. The balance between universality
and specifics may be strongly constrained by the specific materials
instead of being determined by the more subjective purposes. Therefore,
one could further classify the correlated many-body systems according
to the relevant physical degrees of freedom and their interactions.

With the above thoughts, we turn to the multiflavor Mott insulators. 
As we have described in the abstract, these multiflavor Mott systems carry a large and 
active local Hilbert spaces with multiple flavors local states~\cite{kitp}. 
% We will use the former terminology in this review.
One traditional example of such multiflavor Mott insulating systems 
is the one involving both active spin and orbital degrees of freedom
where the spin and orbital states play the role of flavors and
the Kugel-Khomskii spin-orbital superexchange model was proposed
by Kliment Kugel and Daniel Khomskii~\cite{kugel1982}.
The advance that Kugel and Khomskii made beyond the Anderson's mechanism~\cite{PhysRev.79.350}
of the superexchange spin interaction was to include
the orbitals and treat them on the equal footing 
as the spin. 
 Because of the complicated expression and the orbital involvement,
the Kugel-Khomskii model did not receive a significant attention over
the past few decades. Nevertheless, the Kugel-Khomskii physics is
realistic and relevant for many physical systems~\cite{TMO,khomskii2014,oles2017orbital,Andrzej2023,nagaosa,PTPS.160.155}.
In the recent years, the orbital degrees of freedom and the orbital
related physics are getting more attention.
This is due to many factors, such as the rising of topological
materials~\cite{RevModPhys.82.3045,RevModPhys.83.1057,
RevModPhys.93.025002,RevModPhys.90.015001} 
(that often require the spin-orbit coupling to generate the topological bands), 
the spin-orbit-coupled correlated materials where the spin-orbit coupling
is the key ingredient~\cite{annurev-conmatphys-020911-125138},
the experimental progress including the
resonant X-ray scattering measurement that allows the experimental
detection of the orbital structures and excitations~\cite{RevModPhys.83.705,RIXS_orbital},
and so on. Therefore, {it is timing to revisit the Kugel-Khomskii physics.} 
Part of this review aims to suggest the broad applicability of the Kugel-Khomskii model 
in the multiflavor Mott insulating quantum materials  
beyond the original spin-orbital disentangled context.
By ``disentangled'', the spins and the orbitals are independent local variables. 
The new territory for the Kugel-Khomskii 
physics is proposed to be the multiflavor Mott insulators. 
This includes, but is not restricted to,
the breathing materials and/or cluster and molecular magnets,
the rare-earth magnets
with weak crystal fields, the spin-orbital-entangled ${J=3/2}$
local moments of transition metal compounds, {\sl etc}.
We will explain the realization of the multiflavor local Hilbert space in these 
 Mott insulating systems. 
Beyond the quantum materials' context, the ultracold atom
systems, such as the alkaline atoms and the alkaline-earth atoms on
the optical lattices, could serve as candidate systems to realize the
Kugel-Khomskii physics together with their own merits with the exact
high symmetry groups that are more proximate
to the color-like degree of freedom in particle physics
and will be discussed in the second half
of the review.

The observation is that, despite the proposed multiflavor Mott 
insulators do not explicitly contain the orbitals as the original Kugel-Khomskii context, there exist
emergent orbital-like degrees of freedom that play the
role of the orbitals. For the breathing magnets and/or
cluster Mott systems~\cite{PhysRevB.93.245134,PhysRevB.97.035124,PhysRevLett.113.197202,PhysRevB.90.060414,PhysRevLett.116.257204,PhysRevB.94.075146,nikolaev2020quantum,Sheckelton2012,Heung2014,PhysRevResearch.2.043424,PhysRevLett.112.027202,dissanayake2021understanding},
it is the degenerate ground states of the local clusters that function 
as the effective orbitals. The effective Kugel-Khomskii physics appears
when one considers the residual interactions between the degenerate
ground states of the local clusters. These interactions lift the  
remaining degeneracy and create the many-body ground states.
{It is worthwhile to mention that, many of the correlated mior{e} systems
are actually in the cluster Mott insulating regime~\cite{PhysRevX.8.031089,Mak2022}. }
For certain rare-earth magnets~\cite{PhysRevLett.98.157204,
PhysRevB.62.6496,PhysRevB.84.140402,PhysRevB.87.094410,
PhysRevB.90.014429,PhysRevB.99.224407}, although the
orbitals are present implicitly, they are strongly entangled with the
spin, and then what is meaningful is the local
``$J$'' moment. For this case, it is the degenerate energy levels
that can be treated as effective orbital degrees of freedom.
For the transition metal compounds with the ${J=3/2}$ local
moments that are dubbed ``${J=3/2}$ Mott insulators''~\cite{PhysRevB.82.174440,
PhysRevB.101.054439,PhysRevLett.118.217202,
PhysRevB.104.165150,PhysRevB.102.180401,PhysRevLett.121.097201,Heung2014},
one can group the four local states into two fictitious orbitals with one
spin-1/2 moment and then naturally describe the interaction as the
Kugel-Khomskii model. We carefully derive the superexchange model in terms
of the effective spin and orbitals and express the model
in the form of the Kugel-Khomskii interaction. The correspondence between
the microscopic multipolar moments in the ${J=3/2}$ language and
the effective spin-orbital language is established.
This correspondence may be useful for the mutual feedback
between the understanding from different languages and views.
This ${J=3/2}$ local moments and the effective Kugel-Khomskii
physics can be broadly applied to many $4d$/$5d$ Mott systems
such as the Mo-base, Re-based, Os-based double perovskites~\cite{PhysRevB.82.174440},
and can even be relevant to certain $3d$ transition metal
compounds such as vanadates with the $V^{4+}$ ions~\cite{TMO,oles2009spinorbital,Andrzej2023}.
We further discuss the physical consequences
from these effective Kugel-Khomskii description 
for the multiflavor Mott insulators.

We devote the second half of the review to the ultracold atom system.
The ultracold atom system has become a new frontier of
condensed matter physics and provides new opportunities
and platforms for exploring the emergent correlation physics.
The Kugel-Khomskii physics turns out to be particularly relevant
for many alkali and alkaline-earth atoms
that possess large hyperfine spins and thus a large local Hilbert space~\cite{wu2003,wu2006,Gorshkov2010}.
Like the multiflavor Mott insulating quantum materials in the previous paragraphs, 
the physical picture of these large hyperfine spins is
very different from that of traditional magnets with large spins.
Thus, we can discuss some of the common features shared 
by the multiflavor Mott insulating quantum materials
and the ultracold alkali and alkaline-earth atomic Mott insulators. 
Here, the traditional magnets refer to the conventional 
$3d$ transition metal compounds without quenched orbitals~\cite{TMO}
and are used to distinguish from the multiflavor Mott insulating quantum materials.
In these traditional magnets, large spins arise from the Hund's coupling,
and the spins of the localized electrons on the same lattice site are
aligned to form a large spin. 
The leading order contribution to the couplings between different
sites is the Anderson's superexchange of a single pair of electrons, such that
the fluctuation with respect of an ordered moment would be just ${\pm 1}$.
This is the physical origin of the $1/S$-effect, in other words,
as $S$ is enlarged, the system evolves towards the classical direction.
In many alkali and alkaline-earth atom systems, however, the situation
is quite different. 
%This is also common in the above ``${J=3/2}$ Mott insulator''.
The energy scale is far below the atomic ionization energy,
and exchanging a pair of fermions can completely shuffle
the spin configuration among the ${2S+1}$ spin states.
The spin states are much more delocalized in their Hilbert
space. Thus, a large spin here behaves more like a large
number of flavors or colors, which strongly enhances the
quantum fluctuations. 
Since these ultracold alkali and alkaline-earth atoms naturally
support high symmetries beyond the SU(2) for the traditional magnets, 
 it is more appropriate to adopt the perspective
of high symmetries (e.g., SU($N$) and Sp($N$) with ${N=2S+1}$)
with the color-like degrees of freedom for these atomic states~\cite{wu2003,wu2006,controzzi2006,wu2010,wu2012,killian2010,taie2012}.
Such high symmetries often require some fine-tuning for the multiflavor Mott insulating quantum materials except
the more recent twisted multilayer graphenes and transition metal dichalcogenide heterostructures. Nevertheless,  
based on the spirit of Kugel-Khomskii physics, exchanging a pair of electrons for example in the ``${J=3/2}$ Mott insulator''
is also understood to be able to shuffle the 
effective spin and orbital states and make the system more delocalized 
in the Hilbert space with an enhanced quantum fluctuations~\cite{PhysRevB.82.174440}.

This new perspective from these ultracold atom systems provides
an opportunity to explore the many-body physics closely related
to the Kugel-Khomskii-like physics for ${N=4}$, and some aspects
with high symmetries may even be connected to the high-energy physics.
As an early progress, an exact and generic symmetry of Sp(4), or, isomorphically
SO(5), was proved for the spin-${3}/{2}$ alkali fermion
systems~\cite{wu2003,wu2006}.
Under the fine-tuning, the Sp(4) symmetry can be augmented to SU(4).
Later, the SU($N$) symmetry has also been widely investigated in
the alkaline-earth fermion systems due to the vanishing electron spins
and the sole nuclear spin contribution~\cite{cazalilla2009,Hermele2009,Gorshkov2010}.
For instance, the alkaline-earth-like atom $^{173}$Yb with the spin ${S= 5/2}$~\cite{taie2012}
has 6 components and thus SU(6) symmetry, and the $^{87}$Sr atom with ${S = 9/2}$~\cite{killian2010}
has 10 spin components and thus SU(10) symmetry, respectively. Thus,
we will review the properties of quantum magnetism with the ultracold atoms
possessing the Sp($N$) and SU($N$) symmetries with ${N=4}$.
These are the ultracold-atom versions of the Kugel-Khomskii physics.
They are characterized by various competing orders due to the strong
quantum fluctuations. As an exotic example, they can exhibit
the ``baryon-like" physics. In an SU(4) quantum magnet,
quantum spin fluctuations are dominated by the multi-site correlations,
whose physics is beyond the two-site one of the SU(2) magnets as
often studied in the condensed matter systems. It is exciting that
in spite of the huge difference of the energy scales, the large-spin
cold fermions can also exhibit similar physics to quantum chromodynamics (QCD).

The remaining parts of this review are organized as follows.
In Sec.~\ref{sec2}, we start with a brief introduction of the
original proposal of the Kugel-Khomskii superexchange model
in the multiflavored Mott insulating systems with active orbitals.
In Sec.~\ref{sec3}, we explain various quantum materials' realizations
of the Kugel-Khomskii physics and the status of the effective orbitals.
In Sec.~\ref{sec4}, we turn the attention to the ``${J=3/2}$ Mott
insulators'' and establish the Kugel-Khomskii physics.
In Sec.~\ref{sec5}, we review the ultracold atoms on optical lattices
and discuss the high-symmetry models and the emergent physics
with the alkaline and alkaline earth atoms. 
Finally, in Sec.~\ref{sec6}, we summarize this review.

\section{Kugel-Khomskii exchange model}
\label{sec2}

As we have remarked in Sec.~\ref{sec1}, a conventional and representative
example of the multiflavor Mott insulators with a large and active local Hilbert 
space is the one involving active orbital degrees of freedom where the spin 
and orbital states play the role of the multiflavored local Hilbert space. 
Thus, we start the review with these spin-orbital-based Mott insulators 
and their superexchange interactions.

The Anderson superexchange interaction for the spin degrees of freedom
in the Mott insulators is widely accepted as the major mechanism for the
antiferromagnetism~\cite{PhysRev.79.350}. Anderson's treatment was
perturbative. The virtual exchange of the localized electrons from the neighboring
sites through the high-order perturbation processes generates a Heisenberg
interaction between the local spin moments. Anderson's original work was 
based on a single-band Hubbard model where only one orbital related band  
is involved, and Anderson's treatment could be well adjusted to include the
orbitals from the intermediate anions. Insights from these calculations were
summarized as the empirical Goodenough-Kanamori-Anderson (GKA)
rules~\cite{goo63,1957PThPh..17..177K}.

For the Hubbard model with multiple orbitals at the magnetic ions, the orbital
necessarily becomes an active degree of freedom in addition to the spin once
the system is in the Mott insulating phase with localized electrons on the lattice
sites~\cite{kugel1982}. In this case, more structures are involved in the local moment
formation that generates a large local Hilbert space with both spins and orbitals
in the Mott regime,
and the original Anderson's spin exchange model cannot be directly applied 
to the multiflavor Mott insulators with the spin and orbitals here. 
With more active degrees of freedom in the large local physical Hilbert space,
the GKA rules can no longer provide the nature and the magnitude of the exchange
interactions, even the signs of the exchange interactions cannot be determined.
These exchange interactions sensitively depend on the orbital configurations
on each lattice site. Moreover, as the orbital is an active and dynamical degree
of freedom here, the orbitals are intimately involved in the superexchange processes.
Thus, in addition to the exchange of the spin quantum numbers,
the perturbative superexchange processes in these multiflavor Mott insulators
are able to exchange the orbital quantum numbers. In reality,
both the spin and the orbital quantum numbers can
be exchanged separately or simultaneously. Taking together,
the full exchange Hamiltonian would involve pure spin exchange,
pure orbital exchange, and the mixed spin-orbital exchange~\cite{kugel1982}.
This spin-orbital exchange model is nowadays referred as
``Kugel-Khomskii model''.

Closely following Kugel and Khomskii~\cite{kugel1982},
we present an illustrative derivation of the Kugel-Khomskii spin-orbital exchange model 
from a two-orbital Hubbard model. The Hubbard model is given~\cite{kugel1982,kugel1973crystal},
\begin{eqnarray}
H &=& \sum_{\langle ij \rangle }
t^{\alpha\beta}_{ij} a^{\dagger}_{i\alpha\sigma} a^{}_{j\beta\sigma}
+ \frac{U}{2}   \sum_{i} n_{i\alpha\sigma} n_{i\beta\sigma'}
(1 - \delta_{\alpha\beta}\delta_{\sigma\sigma'})
\nonumber \\
&-& \frac{1}{2} \sum_{i,\alpha \neq \beta} J_{\text{H}}^{}
( a^{\dagger}_{i\alpha\sigma} a_{i\alpha\sigma'}^{} a^{\dagger}_{i\beta\sigma'} a_{i\beta\sigma}^{}
+ a^{\dagger}_{i\alpha\sigma}   a^{}_{i\beta\sigma} a^{\dagger}_{i\alpha\sigma'}a_{i\beta\sigma'}^{} ),
\label{eq:twobandHubb}
\end{eqnarray}
where ${\alpha,\beta = 1,2}$ label the two orbitals, $\sigma,\sigma'$ label the spin
quantum number, and the final term takes care of the inter-orbital pair hopping and
the Hund's coupling ($J_{\text{H}}$). 
The orbital degeneracy is assumed, and an isotropic
and diagonal hopping ${t_{11} = t_{22}  =t, t_{12} =0}$ is further assumed.
In reality, the isotropic and diagonal hoppings are not guaranteed, and the
orbital degeneracy is not quite necessary. A standard perturbation treatment
yields the conventional Kugel-Khomskii model with
\begin{eqnarray}
H_{\text{KK}} = \sum_{\langle ij \rangle} J_1 \, {\boldsymbol S}_i \cdot {\boldsymbol S}_j
+ J_2 \, {\boldsymbol{\tau}}_i \cdot {\boldsymbol{\tau}}_j
+ 4J_3\,  ({\boldsymbol S}_i \cdot {\boldsymbol S}_j )\,
( {\boldsymbol{\tau}}_i \cdot {\boldsymbol{\tau}}_j ),
\label{eq2}
\end{eqnarray}
where we have the relations between the electron billinear operators and the 
spin-orbital operators,
\begin{eqnarray}
\sum_{\sigma} a^{\dagger}_{i,1,\sigma} a^{}_{i,1,\sigma} &=& \frac{1}{2} + \tau^z_i  , \\
\sum_{\sigma} a^{\dagger}_{i,2,\sigma} a^{}_{i,2,\sigma} &=& \frac{1}{2} - \tau^z_i  , \\
\sum_{\sigma} a^{\dagger}_{i,1,\sigma} a^{}_{i,2,\sigma} &=& \tau^+_i , \\
\sum_{\sigma} a^{\dagger}_{i,2,\sigma} a^{}_{i,1,\sigma} &=& \tau^-_i ,
\end{eqnarray}
and
\begin{eqnarray}
\sum_{\alpha} a^{\dagger}_{i,\alpha,\uparrow} a^{}_{i,\alpha,\uparrow} &=& \frac{1}{2} + S^z_i  , \\
\sum_{\alpha} a^{\dagger}_{i,\alpha,\downarrow} a^{}_{i,\alpha,\downarrow} &=& \frac{1}{2} - S^z_i  , \\
\sum_{\alpha} a^{\dagger}_{i,\alpha,\uparrow} a^{}_{i,\alpha,\downarrow} &=& S^+_i , \\
\sum_{\alpha} a^{\dagger}_{i,\alpha,\downarrow} a^{}_{i,\alpha,\uparrow} &=& S^-_i ,
\end{eqnarray}
and the exchange couplings are given as
\begin{eqnarray}
&& J_1 = \frac{2t^2}{U} (1 - \frac{J_{\text H}^{}}{U}), \\
&& J_2=J_3 =\frac{2t^2}{U} (1 + \frac{J_{\text H}^{}}{U}).
\end{eqnarray}
{Here ${\boldsymbol S}_i$ and ${\boldsymbol{\tau}}_i$ refer to the 
electron spin and the orbital pseudospin, respectively. In particular, when $\tau^z=1/2$ ($-1/2$), 
the orbital 1 (2) is occupied. }
Because the spin and orbitals are disentangled, the spin sector retains
the SU(2) rotational symmetry.
The orbital SU(2) symmetry in Eq.~\eqref{eq2} is accidental
and is due to the special choice of the hopping parameters. Via a fune-tuning of the hoppings and the interactions,
the Kugel-Khomskii model could have a larger symmetry such as
SU(4)~\cite{PhysRevB.60.6584},
and the limit with a higher
symmetry may provide a new solvability of this complicated
but realistic model.

In the above derivation of the Kugel-Khomskii model, the orbital degeneracy is not
actually required. As long as the crystal field splitting between the orbitals is small
or comparable to the exchange energy scale, one needs to seriously include the orbitals
into the description of the correlation physics for the local moments in the Mott regime.
This would clear up the unnecessary constraint of the Kugel-Khomskii physics 
for the Mott systems to have an explicit orbital degeneracy. A perfect orbital degeneracy
requires a high crystal field symmetry and is not quite common. The clearance 
of this constraint would already expand the applicability of the Kugel-Khomskii 
physics in the transitional metal compounds.

From the above example, one could readily extract some of the essential
properties for the Kugel-Khomskii model and the related physics
for the multiflavor Mott insulators with active spins and orbitals.
As the orbitals have the spatial orientations in the real space, the electron hopping
between the orbitals from the neighboring sites depends strongly on the
bond orientation and the orbital configurations. The resulting Kugel-Khomskii
model is anisotropic in the orbital sector, and the exchange interaction depends
on the bond orientations~\cite{kugel1982}. Because the spin and the orbital are disentangled
in the parent Hubbard model and the spin-orbit coupling is not considered,
the spin interaction in the spin sector, however, remains isotropic in the
Kugel-Khomskii model and is described by the conventional Heisenberg interaction.
For the relevant transition metal compounds, the common orbital degrees of
freedom can be $e_g$ and $t_{2g}$ orbitals. In the case of the $e_g$ orbitals,
an effective pseudospin-1/2 operator is often used to label the orbital state~\cite{kugel1982},
and this corresponds to the situation in the above example. For the $t_{2g}$
orbitals, a pseudospin-1 operator is used to operate on the three $t_{2g}$
orbitals~\cite{PhysRev.171.466,PhysRevB.78.094403},
and the corresponding Kugel-Khomskii model is much more involved than the $e_g$ 
case. The local physical Hilbert space is significantly enlarged by the spin and orbital
states. The Kugel-Khomskii model operates quite effectively
in this enlarged local Hilbert space. As it was mentioned more generally in Sec.~\ref{sec1}, 
this operation is quite different from the spin-only Mott insulators with a large-$S$ 
local spin moment where the orbital degree of freedom is quenched. 
Although the physical Hilbert space is enlarged with a large-$S$ local spin moment, 
the simple Heisenberg spin model merely changes the spin quantum number 
by 1 with one operation. In contrast, the Kugel-Khomskii spin-orbital model is
more effectively in delocalizing the spin-orbital states in the enlarged
physical Hilbert space and thus enhances the quantum fluctuations.
The fluctuation-driven orders and/or exotic quantum states can
be stabilized in this fashion~\cite{PTPS.160.155,RevModPhys.87.1,PhysRevLett.78.2799}.

\section{Modern quantum material realization of emergent Kugel-Khomskii physics}
\label{sec3}

In the previous section, we have demonstrated that
the original Kugel-Khomskii model can be derived from an extended
Hubbard model with multiple orbitals for the multiflavor Mott insulators. 
The orbital degeneracy is often assumed but not necessarily required. 
If the orbital separation is of the energy scale as the superexchange energy scale, 
then one should carefully include these orbitals into our modeling. 
Depending on the electron filling, the spin moment can be spin-1/2 and spin-1,
and the pseudospin for the orbital sector can be pseudospin-1/2
(for $e_g$ degeneracy, or two-fold $t_{2g}$ degeneracy) and
pseudospin-1 moment (for three-fold $t_{2g}$ degeneracy). The spin and
orbitals are disentangled in this model.

A well-known example would be the Fe-based superconductors~\cite{annurev-conmatphys-033117-054137,annurev-conmatphys-020911-125055,annurev-conmatphys-070909-104041}.
Although the parent materials behave mostly like a bad metal
and are thus modeled by an extended Hubbard model with multiple
electron orbitals, many important physics such as the magnetic excitations
and spectra may be better understood from the local moment
picture. This was used to interpret the magnetic excitations
in FeSe that is believed to be the most ``Mott''-like Fe-based
superconducting system~\cite{Bohmer2017,Kreisel2020}.
The active orbital degree of freedom in FeSe,
however, has not been included into the analysis of the
magnetic properties and the excitations~\cite{WangFa2015}.
Thus, FeSe can be a good application of the original
Kugel-Khomskii model for the understanding of the magnetism,
the orbital physics and the nematicity~\cite{PhysRevB.85.104510,
PhysRevX.8.031033,science1202226}.

While the Kugel-Khomskii model is proposed for
the multiflavor Mott insulators with the disentangled spin and orbitals, 
we explain its broad application
to other systems below.

\begin{figure}[t]
\includegraphics[width=7.5cm]{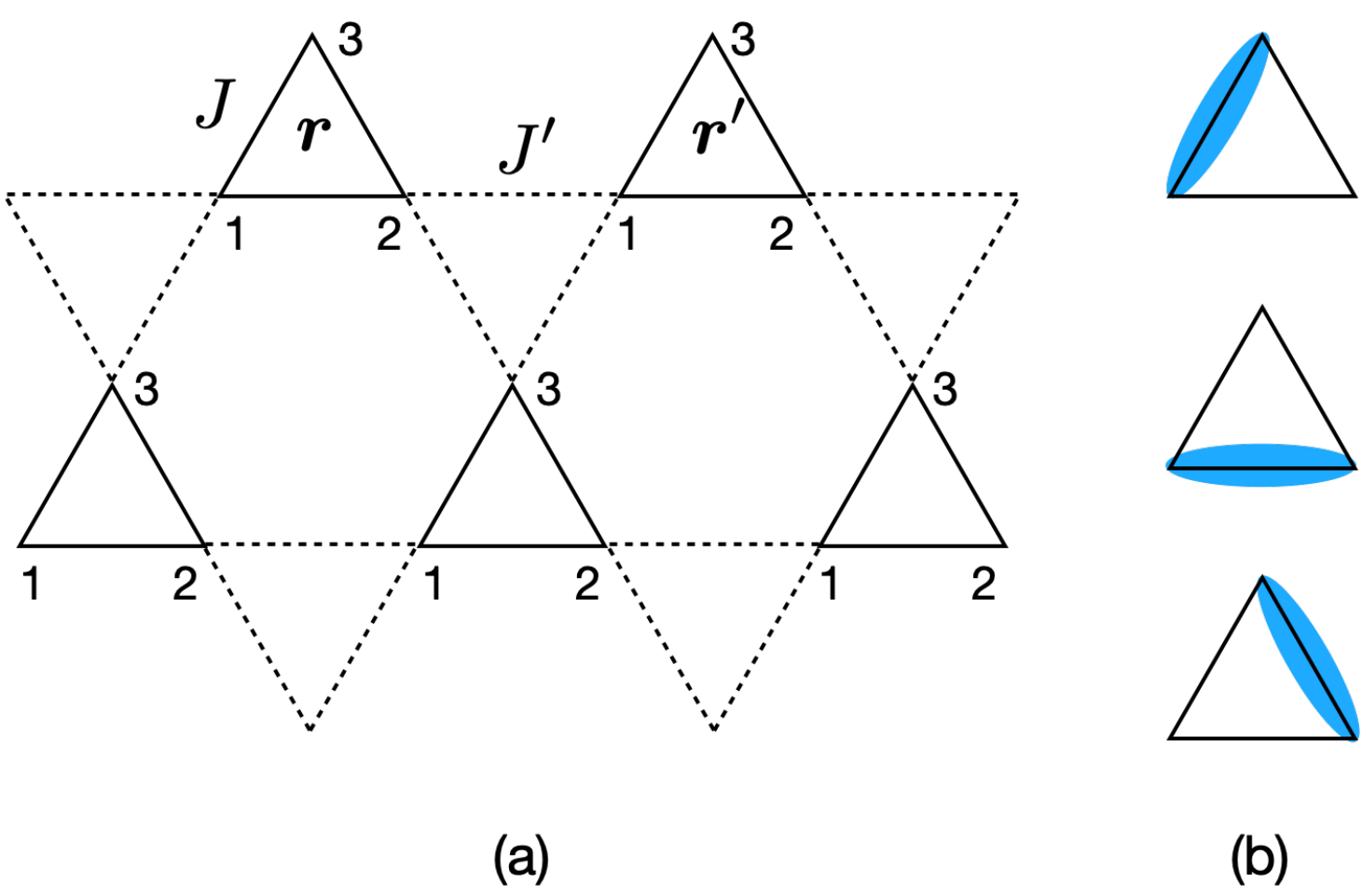}
\caption{(a) The breathing kagom\'{e} lattice structure with an
alternating exchange coupling on the triangular cluster.
(b) The three configurations of the spin singlet occupation
on the triangular cluster. The (blue) dimer refers to the
spin singlet of two spins, and the uncovered site is the
dangling spin moment. Two of the three configurations
are linearly independent. There are totally four ground
states of the triangular cluster after including the two
fold degeneracy of the dangling spin.
}
\label{fig1}
\end{figure}

\subsection{Breathing magnets and cluster magnets}

Breathing magnets (and cluster magnets) represent a new family
of magnetic materials whose building blocks are not the magnetic
ions, and can find their applications in
many organic magnets~\cite{C1CS90019F,GATTESCHI19922092}
and even inorganic compounds~\cite{fchem.2018.00461}.
Instead, the systems consist of the magnetic cluster units as
the building blocks, and these magnetic cluster units provide the
elementary and local degrees of freedom for the magnetism. To
understand the physics of these systems, one ought to first
understand the local physics on the cluster unit and find the
relevant low-energy states. 
From the perspective of the multiflavor Mott insulators, the local states of the
magnetic cluster unit contribute to the multiple flavors of states in the local Hilbert space. 
The many-body model for the system
should be constructed from these relevant local low-energy states.
To illustrate the point above,
we notice that the early spin liquid candidate
$\kappa$-(ET)$_2$Cu$_2$(CN)$_3$ can actually be placed
into the category of the cluster magnets~\cite{PhysRevLett.91.107001}.
In $\kappa$-(ET)$_2$Cu$_2$(CN)$_3$, each (ET)$_2$
molecular dimer hosts odd number of electrons. As
the molecular dimers form a triangular lattice,
it was proposed that this model realizes the
triangular lattice Hubbard model at the half filling.
The basis of the Hamiltonian is the Wannier functions associated
with the antibonding states of the highest occupied molecular
orbitals on each (ET)$_2$ dimer~\cite{JPSJ.78.083710}.
More generally, the local energy levels of the magnetic
clusters should be understood or classified from the
irreducible representation of the local symmetry group,
and the effective orbital degrees of the freedom on the cluster
is then interpreted as the local basis of the irreducible
representation~\cite{PhysRevB.93.245134,PhysRevB.97.035124,Heung2014}.

To deliver the idea of the effective orbital degree of freedom,
we start from the breathing kagom\'e lattice (see Fig.~\ref{fig1}),
and assume the simple Heisenberg interactions with alternating couplings,
\begin{eqnarray}
H = J \sum_{\langle ij \rangle \in \bigtriangleup
} {\boldsymbol S}_i \cdot {\boldsymbol S}_j
+ J' \sum_{\langle ij \rangle \in \bigtriangledown}
{\boldsymbol S}_i \cdot {\boldsymbol S}_j ,
\end{eqnarray}
where ``$\bigtriangleup$'' (``$\bigtriangledown$'')
refers to the up (down) triangles.
The word ``breathing'' refers to the fact that the
 ``$\bigtriangleup$'' triangles are of different size from the
 ``$\bigtriangledown$'' triangles and was first used in the
 context of the breathing pyrochlore magnets~\cite{JPSJ.84.043707}.
In the strong breathing limit
with ${J\gg J'}$, one should first consider the local states on the up triangles
and couple these local states together through the $J'$-links on the down triangles.
On each up triangles, there are three spin-1/2 local moments. With the
antiferromagnetic interactions, the ground states have four fold degeneracies.
This can be understood simply from the spin multiplication relation with
\begin{eqnarray}
\frac{1}{2} \otimes \frac{1}{2} \otimes \frac{1}{2}
 = \frac{1}{2} \oplus \frac{1}{2} \oplus \frac{3}{2},
\end{eqnarray}
where the left hand side refers to the three spin local moments on
the up triangle and the right hand side refers to the spin quantum
number of the total spin of the up triangular cluster. The total spin,
${S_{\text{tot}}=3/2}$, is only favored if the interaction is
ferromagnetic. The antiferromagnetic spin interaction favors a total spin
${S_{\text{tot}}=1/2}$, that is realized by forming a spin singlet
between two spins and leaving the remaining spin as a dangling spin
(see Fig.~\ref{fig1}). This would naively lead to three singlet occupation
configurations. It turns out that only two of them are linearly independent.
Counting the spin-up and spin-down degeneracy of the dangling spin, there
are in total four fold ground state degeneracies on each up triangular
cluster. To make connection with the Kugel-Khomskii physics,
one can simply regard the total spin of the up-triangle cluster
as the effective spin, and regard the two-fold degeneracy
of the spin singlet configuration as the effective orbital.
The later has the spirit of an orbital as both are even under time reversal.
The effective spin-orbital states on the up-triangle cluster are listed as follows,
\begin{eqnarray}
&&|{{s^z= \uparrow},{\tau^z=\uparrow }}\rangle
= \frac{1}{\sqrt{3}}
\big[ | {\downarrow_1 \uparrow_2 \uparrow_3 } \rangle
+ e^{i \frac{2\pi}{3}}   | {\uparrow_1 \downarrow_2 \uparrow_3 } \rangle
+ e^{-i \frac{2\pi}{3}}   | {\uparrow_1  \uparrow_2 \downarrow_3} \rangle \big],
\nonumber \\
&&|{{s^z = \uparrow},{\tau^z =\downarrow }}\rangle
= \frac{1}{\sqrt{3}}
\big[ | {\downarrow_1 \uparrow_2 \uparrow_3 } \rangle
+ e^{-i \frac{2\pi}{3}}   | {\uparrow_1 \downarrow_2 \uparrow_3 } \rangle
+ e^{i \frac{2\pi}{3}}   | {\uparrow_1  \uparrow_2 \downarrow_3} \rangle \big],
\nonumber \\
&&|{{s^z= \downarrow},{\tau^z=\uparrow }}\rangle
= \frac{1}{\sqrt{3}}
\big[ | {\uparrow_1 \downarrow_2 \downarrow_3 } \rangle
+ e^{-i \frac{2\pi}{3}}   | {\downarrow_1 \uparrow_2 \downarrow_3 } \rangle
+ e^{i \frac{2\pi}{3}}   | {\downarrow_1  \downarrow_2 \uparrow_3} \rangle \big],
\nonumber \\
&&|{{s^z = \downarrow},{\tau^z =\downarrow }}\rangle
= \frac{1}{\sqrt{3}}
\big[ | {\downarrow_1 \uparrow_2 \uparrow_3 } \rangle
+ e^{-i \frac{2\pi}{3}}   | {\uparrow_1 \downarrow_2 \uparrow_3 } \rangle
+ e^{i \frac{2\pi}{3}}   | {\uparrow_1  \uparrow_2 \downarrow_3} \rangle \big].
\nonumber
\end{eqnarray}
The above four states define the four flavors of the local states for the
local triangular unit if one views the system as the multiflavor Mott insulator 
with the up-triangle cluster as the effective lattice site. 
One then includes the $J'$ interaction between the up-triangular
clusters, and the resulting model is a Kugel-Khomskii model
that is of the following form~\cite{PhysRevLett.81.2356},
\begin{eqnarray}
H_{\text{KK}} &=& \frac{4J'}{9} \sum_{ \langle {\boldsymbol r}{\boldsymbol r}' \rangle }
({\boldsymbol s}_{\boldsymbol r} \cdot {\boldsymbol s}_{\boldsymbol r'})
\big[\frac{1}{2} -     (\alpha_{{\boldsymbol r}{\boldsymbol r}' }^{} \tau^-_{\boldsymbol r}
+  \alpha_{{\boldsymbol r}{\boldsymbol r}' }^{\ast}\tau^+_{\boldsymbol r}   )\big]
\nonumber \\
&&\quad\quad\quad\quad\quad\,\,\,  \times
\big[\frac{1}{2} -     (\beta_{{\boldsymbol r}{\boldsymbol r}'}^{} \tau^-_{\boldsymbol r'}
+  \beta_{{\boldsymbol r}{\boldsymbol r}' }^{\ast}\tau^+_{\boldsymbol r'}   )\big] ,
\end{eqnarray}
where ${\boldsymbol r}$ refers to the center of the up-triangular cluster,
and ${\boldsymbol s}_{\boldsymbol r} $ defines the total spin
on the up-triangular cluster at ${\boldsymbol r}$.
The parameters $\alpha_{{\boldsymbol r}{\boldsymbol r}'}$
and $\beta_{{\boldsymbol r}{\boldsymbol r}'}$ are the
bond-dependent phase factors that are consistent with
the orbital-like nature of the pseudospin ${\boldsymbol{\tau}}$.
It is found that~\cite{PhysRevLett.81.2356},
the factor $\alpha_{{\boldsymbol r}{\boldsymbol r}'}$
 equals to 1, $e^{i4\pi/3}$, or $e^{i2\pi/3}$
when the $J'$-coupled bond connects
two neighboring up-triangles at ${\boldsymbol r}$
and ${\boldsymbol r}'$ from the spin $1,2,3$
on the ${\boldsymbol r}$ triangle, respectively.
 Similarly, the factor $\beta_{{\boldsymbol r}{\boldsymbol r}'}$
 equals to 1, $e^{i4\pi/3}$, or $e^{i2\pi/3}$
when the $J'$-coupled bond connects
two neighboring up-triangles at ${\boldsymbol r}$
and ${\boldsymbol r}'$ to the spin $1,2,3$
on the ${\boldsymbol r}'$ triangle, respectively.
Since the centers of the up-triangular clusters form a triangular lattice,
the emergent Kugel-Khomskii model is then defined on the triangular lattice.

Apart from the breathing kagom\'{e} magnet here, the breathing pyrochlore
magnet can also be understood in a similar fashion. A recent interest of
the breathing pyrochlore magnet is Ba$_3$Yb$_2$Zn$_5$O$_{11}$
that is in the strong breathing limit. Due to the strong spin-orbit
coupling of the Yb $4f$ electrons, the local Hamiltonian on the
smaller tetrahedron is not a simple Heisenberg model. Nevertheless,
the understanding from the irreducible representation of the
tetrahedral group should still be applicable and has already been
applied to the experiments~\cite{dissanayake2021understanding}.
The similar breathing structure occurs in the lacunar spinels~\cite{Kim_2014,PhysRevB.104.224425} 
where many distinct and interesting properties emerge from the 
clusterization of electrons and the effective $J$-moments on the tetrahedra.

\subsection{Rare-earth magnets with weak crystal field}

\begin{figure}[t]
 \includegraphics[width=7.2cm]{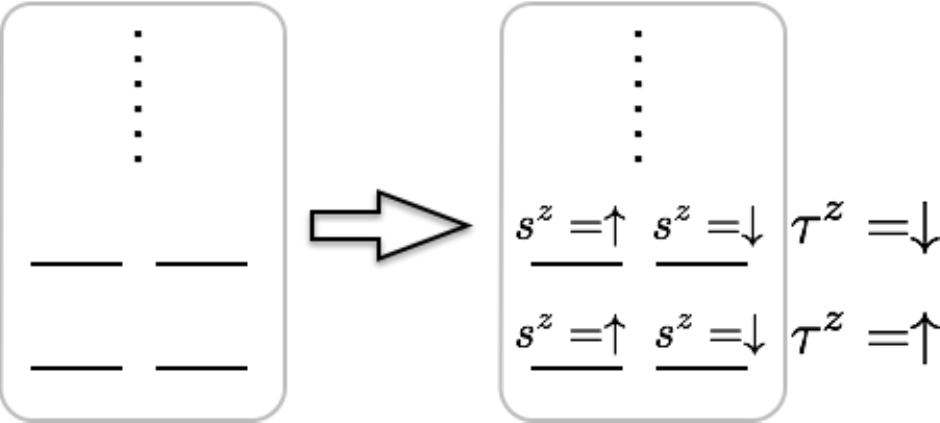}
\caption{The assignment of the effective spin and orbital to the
local crystal field states of rare-earth moments. This applies to
the Tb$^{3+}$ ion in Tb$_2$Ti$_2$O$_7$, Tb$_2$Sn$_2$O$_7$
and others.
The left depicts the crystal field energy levels of the Tb$^{3+}$ ion,
and the right assigns the effective spin and orbital states.
}
\label{fig2}
\end{figure}

It is a bit hard to imagine that the rare-earth magnets
are described by the Kugel-Khomskii model.
Usually, the rare-earth local moments are described
by some effective spin-1/2 degrees of freedom,
and this two-fold degeneracy is protected by time reversal
symmetry and Kramers theorem for the Kramers doublet,
and by the point group symmetry for the non-Kramers doublet.
For the rare-earth local moments, the orbital degrees
of freedom have already been considered from the atomic
spin-orbit coupling that entangles the atomic spin with
the orbitals and leads to total moment ``$J$''. The
effective spin-1/2 doublet arises from the crystal
field ground state levels of the total moment $J$.
The low-energy magnetic physics is often understood
from the interaction between the effective spin-1/2
local moments. This paradigm works rather well
for the pyrochlore rare-earth magnets and the triangular lattice rare-earth 
magnets~\cite{PhysRevB.78.094418,PhysRevB.94.035107,PhysRevLett.112.167203,PhysRevB.83.094411,PhysRevLett.109.167201}.
The reason for the success of the paradigm
is due to the large crystal field gap between
the ground state doublet and the excited ones
in the relevant materials. If this precondition breaks
down, then we need to think about other resolution.

For the Tb$^{3+}$ ion in Tb$_2$Ti$_2$O$_7$ and 
Tb$_2$Sn$_2$O$_7$~\cite{PhysRevLett.98.157204,PhysRevB.62.6496,PhysRevB.90.014429,PhysRevB.87.094410,PhysRevB.84.140402,PhysRevB.89.134410,PhysRevB.85.054428,PhysRevB.76.184436},
it is known that the crystal field energy gap between the ground
state doublet and the first excited doublet is not very large
compared to the Curie-Weiss temperatures in these systems (see Fig.~\ref{fig2}). 
Thus the ground state doublet description is insufficient to capture
the low temperature magnetic properties. This regime is quoted
as ``weak crystal field magnetism'' in Ref.~\onlinecite{PhysRevB.99.224407}.
Both the ground state doublet and the first excited doublet in the left energy diagram 
of Fig.~\ref{fig2}
should be included into the microscopic model, and these low-energy doublets 
contribute to the multiple flavors of local Hilbert space 
in this rare-earth version of the multiflavor Mott insulator. 
To think along the line of the Kugel-Khomskii physics,
we assign the energy levels with the effective spin and the effective 
orbital configurations according to Fig.~\ref{fig2}. 
Here the two effective orbitals are separated
by a crystal field energy gap. The exchange interaction between
the local moments would be of the Kugel-Khomskii-like. 
We expect other rare-earth magnets beyond Tb$_2$Ti$_2$O$_7$
and Tb$_2$Sn$_2$O$_7$ could share a similar physics from the
perspective of Kugel-Khomskii and the multiflavor Mott insulators.

\subsection{${J=3/2}$ Mott insulator}

\begin{figure}[t]
\includegraphics[width=7.2cm]{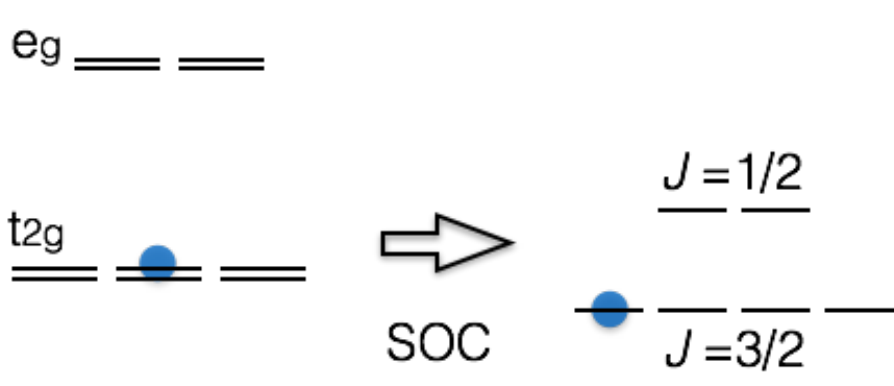}
\caption{The splitting of the {\sl single} electron state under the spin-orbit coupling.
The left is the crystal field scheme in the octahedral environment with the cubic symmetry.
The right is the crystal field scheme under the inclusion of the spin-orbit coupling
(SOC). The energy separation between the ${J=1/2}$ doublet and the ${J=3/2}$
quadruplet is set by the SOC. }
\label{fig3}
\end{figure}

What is ``${J=3/2}$ Mott insulator''? To present this notion,
we begin with the notion of ``${J=1/2}$ Mott insulator'' that
seems to be quite popular in recent
years~\cite{PhysRevB.78.094403,PhysRevLett.105.027204,PhysRevLett.102.017205,PhysRevLett.101.076402}.
 The ${J=1/2}$ Mott insulator
was proposed to be relevant to various iridates,
$\alpha$-RuCl$_3$~\cite{PhysRevB.90.041112},
and even the Co-based $3d$ transition metal
compounds~\cite{PhysRevLett.125.047201,PhysRevB.97.014407,Motome2020,PhysRevB.97.014408,Coldea2021}.
This can be understood from the Ir$^{4+}$ ion under the octahedral
crystal field environment~\cite{PhysRevB.78.094403,PhysRevLett.102.017205}. The $t_{2g}$ and $e_g$
levels for single electron states are splited by a large crystal field gap.
When the spin-orbit coupling (SOC) is switched on,
the $t_{2g}$ orbital is entangled with the spin degree
of freedom, leading to an upper ${J=1/2}$ doublet and
a lower ${J=3/2}$ quadruplet (see Fig.~\ref{fig3}). The Ir$^{4+}$ ion
has a $5d^5$ electronic configuration such that
the lower quadruplet is fully filled and the upper doublet is
half-filled. In the Mott insulating phase, the local
moment is simply described by the spin-orbit-entangled
${J=1/2}$ doublet. The candidate materials are often referred
as ``${J=1/2}$ Mott insulators''.
As the orbital is implicitly involved into the moment, the exchange interaction
inherits the orbital character and depends on the
bond orientations and the components of the moments. Thus,
the exchange interaction is usually not of Heisenberg like.
A consequence of this anisotropic interaction is the Kitaev interaction
that was popular and led to the development of the field of Kitaev materials.
The major advantage of ``${J=1/2}$ Mott insulators'' on the model side
is to provide extra interactions beyond the simple Heisenberg interaction,
and these extra interactions could provide more opportunities to realize
interesting quantum phases and orders.
While it may be fashionable to name these extra interactions as Kitaev interactions
and/or others, we stick to the old convention by Moriya so that the
comparison can be a bit more insightful. The exchange interaction
for
the spin-1/2 operators is always pairwise and quadratic.
According to Moriya~\cite{PhysRev.120.91,TMO},
these interactions are classified as
(symmetric) Heisenberg interaction, (antisymmetric)
Dzyaloshinskii-Moriya interaction, and (symmetric) pseudo-dipole interaction.

The ``${J=3/2}$ Mott insulator'' is realized when one single electron
is or three electrons are placed on the lower $t_{2g}$ quadruplet and
the system becomes Mott insulating~\cite{PhysRevB.82.174440,
PhysRevB.101.054439,PhysRevLett.118.217202,
PhysRevB.104.165150,PhysRevB.102.180401,PhysRevLett.121.097201,Heung2014}
as well as the lacunar spinels in the cluster localization limit~\cite{Kim_2014}.
Despite the popularity of the ``${J=1/2}$ Mott insulators'', the
``${J=3/2}$ Mott insulators'' did not receive much attention.
We do not have any bias towards either of them, and
simply address and explain what the nature could
provide to us. What are the new features of the
``${J=3/2}$ Mott insulators'' from the model side?
First of all, ${J=3/2}$ local moments provide a larger local Hilbert space with more flavors of local states
than ${J=1/2}$ and allow more possibilities for interesting and unusual quantum
phases and orders, thereby making the ``${J=3/2}$ Mott insulator'' an perfect
example of the multiflavor Mott insulator. It was conventionally believed 
that large spin local moments like the spin 3/2 tend to behave more classically. 
This conventional belief,
however, does not really apply to ${J=3/2}$ Mott insulators. 
More specifically,
this would come to our second point below. 
The conventional Heisenberg model does not generate strong quantum
fluctuations as it only changes the spin quantum number by $\pm$1, and
thus the large spin magnets with a simple Heisenberg model usually
behave rather classically. For the ${J=3/2}$ Mott insulators, 
more operators beyond ${J^x, J^y, J^z}$ are generated 
in the superexchange processes and interactions due to 
the inclusion of the orbital degrees of freedom in the
${J=3/2}$ local moments via the SOC. These operators are actually
generators of the SU(4) group that are ``$4\times 4$'' $\Gamma$
matrices. These $\Gamma$ matrix operators allow the system to fluctuate
more effectively between different spin states and generate stronger  
quantum fluctuations. This point is similar to the one that has been
invoked more generally in the introduction section (see Sec.~\ref{sec1}).
Thirdly, similar to the ``${J=1/2}$ Mott insulators'',
the exchange interaction of the ``${J=3/2}$ Mott insulators''
is highly anisotropic and depends on the bond orientation.

Finally, we make a remark that the $\Gamma$ matrix model
for the ``${J=3/2}$ Mott insulators'' can be regarded 
as a parent model for various anisotropic models 
for the effective spin-1/2 local moments. 
This is understood by introducing the single-ion anisotropic
term and reducing the $\Gamma$ matrix model into
the low-energy manifold favored by the single-ion spin term.   
This process may be made an analogy with the models for the
free-electron band structure topology.
The Luttinger model~\cite{PhysRev.97.869,PhysRevB.103.165139,PhysRevX.8.041039,PhysRevResearch.2.023416,PhysRevLett.111.206401,PhysRevLett.119.266402,PhysRevB.99.054505,PhysRevLett.120.057002},
that uses the $\Gamma$ matrices for the $k.p$ theory 
and gives rise to the Luttinger semimetal
with a quadratic band touching in three dimensions, can be regarded as
a parent model for generating other models for
three-dimensional (3D) topological insulators and 3D Weyl semimetal
 upon introducing strains and magnetism.

\begin{figure}[t]
\includegraphics[width=8.4cm]{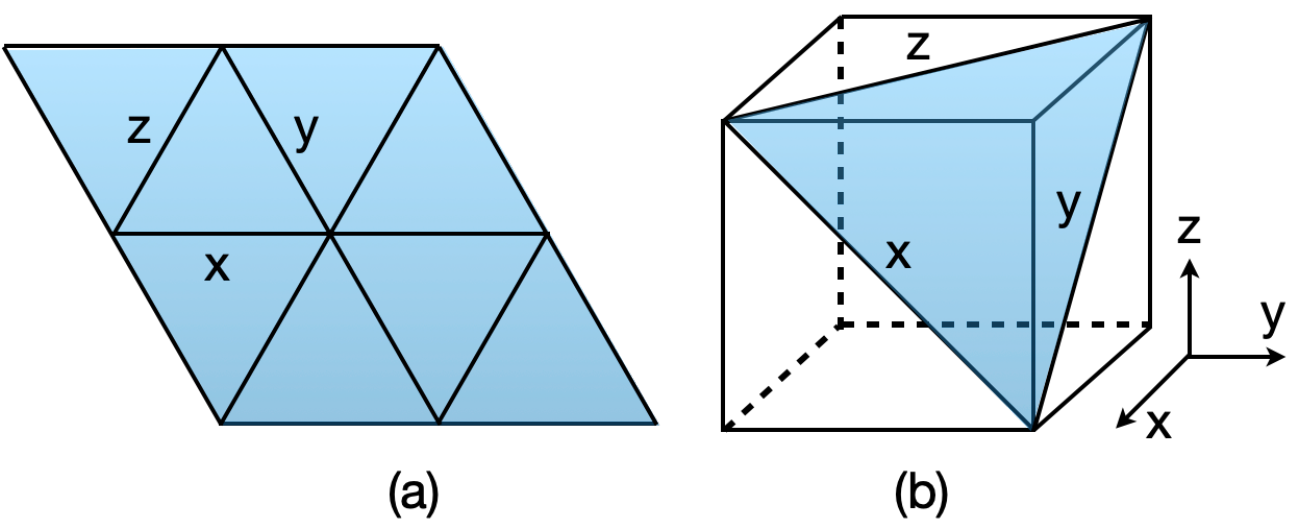}
\caption{(a) The triangular lattice with ``$x,y,z$''
bond assignment for the nearest neighbors.
(b) The [111] layer of the transition metal oxide interfaces.
The coordinate system defines the coordinate system for the spin 
and the orbitals. 
}
\label{fig4}
\end{figure}

In the next section, we will focus on the ${J=3/2}$ Mott insulator
on the triangular lattice structure and explicitly derive the
Kugel-Khomskii model for this multiflavor Mott insulator.

\subsection{Some general remarks for and beyond the emergent Kugel-Khomskii physics} 
\label{sec3d}

%%%%%%%%%%%%%%%%%%%%%%%%%%%%%%%%%%%%%%%%

Before we expand the discussion on the ${J=3/2}$ Mott insulator, we 
here give a few remarks based on the emergent Kugel-Khomskii physics
for the multiflavor Mott insulators. The identification and separation of 
the effective spin and orbital degrees of freedom in these systems 
already indicates the distinct physical properties of these effective degrees 
of freedom. For instance, the effective spin is odd under time reversal 
and is usually related to the magnetic moment, while the effective
orbital is often even under time reversal. In the case of the ordering
phenomena, the effective spin and the effective orbital do not have 
to order at the same time~\cite{PhysRevLett.92.167201,PhysRevLett.89.227203,Khomskii_2003}, 
which is already feasible from the Ginzburg-Landau sense~\cite{PhysRevB.82.174440}.  
Thus, one could have separate ordering transitions 
with different transition temperatures for the effective spin and orbital, respectively.

Due to the discrete lattice symmetry, the orbital ordering is often associated
with the change of the crystal structure, leading to the lattice distortion and the phonon anomaly~\cite{PhysRevLett.89.227203,Khomskii_2003}. 
This can be a useful experimental diagnosis of the effective orbital ordering. 
Like the traditional Kugel-Khomskii physics, the orbital order would reconstruct the 
spin model, the orientations of the orbital ordering then give rise to the spatially-anisotropic 
spin models that are often quite rare for the conventional magnets~\cite{PhysRevLett.101.266405,PhysRevLett.126.106401,Lee_2011}. 
For example, the quasi-one-dimensional Heisenberg chain physics in KCuF$_3$ was 
expected to arise from the orbital order that enhances the spin coupling along one lattice direction~\cite{PhysRevLett.101.266405}.
Such dimensional reduction is particularly useful to link the low-dimensional physics with the high-dimensional models and systems. 
The one-dimensional spin-1 Haldane-chain physics (and the symmetry protected topological order) was invoked 
for FeSe upon nematic or orbital ordering~\cite{Wang_2015} and certain spin-1 honeycomb lattice Kugel-Khomskii model~\cite{Savarync}.

Even in the ordered phases, the elementary excitations for these multiflavor Mott insulators are fundamentally different from 
the conventional magnets in the large-$S$ limit. 
The latter is the simple spin-wave excitation. In contrast, the former 
is characterized by many branches of spin-wave-like excitations. 
For the conventional magnets with the large-$S$ moment, the simple 
quadratic spin interaction merely creates the magnon excitation with 
the $\pm 1$ change of the spin quantum number and thus supports 
one spin-wave branch for each magnetic sublattice. 
In the multiflavor Mott insulators, the effective model 
could excite the system from the ordered flavor of state to all other flavor of states  
 if the system is simply ordered to one flavor of state~\cite{PhysRevB.82.174440}.  
 For the emergent Kugel-Khomskii model with the spin-1/2 and the pseudospin-1/2 moments, there would be 
 three branches of spin-wave-like excitations for each sublattice~\cite{PhysRevB.60.6584}. 
 These excitations
 are sometimes referred as the flavor-wave excitations.

It was previously argued that, the more exotic quantum states could be stabilized by the enhanced quantum fluctuations 
due to the enlarged local Hilbert space and the involved multiflavor interactions for the multiflavor Mott insulators. 
It has been shown that, the flavor ordered stripe state~\cite{PhysRevLett.105.265301}, 
the valence bond crystal~\cite{PhysRevB.84.174441,PhysRevLett.113.127204,PhysRevLett.107.215301}, 
Dirac spin liquid~\cite{PhysRevLett.121.097201,yao2021intertwining}, 
spinon Fermi surface state~\cite{PhysRevA.93.061601} and 
chiral spin liquids~\cite{Hermele2009,PhysRevB.84.174441,PhysRevLett.113.127204,yao2021intertwining,PhysRevResearch.3.023138} 
could be stabilized by the enhanced quantum fluctuations and the large symmetry group like SU($N$).
The candidate systems are 
 Ba$_3$CuSb$_2$O$_9$~\cite{PhysRevB.90.094422,Katayama_2015,PhysRevB.96.115116}, 
 the ${J=3/2}$ Mott insulators like ZrCl$_3$~\cite{PhysRevLett.121.097201,PhysRevB.104.224436}, 
 the double moir\'{e} layers 
 with transition metal dichalcogenides or graphenes, as well as the ultracold atom contexts with the more exact symmetries
 that will be discussed in Sec.~\ref{sec5}. The double moir\'{e} layers 
 with transition metal dichalcogenides or graphenes have the enlarged symmetry like SU(4) or SU(8) from their
 spin, valley and layer freedom~\cite{PhysRevResearch.3.023138,Zhang_2021,zhang2020electrical}. 
 Some of the results for the multiflavor Mott insulator of 
 this context overlap with the ones from the ultracold atoms,
 and we do not give much discussion here. 
 
 In general, the multiflavor Mott insulator provides a feasible platform to study the exotic quantum states. 
To manipulate and probe these exotic quantum states and excitations, 
one necessarily has to understand the role of different degrees of freedom and how they are coupled
to the external probes such as electric field, magnetic field, strains, electric currents
and so on. The identification and separation of the degrees of freedom becomes much more meaningful 
for this practical purpose. 
For example, the selective coupling with the spin/magnetic excitations instead of the orbital modes or non-magnetic 
excitations in the neutron scattering and NMR measurements is crucial to understand the excitation structures 
for the conventional Kugel-Khomskii systems and other multiflavor Mott insulators~\cite{PhysRevB.98.045119}. 
The presence of the layer degrees of freedom in the multiflavor Mott insulators
for the double moir\'{e} layers with transition metal dichalcogenides or graphenes
allows the design of dragging type of transport measurements for the interlayer 
transport behaviors in the exotic quantum phases~\cite{zhang2020electrical}.

\section{Emergent Kugel-Khomskii physics in ${J=3/2}$ Mott insulator}
\label{sec4}

Even though the ${J=3/2}$ Mott insulator widely exists in many materials,
we here consider a ${J=3/2}$ Mott insulator in the triangular lattice for the
specific demonstration purpose. 
%This is expected to be relevant for the material ZrCl$_3$ where the Zr$^{3+}$ ions form a honeycomb lattice~\cite{PhysRevLett.121.097201}. 
It is relevant for the [111] interface of the transition metal oxide heterostructure. 
As we have explained in the previous section, the requirement 
for the magnetic ions is to have a $4d^1$ or $5d^1$ electron 
configuration in an octahedral environment where the SOC is active.
To build up a physical model for this triangular lattice 
${J=3/2}$ Mott insulator, we first specify the degree of freedom. 
Before including the effect of the SOC, the local moment
is described by a localized spin-1/2 electron on the three-fold
degenerate $t_{2g}$ orbitals at each magnetic ion site.
The interaction between the local moments is understood
as a standard Kugel-Khomskii superexchange model
for the $t_{2g}$ systems. Including the atomic SOC,
we express the full model as
\begin{eqnarray}
H = H_{\text{KK}} + H_{\text{SOC}},
\label{eq3}
\end{eqnarray}
where the first term describes the standard Kugel-Khomskii superexchange interaction,
and the second term is the on-site atomic SOC.
For the $z$ bond in Fig.~\ref{fig4}(a), the
superexchange interaction has the following form,
\begin{eqnarray}
H_{\text{KK}}^z = \sum_{\langle ij \rangle}
J\, \Big( {\boldsymbol S}_{i,xy} \cdot  {\boldsymbol S}_{j,xy}
 + \frac{1}{4} n_{i,xy} n_{j,xy} \Big),
 \label{eqsev}
\end{eqnarray}
where ${\bf S}_{i,xy}$ defines the electron spin on the $xy$ orbital
with ${{\bf S}_{i,xy} = {\bf S}_i \, n_{i,xy}}$, and
$n_{i,xy}$ defines the electron occupation number on the $xy$ orbital.
As the $xy$ orbital gives a dominant $\sigma$-bonding for the $z$ bond, 
the superexchange interaction is primarily given by the exchange process along this $\sigma$-bonding.
The electron hoppings to other orbitals or between other orbitals are expected to be 
quite weak and are neglected here. 
To keep things simple, the other Kanamori parameters such as the Hund's coupling and the pair hopping
are further neglected as they are comparably weaker than the Hubbard interaction.
These approximations result in the primary contribution in the above equation for our demonstration
purpose.
The exchange interactions on the $x$ and $y$ bonds can be written down from
a simple cubic permutation.
Interestingly, the Kugel-Khomskii superexchange interaction for the disentangled spin-orbital states in 
Eq.~\eqref{eqsev} is not quite related to the emergent Kugel-Khomskii physics 
for the spin-orbit-entangled ${J=3/2}$ quadruplets that is shown below.

The atomic SOC has the expression,
\begin{eqnarray}
H_{\text{SOC}} = \sum_i -\lambda\, {\boldsymbol l}_i \cdot {\boldsymbol S}_i,
\end{eqnarray}
where ${\boldsymbol l}_i$ is the effective orbital angular
momentum for the $t_{2g}$ orbital with ${l=1}$, and
${\boldsymbol S}_i$ is the spin-1/2 operator for the
localized electron.
As it is shown in Fig.~\ref{fig3}, the $H_{\text{SOC}}$ term splits the 6-fold spin and orbital degeneracy of the
single-electron states into the upper doublet with ${J=1/2}$ and the lower quadruplets with ${J=3/2}$.
To demonstrate how the Kugel-Khomskii physics emerges in this context,
we need to compare the energy scales of SO coupling $\lambda$ and the
superexchange $J$.
 In many $4d$/$5d$ systems, the atomic SOC is
often of similar energy scales as the electron
bandwidth and the electron correlation.
In the Mott insulating regime, however, what is available for
a meaningful comparison is the superexchange coupling.
This coupling is usually much weaker than the
atomic SOC for the $4d$/$5d$ materials.
In the sense of perturbation theory, the atomic SOC
is treated as the main Hamiltonian and the superexchange
is regarded as the perturbative term. For the ${J=3/2}$ 
Mott insulator with one electron per site, the atomic SOC
entangles the orbital angular momentum ${\boldsymbol l}$
with the spin ${\boldsymbol S}$, leading
to a ${J=3/2}$ quadruplet for each site. The superexchange
interaction then operates on the degenerate manifold
of the ${J=3/2}$ local moments.

Following the spirit of the degenerate perturbation theory, we project 
the superexchange model on the degenerate manifold,
\begin{eqnarray}
H_{\text{eff}}  =\prod_i   \sum_{m_i }
|  m_i^{}   \rangle \langle    m_i^{} |  \cdot
 H_{\text{KK}} \cdot  \prod_j
  \sum_{m_j} |   m_j^{} \rangle \langle  m_j^{} |  ,
\end{eqnarray}
where $m_i$ ($m_j$) is the quantum number of the $J^z_i$ ($J^z_j$) operator and takes
the values of $\pm 1/2, \pm3/2$. As we explained
in the previous section, the effective Hamiltonian
can be expressed into a quadratic form in terms of the
$\Gamma$ matrices at each site. The $\Gamma$-matrix
expression, however, hides the original physical meaning.
In the following, we first express the effective model in the
$J$-basis, and then explain the emergent Kugel-Khomskii
physics by expressing it into the form of Kugel-Khomskii
interaction. In terms of the ${\boldsymbol J}$ operators,
the effective model has the expression, 
\begin{eqnarray}
H_{\text{eff}}^z = \sum_{\langle ij \rangle} J \,
\Big(\tilde{\boldsymbol S}_{i,xy} \cdot \tilde{\boldsymbol S}_{j,xy}
+ \frac{1}{4}  \tilde{n}_{i,xy} \cdot \tilde{n}_{j,xy} ) ,
\end{eqnarray}
where we have
\begin{eqnarray}
 \tilde{S}_{i,xy}^x &=&  P_{\frac{3}{2}} \, S_{i,xy} ^x \, P_{\frac{3}{2}} = \frac{J^x_i}{4}
 -\frac{J^z_i J^x_i J^z_i}{3}, \\
 \tilde{S}_{i,xy}^y &=& P_{\frac{3}{2}} \, S_{i,xy} ^y \, P_{\frac{3}{2}}
 = \frac{J^y_i}{4} - \frac{J^z_i J^y_i J^z_i}{3} , \\
 \tilde{S}_{i,xy}^z &=& P_{\frac{3}{2}} \, S_{i,xy} ^z \, P_{\frac{3}{2}}
 = \frac{3J^z_i}{4} - \frac{J^z_i J^z_i J^z_i}{3} , \\
 \tilde{n}_{i,xy}     &=& P_{\frac{3}{2}} \, n_{i,xy}  \,  P_{\frac{3}{2}}
 = \frac{3}{4} - \frac{(J^z_i)^2}{3}   .
\end{eqnarray}

Unlike the simple Heisenberg model that only involves the linear spin operators,
the effective model involves the spin products with two or three
``${\boldsymbol J}$'' operators. These operators are high order magnetic
multipolar moments and are able to switch the local spin state from one $J$
state to any other states, and thus quantum fluctuations are strongly enhanced.
In terms of the ``${\boldsymbol J}$'' operators, the effective model is
rather difficult to be tackled with, and the conventional wisdom cannot
provide more physical intuition.  Instead, we turn to the perspective of the emergent
Kugel-Khomskii physics where the previous knowledge and theoretical techniques about the Kugel-Khomskii model can be adapted~\cite{TMO,khomskii2014,PhysRevLett.91.087206,PhysRevB.56.R14243,PTPS.160.155,PhysRevLett.93.077208}. 
For this purpose,
we merely need to show the Kugel-Khomskii structure and
reduce the effective model into the standard Kugel-Khomskii form.

We first make the following mapping between
the effective spin states and the
fictitious spin and orbital states,
\begin{eqnarray}
|J^z_i = + \frac{3}{2}\rangle  & \equiv & | s^z_i=+\frac{1}{2}, \tau^z_i =+ \frac{1}{2} \rangle , \\
|J^z_i = + \frac{1}{2}\rangle  & \equiv & | s^z_i=+\frac{1}{2}, \tau^z_i =- \frac{1}{2}  \rangle ,
\label{eq13} \\
|J^z_i = - \frac{1}{2} \rangle  & \equiv & |  s^z_i=-\frac{1}{2}, \tau^z_i =- \frac{1}{2} \rangle ,
\label{eq14} \\
|J^z_i = - \frac{3}{2}\rangle  & \equiv & | s^z_i=-\frac{1}{2}, \tau^z_i =+ \frac{1}{2}  \rangle .
\label{eq15}
\end{eqnarray}
To distinguish from the physical spin ``${\boldsymbol S}$'',
we use the little ``$s$'' to label the fictitious spin, and use
``$\tau$'' to label the fictitious orbital. Although $\tau$ is labeled
as the ``orbital'', the transformation under the time reversal
differs from the usual orbital degree of freedom.
This is because ${{\mathcal T} | J_i^z = m \rangle
= i^{2m} | J_i^z = - m \rangle }$ and ${\mathcal T}$
does not modify the orbital degree of freedom for the usual
real orbital wavefunctions.
Here, the ${| J_i^z = 3/2 \rangle}$ and ${| J_i^z = 1/2 \rangle}$
have a different sign structure under the time reversal operation.
This is the case even if we switch the assignment in
Eq.~\eqref{eq14} and Eq.~\eqref{eq15}.
With the assignment in the above equations, the original ${\boldsymbol J}$
related operators can be expressed as
\begin{eqnarray}
 \tilde{S}_{i,xy}^x &=&  \frac{1}{3} s_i^x \, (1-2 \tau^z_i)  , \\
 \tilde{S}_{i,xy}^y &=&  \frac{1}{3} s_i^y \, (1-2 \tau^z_i), \\
 \tilde{S}_{i,xy}^z &=&   \frac{1}{3} s_i^z\, (1-2 \tau^z_i), \\
 \tilde{n}_{i,xy}     &=&  \frac{1}{3} (1-2 \tau^z_i).
\end{eqnarray}

Collecting all the interactions, we obtain the emergent Kugel-Khomskii model,
and the interaction on the other bonds can be generated likewise. This interaction
now carries the basic features of the conventional Kugel-Khomskii model with the following expression,
\begin{eqnarray}
H_{\text{eff}}^z &=& \sum_{\langle ij \rangle} \frac{4J}{9} \Big({\boldsymbol s}_i \cdot {\boldsymbol s}_j
+ \frac{1}{4} \Big)
\Big(\frac{1}{2} - \tau_i^z\Big)\Big(\frac{1}{2} - \tau_j^z \Big).
\end{eqnarray}
The couplings on the remaining bonds can be obtained 
by the symmetry transformation. The understanding from the Kugel-Khomskii 
physics that was discussed in Sec.~\ref{sec3d} can then be applied. 
As the model from the ${J=3/2}$ Mott insulators can be regarded as the parent model
for many anisotropic spin-1/2 models, one can show that, with the strong easy-plane single-ion
anisotropy like $+D (J^x_i +J^y_i + J^z_i)^2$, the effective spin-1/2 model 
carries a Kitaev-like interaction in addition to other anisotropic interactions.

%%%%%%%%%%%%%%%%%%%%%%%%%%%%%%%%%%%%%%%%%%%%%%%%%%%%%%%%%%
\section{SU($N$) and Sp($N$) magnetism with ultra-cold fermions}
\label{sec5}

Fermion systems with large internal degrees of freedom not only exist in quantum material as multiflavor Mott insulators, 
but also include large-spin alkali and alkaline-earth fermions.
These atoms often exhibit large hyperfine spins, whose ${2S+1}$ hyperfine atomic levels certainly form a high representation of the SU(2) group.
As we have mentioned in the introduction (see Sec.~\ref{sec1}), theoretically it was proposed to study such systems from the perspective of
 {{\sl exact}} high symmetries such as SU($N$) and Sp($N$) \cite{wu2003,wu2006a,Gorshkov2010}, 
  in which the high symmetries greatly enhance the connectivity among different internal components. 
Consequently, they share very similar features in quantum magnetism in their Mott-insulating states to the  
Kugel-Khomskii systems despite the huge difference between their energy scales
\cite{wu2003,wu2006,chen2005,wu2005a,controzzi2006, xu2008,
cazalilla2009,Hermele2009,PhysRevResearch.3.023138,Gorshkov2010,wu2010,hung2011,wu2012}.
{The unprecedented exact high symmetries of the ultracold alkaline and alkaline-earth atom systems
allow us to access many unconventional theoretical models and 
quantum phases that are a bit difficult to realize 
in quantum materials and only occur under the theoretical idealization. 
Unlike the previous discussion about the multiflavor Mott insulators
and the Kugel-Khomskii physics in quantum materials, our review below will explore the models and the 
consequences from the high symmetries
in the Mott insulating regime for the ultracold atoms. }

For the alkaline-earth atoms, their hyperfine-spins completely come from the nuclear spins since their atomic shells are filled.
Interactions from atomic scatterings are independent from spin components since the nuclei are deeply inside.
This is the microscopic reason of their SU($N$) interactions.
Nevertheless, for the alkali high spin atoms, generally speaking, their
interactions are spin-dependent.
In this case, the SU($N$) symmetry is not generic, and the Sp($N$)
symmetry is the next highest ~\cite{wu2003,wu2006,wu2010,wu2012,cazalilla2009}.
As a concrete example, below we will use the simplest case of large spin ultra-cold fermions, i.e., spin-${3}/{2}$ fermions
(e.g. $^{132}$Cs, $^9$Be, $^{135}$Ba, $^{137}$Ba, $^{201}$Hg).
It can be proved that such systems possess an exact and
generic symmetry of Sp(4), or, isomorphically SO(5).
Various many-body orders including quantum magnetism in different
tensor channels, pairing and density-wave orders 
\cite{wu2003,wu2006}.
The generic Sp(4) symmetry here plays the role of the SU(2) symmetry in the
spin-${1}/{2}$ systems.

 Such a high symmetry group of the ultracold atom systems 
requires a fine-tuning of the parameters
for the Kugel-Khomskii model in quantum materials. For instance, the symmetry can be tuned to
SU(4) for the $e_g$ orbitals. 
In this limit, the Kugel-Khomskii model in the quantum materials is connected to 
the SU($N=4$) model in the ultracold atom context in the narrow sense of the model
relation.  
For the $t_{2g}$ orbitals in quantum materials, more complicated 
operators are involved in the Kugel-Khomskii model~\cite{kugel1982}, and less symmetries are explicitly
present. 
On the contrary, there were efforts~\cite{Gorshkov2010} 
that introduce two atomic orbitals for the alkaline
earth atoms such that the resulting model is the SU($N>2$) extension 
of the original the Kugel-Khomskii model where the SU(2) spin is replaced by the
 SU($N>2$) spin while the orbital sector of the electrons is replaced by the atomic orbitals 
on the optical lattices.

We also review the ``baryon-like" quantum magnetism.
In quantum chromodynamics (QCD), quarks form the fundamental representation of the SU(3) group.
Three quarks of all the $R$, $G$, $B$ components form a baryon,
a color singlet, while, two quarks cannot form a color singlet.
Similarly, the SU($N$) fermion systems are characterized
by the $N$-fermion correlations, which leads to the muli-fermion SU($N$) singlet clustering ordering \cite{wu2005a,rapp2007,lecheminant2008}.
Similarly, the physics of an SU($N$) quantum magnet can also be
beyond the SU(2) case which is often studied condensed matter systems.
In spite of the huge difference of energy scales, the large-spin cold fermions 
could exhibit similar physics to that in QCD \cite{bossche2000,mishra2002,chen2005,xu2008}.

 %--------------------------------------
\subsection{The spin-{3}/{2} Hubbard model - the generic Sp(4) symmetry}

We generalize the spin-{1}/{2} Hubbard model to spin-{3}/{2} case. 
Only the standard spin SU(2) symmetry is assumed at the beginning, and the hidden
Sp(4) symmetry will be explained. 
The free-fermion part $H_0$ reads,
\bea
H_0&=&-t \sum_{\langle ij \rangle, \sigma} \left(\psi^{\dag}_{i \sigma}
\psi_{j \sigma}+h.c.\right)-\mu \sum_{i} n(i),
\label{eq:sp32h0}
\eea
where $\psi_\sigma$ is the 4-component fermion spinor operator and
$\sigma$ represents the eigenvalues of $s_z$ with
${\sigma=\pm {3}/{2}, \pm {1}/{2}}$;
$\langle ij \rangle$ denotes the nearest neighboring bonding;
${n=\psi^{\dag}_{\sigma} \psi_{\sigma}}$ is the particle-number operator.
If two spin-${3}/{2}$ fermions are put on the same site, their total spin is either $0$ (singlet) or 2 (quintet).
Then the onsite interactions are represented in the spin SU(2) language as
\begin{equation}
H_{int}=U_0 \sum_i P^{\dag}_{0}(i)P_{0}^{}(i)
+U_2 \sum_{i,-2\le m\le 2}P^{\dag}_{2m}(i)P_{2m}^{}(i) ,
\label{eq:sp32hint}
\end{equation}
where $P^{\dag}_{0}$ and $P^{\dag}_{2,m}$ are the pairing operators
in the singlet and quintet channels, respectively.
Their interaction strengths are denoted as $U_0$ and $U_2$,
respectively.

In fact, the Hamiltonian of Eqs.~\eqref{eq:sp32h0}) plus ~\eqref{eq:sp32hint}) 
possesses an Sp(4) symmetry, which is the double covering group of SO(5).
 For simplicity, we will use Sp(4) and SO(5) interchangeably below.
The Sp(4) algebra is constructed below.
In addition to spin and charge, 
spin quadrupole and octupole operators are necessary 
in the spin-${3}/{2}$ system.
As shown in Appendix \ref{sect:gamma}, the five spin 
quadrupole matrices are actually the Dirac  $\Gamma$-matrices
denoted as $\Gamma^a (1\le a\le 5)$, and their commutators
are defined as $\Gamma^{ab}=\frac{i}{2}[\Gamma^a, \Gamma^b] 
(1\le a< b \le 5)$.
The spin-quadrupole operators $n_a$ are defined as
\bea
n_a=\frac{1}{2}\psi^\dagger_\alpha \Gamma^a_{\alpha\beta}
\psi_\beta^{}, ~~~ (1\le a\le 5).
\label{eq:generators1}
\eea
On the other hand, spin and spin octupole operators, in
total 10 operators, can be reorganized as
\cite{wu2003,wu2006a}
\bea
L_{ab} = -\frac{1}{2}\psi^\dagger_\alpha
\Gamma^{ab}_{\alpha\beta} \psi_\beta.
\label{eq:generators2}
\eea

$L_{ab}$'s span the Sp(4), or, isomorphically, SO(5) group, and $n_a$'s transform as a 5-vector of SO(5). 
$L_{ab}$ and $n_a$ together span the algebra of the SU(4)
group, which is isomorphic to SO(6) group. Hence, Sp(4) is a subgroup of SU(4).
Among them, three diagonal operators commute with each other:
\bea
L_{15}&=&
\frac{1}{2}(n_{\frac{3}{2}}+n_{\frac{1}{2}}-n_{-\frac{1}{2}}
-n_{-\frac{3}{2}}),  \nn \\
L_{23}&=& \frac{1}{2}(n_{\frac{3}{2}}-n_{\frac{1}{2}}
+n_{-\frac{1}{2}}-n_{-\frac{3}{2}}),  \nn \\
n_4&=&\frac{1}{2}(n_{\frac{3}{2}}-n_{\frac{1}{2}}-n_{-\frac{1}{2}}
+n_{-\frac{3}{2}}).
\eea 

Now it is ready to reveal the hidden Sp(4) symmetry.
$\psi_\alpha$ is a 4-component spinor of SU(4) and $n$ 
is an SU(4) scalar. Hence, $H_0$ is in fact SU(4) invariant, 
of course, Sp(4) invariant.
It is enlightening to formulate Eq.~\eqref{eq:sp32hint} as
\bea
H_{int}=\frac{3U_0+5U_2}{16} \sum_{i} \left[n(i)-2 \right]^2
+\frac{U_0-U_2}{4} \sum_{i,1\le a\le 5}n^2_{a}(i). \ \ \, \ \ \
~~ \label{eq:hubbard_so5}
\eea
Since the $n^a$ operators form a 5-vector, Eq.~\eqref{eq:hubbard_so5}
is also Sp(4) invariant.

%-----------------------------------------------------------
\subsection{The superexchanges at quarter-filling}

Under sufficiently strong repulsive interactions,  spin-${3}/{2}$
Hubbard model will enter the  Mott-insulating state even at ${1}/{4}$-filling, i.e., one fermion per site.
In principle, the magnetic superexchange interaction can be expressed
in terms of the bi-linear, bi-quadratic and bi-cubic SU(2) Heisenberg
terms of $\mathbf S_i \cdot \mathbf S_j$,
$(\mathbf S_i \cdot \mathbf S_j)^2$,
$(\mathbf S_i \cdot \mathbf S_j)^3$, respectively.
Nevertheless, the superexchange exists in the bond-spin singlet
and quintet channels, respectively.
Hence, only two of these three terms are independent, which
corresponds to the Sp(4) symmetric case.

The above situation is similar to the spin-1 Heisenberg model with
both bi-linear and bi-quadratic terms.
Its general form is controlled by the parameter angle $\theta$,
\bea
H_{spin 1}=J \sum_{\avg{ij}}  \left( \cos\theta \, (\mathbf{S}_i \cdot
\mathbf{S}_j) +\sin \theta \, (\mathbf{S}_i \cdot \mathbf{S}_j)^2
\right).
\label{eq:biquad}
\eea
Eq.~\eqref{eq:biquad} exhibits two different types of SU(3) symmetries,
a ``uniform" one and a ``staggered" one.
The ``uniform" one appears at $\theta=\pi/4$ and $5\pi/4$, i.e., every site lies in the fundamental representation.
The ``staggered" one appears at $\theta=\pi/{2}$ and ${3}\pi/{2} $,
in which two sublattices lie in the fundamental and anti-fundamental representations, respectively.
They are explained in Appendix \ref{sect:append}.

It is enlightening to express the superexchange model of spin-${3}/{2}$ fermions 
in the explicitly Sp(4) invariant form~\cite{chen2005},
\bea
\label{eq:spinchain_so5}
H_{ex}=\sum_{\langle ij \rangle} \Big\{
J_L \sum_{1\le a < b \le 5}
L_{ab}(i)L_{ab}(j)+ J_N \sum^5_{a=1} n_a(i)n_a(j) \Big \}.~~
\eea
Each site lie in the fundamental representation of Sp(4),
i.e., a 4-component spinor.
${J_L=(J_0+J_2)/4}$ and ${J_N=(3J_2-J_0)/4}$, and
$J_0$ and $J_2$ are exchange strength in the singlet
and quintet channels, respectively.
At the level of the 2nd order perturbation theory, they are
\bea
J_0=4t^2/U_0, \ \ \, J_2=4t^2/U_2.
\eea

Sp(4) is a rank-2 Lie algebra.
Hence, it has three good quantum numbers, i.e., one more
compared to those of SU(2).
They are defined as
\bea
&&C=\sum_{1\le a<b \le 5}\Big\{\sum_{i} L_{ab} (i) \Big \}^2,  \\
&&L^{tot}_{15}=\sum_i L_{15}(i), \ \ \,
L^{tot}_{23}=\sum_i L_{23}(i).
\eea
$C$ is the Sp(4) Casimir in analogy to total spin square of
an SU(2) system;
$L^{tot}_{15}$ and $L^{tot}_{23}$ are the analogies
to the total $S_z$.

\subsection{The uniform and staggered SU(4) symmetries}

Similar to the spin-1 model of Eq.~\eqref{eq:biquad}, Eq.~\eqref{eq:spinchain_so5}  
exhibits two different types of
SU(4) symmetries.
The first case takes place at ${J_0=J_2}$, or, equivalently, ${U_0=U_2}$,
denoted as SU(4)$_A$ below. Eq.~\eqref{eq:spinchain_so5} is reduced
to the SU(4) Heisenberg model
\bea
H_A=J \sum_{\langle i,j \rangle} \Big\{
L_{ab}(i)L_{ab}(j)+n_a(i)n_a(j) \Big \},
\eea
for which each site lies in the fundamental representation
of SU(4), and ${J=J_0/2=J_2/2}$.
$H_A$ is equivalent to the SU(4) Kugel-Khomskii type model \cite{kugel1982,sutherland1975,li1998,bossche2000,bossche2001}
as reviewed in the previous sections, which is similar to the
``uniform" case of SU(3).

The second SU(4) symmetry is similar to the ``staggered" case of
SU(3), which takes place in a bipartite lattice and
in the limit of ${U_2\rightarrow +\infty}$, i.e., ${J_2=0}$.
To see this explicitly, the particle-hole transformation
${\psi_\alpha \rightarrow R_{\alpha\beta} \psi_\beta^\dagger}$
is performed to one sublattice, and the other sublattice is left
unchanged, where $R$ is the charge conjugation matrix
\bea
R= \left (
\begin{array} {cc}
0 &i\sigma_2 \\
i\sigma_2 & 0 \\
\end{array}
\right ).
\label{eq:R4}
\eea
Under this operation, the fundamental representation of SU(4) transforms
to its anti-fundamental representation whose Sp(4) generators and vectors
become ${L'_{ab}=L_{ab}}$ and ${n'_{a}=-n_{a}}$.
The Sp(4) generators remain invariant under
the particle-hole transformation is due to its pseudo-reality.
Then Eq.~\eqref{eq:spinchain_so5} can be recast to
\bea
H_B=J^\prime \sum_{\langle i,j\rangle}
\left\{ L'_{ab}(i)L_{ab}^{}(j)
+n'_a(i)n^{}_a(j) \right \},
\label{eq:SU4B}
\eea
where $J^\prime=J_0/4$.
Eq.~\eqref{eq:SU4B} is SU(4) invariant again.

%-----------------------------------------------------------------
\begin{figure}[t]
\includegraphics[width=0.6\linewidth]{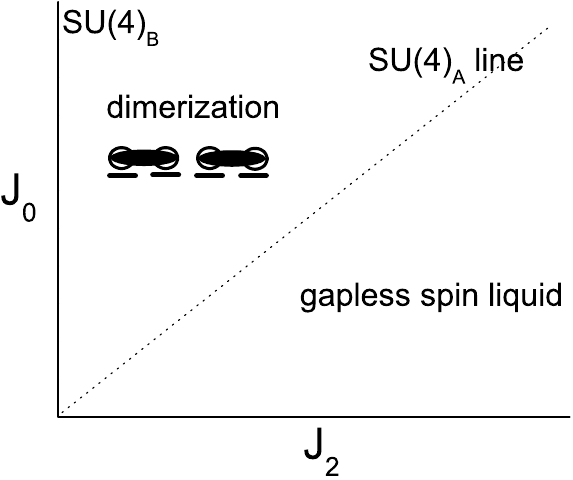}
\caption{Phase diagram of the 1D Sp(4) spin chain in terms of $J_0$
and $J_2$ with ${\tan\theta=J_0/J_2}$.
The SU(4)$_A$ and SU(4)$_B$ symmetries are located along the lines of
${\theta=45^\circ}$ and $90^\circ$, respectively.
The dimerized spin Peierls phase appears at ${90^\circ\ge\theta>45^\circ}$,
and the gapless spin liquid phase is located at ${45^\circ\ge \theta\ge 0^\circ}$.
Figure is adapted from Ref.~\onlinecite{hung2011}.
} \label{fig:phase}
\end{figure}
%-------------------------------------------------------------------

These two SU(4) symmetries imply fundamentally different physics.
Both of them have high energy analogies:
The ``uniform" SU(4)$_A$ is baryon-like, while that of SU(4)$_B$
is meson-like.
In the former case, every site belongs to the fundamental representation.
At least 4 sites are required to form an SU(4) singlet
with the wavefunction represented by
\bea
\frac{1}{\sqrt{4!}}\epsilon_{\alpha\beta\gamma\delta}
\psi^\dagger_\alpha(1)\psi^\dagger_\beta(2)
\psi^\dagger_\gamma(3)\psi^\dagger_\delta(4) |\Omega\rangle,
\eea
where
$1\sim 4$ are site indices,
$\epsilon_{\alpha\beta\gamma\delta}$ is the rank-4 fully antisymmetric
tensor, and $\Omega$ is the vacuum.
Hence, dramatically different from the SU(2) case, two sites across a
bond cannot form a singlet.
Quantum magnetism based on $H_A$ exhibits strong features
of the 4-site correlation, i.e., a baryon-like state.

For the case of SU(4)$_B$, two sites are sufficient to
form an SU(4) singlet via the charge conjugation matrix as
\bea
\frac{1}{2} R_{\alpha\beta}\psi^\dagger_\alpha(1) \psi^\dagger_\beta(2)|\Omega \rangle.
\eea
This can be viewed as a large-$N$ version of the usual spin-$1/2$
SU(2) Heisenberg model.

%----------------------------------------------------------
\subsection{Quantum magnetism of an Sp(4) spin chain }

The Sp(4) quantum magnetism occurs in the strong repulsive
interaction regime at ${1}/{4}$-filling, i.e., one particle per site.
For the 1D system, the bosonizaiton study shows that 
the charge gap opens due to the relevancy of the $4k_f$-Umklapp process at the Luttinger parameter ${0<K_c<{1}/{2}}$ \cite{wu2005a,wu2006a}.
Then the low energy physics is captured by the Sp(4) superexchange process described by Eq.~\eqref{eq:spinchain_so5}.

The 1D phase diagram is sketched in Fig.~\ref{fig:phase}.
A quantum phase transition takes place at the SU(4) symmetric point of 
 ${J_0=J_2}$, or, ${U_0=U_2}$:
When ${J_0>J_2}$ (${U_0<U_2}$), the system develops a spin gap with the
presence of the spin Peierls distortion, while at $J_0\le J_2$ (${U_0\le U_2}$), the system enters a gapless spin liquid phase
and maintains the translation symmetry~\cite{wu2006a}.
The nature of this transition is Kosterlitz-Thouless like.
For convenience, a parameter angle $\theta$ is employed to represent
$J_{0,2}$ as
\bea
J_0=\sqrt 2 \cos\theta, J_2=\sqrt 2 \sin\theta.
\eea
Hence, the gapless spin liquid phase lies at ${0\le \theta \le 45^\circ}$,
while the spin Peirerls phase exhibiting dimerization lies at
${45^\circ <\theta \le 90^\circ}$.

The above field-theoretical analysis was confirmed by the density-matrix-renormalization-group (DMRG) simulations~\cite{hung2011}.
The ground state spin gap $\Delta_{sp}$ is defined as the energy
difference between the ground state and the lowest Sp(4) multiplet.
The DMRG results show that for the cases of ${\theta> 45^{\circ}}$, {\it i.e.}, 
$J_2/J_0 < 1$, $\Delta_{sp}$'s saturate as enlarging
the lattice size, which signatures the opening of spin gap.
On the other hand, $\Delta_{sp}$'s vanish at $\theta \le 45^\circ$
demonstrating gapless ground states.
These results show that the phase boundary is located
at ${\theta=45^\circ}$ with the SU(4)$_A$ symmetry,
on which the system is also gapless.

%---------------------------------------------
\begin{figure}[t]
\centering
\includegraphics[width=0.6\linewidth]{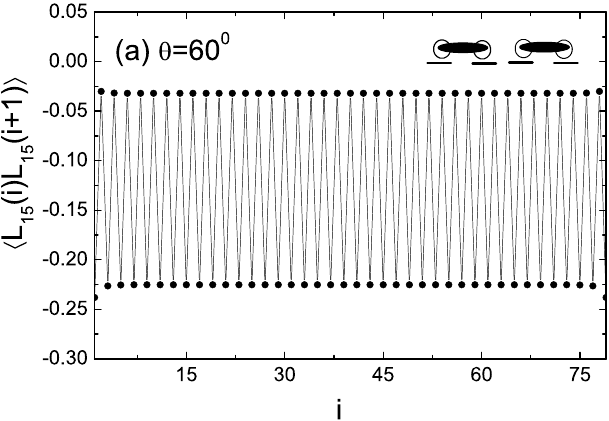}
\centering
\includegraphics[width=0.6\linewidth]{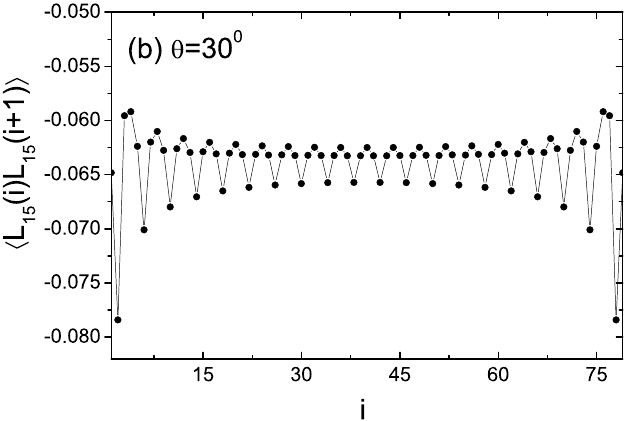}
\caption{The DMRG results of the nearest neighbor correlations of
${\langle L_{15}(i)L_{15}(i+1) \rangle}$ under the open boundary condition.
The parameter values are ${\theta=60^{\circ}}$ in ($a$) and ${\theta=30^{\circ}}$
in ($b$). A 2-site periodicity appears in ($a$) and a 4-site periodicity
with a power-law decay shows up in ($b$). Figures are adapted from Ref.~\onlinecite{hung2011}.
}
\label{fig:nncorr}
\end{figure}
%---------------------------------------------

The spin gapped and gapless phases exhibit different characteristic oscillations.
The DMRG results of the nearest neighbor (NN) correlations of the Sp(4)
generators are calculated with the open boundary condition.
As an example, ${\langle L_{15}(i) L_{15}(i+1)\rangle}$ is shown
in Fig.~\ref{fig:nncorr} and those of other Sp(4)
generators are the same.
In the spin gapped phase, say, at ${\theta=60^\circ}$, a
characteristic 2-site oscillation, corresponding to a dimer pattern,
is pinned by the open boundary condition.
It does not decay as moving to the center of the chain implying
the presence of long-range-ordering in the thermodynamic limit.
This is consistent with the $4k_f$ charge-density-wave ordering,
exhibiting as dimerization.
In contrast, in the gapless region $0\le \theta\le 45^\circ$,
the open boundary induces a power-law decay.
For example, at ${\theta^\circ=30^\circ}$, it approximately
exhibits a 4-site oscillation, and the periodicity is the
same for  $0\le \theta \le 45^\circ$.
This is in agreement with the dominant $2k_f$ spin correlations
in the bosonization analysis.

In the gapless phase, {\it i.e.}, $0\le \theta\le 45^\circ$, 
the decay powers of the two-point correlation functions are discussed below.
The correlation can be expressed asymptotically as
\bea
\langle  O(i_0) O(i) \rangle \propto 
\frac{\cos \frac{\pi}{2} |i_0-i|} {|i_0-i|^\kappa},
\eea
where the operator $O=L$, or, $n$ represents an operator in the
10-generator channel, or, the 5-vector channel of the Sp(4) group, respectively.
The critical exponents for $L$ and $n$ along the SU(4)$_A$ line
(${\theta=45^\circ}$) should be the same, as fitted by
${\kappa\approx 1.52}$, which is in a good agreement with the value of
$1.5$ from the bosonization
analysis  \cite{wu2005a,yamashita1998,azaria1999}.
As away from the SU(4)$_A$ line, the degeneracy between $L$ and $n$
is lifted.
The simulations show that the critical exponent for $L$ shows
${\kappa_L<1.5}$, while that for $n$ exhibits
${\kappa_n>1.5}$.

%------------------------------------------------------
\subsection{The two-dimensional Sp(4) magnetism -- Exact diagonalization
on a $4\times 4$ cluster}

The 2D Sp(4) magnetism based on Eq.~\eqref{eq:spinchain_so5} is very challenging, 
whose global phase diagram remains unsolved.
Especially, in the region of ${0<\theta\le 45^\circ}$, the phase
is dominated by the baryon-type physics which is quite different
from the SU(2) quantum magnetism \cite{li1998,chen2005}.
On the other hand, the physics along the SU(4)$_B$ line, i.e, ${\theta=90^\circ}$, is relatively clear.
Quantum Monte Carlo simulations (QMC) show a long-range N\'eel
ordering with a much weaker N\'eel moment compared to
the SU(2) case \cite{harada2003}.
However, except this special case, the sign-problem appears and
no conclusive results are known.
Along the SU(4)$_A$ line, it is likely that there exist a strong 
plaquette-like correlation as proposed by Bossche {\it et. al.}~\cite{bossche2000,mishra2002}.
Such a state is a baryon-like state, exhibiting the plaquette
type correlations, and each plaquette is a 4-site 
SU(4) singlet state.

Below we review the ground state correlations via the exact
diagonalization (ED) method.
Even though it can only be applied to a small ${4\times 4}$ cluster,
the associated ground state profiles at different values of $\theta$
still yield valuable information to speculate the thermodynamic limit.
Based on the ED result, a phase transition would be expected as follows: 
A N\'eel order-like state breaking the Sp(4) symmetry appears at large values of $\theta$, 
in particular close to $90^\circ$; while an Sp(4) singlet ground state breaking the translation symmetry exits %at momentum $X$ 
as lowering $\theta$ to smaller values.

%----------------------------------------------------------------
\subsubsection{The correlation functions}
\label{sect:mag}

%-----------------------------------------------------------
\begin{figure}[t]
\centering
\includegraphics[width=0.7\linewidth]{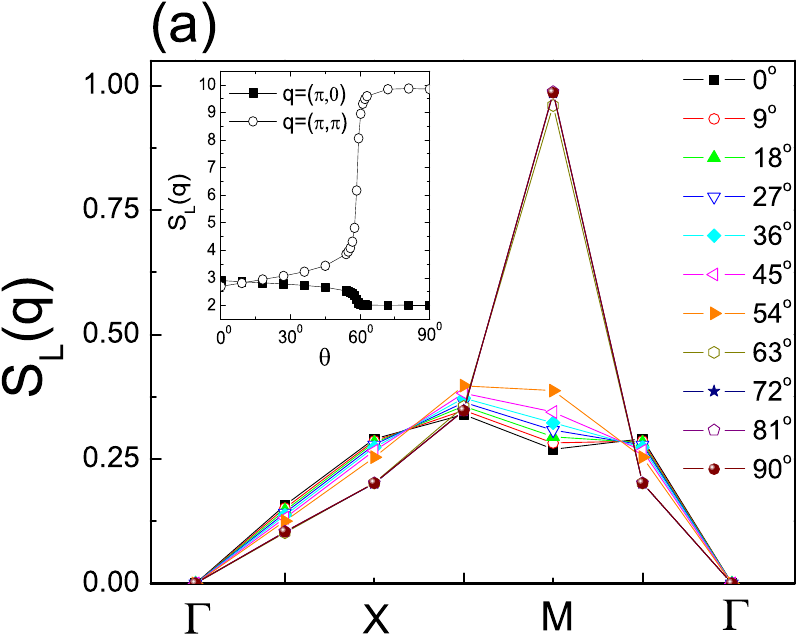}
\centering
\includegraphics[width=0.7\linewidth]{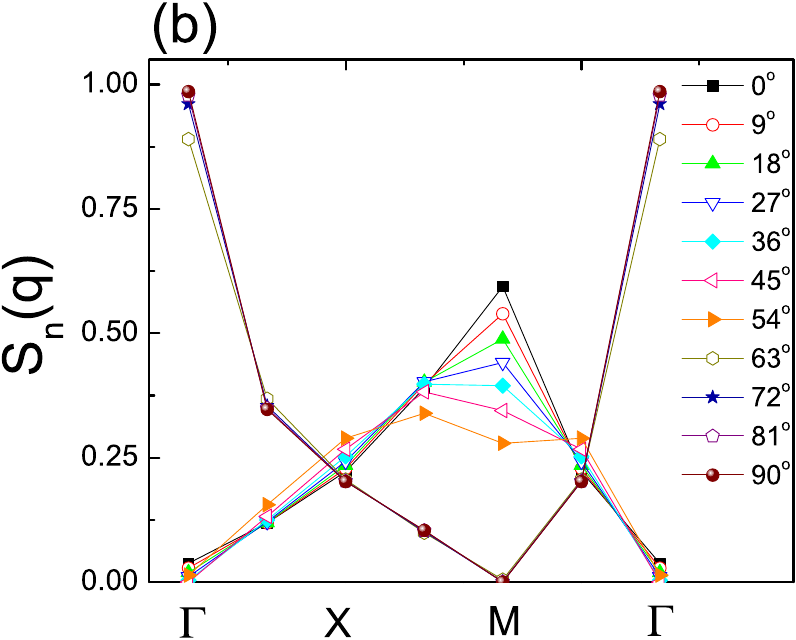}
\caption{ The structure factors for the $4 \times 4$ cluster
with ($a$) $S_L(\mathbf{q})$ in the Sp(4) generator sector and
($b$) $S_n(\mathbf{q})$ in the Sp(4) vector sector.
The expressions of $S_L$ and $S_n$
are defined in Eq.~\eqref{eq:structure}.
The inset in ($a$) is the comparison between $S_L(\pi,0)$
and $S_L(\pi,\pi)$ versus $\theta$. Figures are adapted
from Ref.~\onlinecite{hung2011}.
} 
\label{fig:neel}
\end{figure}
%--------------------------------------------------------------- 

We use structure factors to describe the ground state properties.
A structure factor converges to a finite value in the thermodynamic
limit in the presence of long-range ordering~\cite{harada2003,schulz1992}.
Nevertheless, for a small cluster, the tendency of ordering is
reflected as the peak of a structure factor at a characteristic
momentum. The following structure factors will be employed,
\bea
S_O(\mathbf{q})=\frac{1}{N}\sum_{i,j} e^{i\mathbf{q}\cdot
(\mathbf{r}_i-\mathbf{r}_j)}
\avg{G|O(i) O(j)|G},
\label{eq:structure}
\eea
where $O=L$ referring to an operator of the Sp(4) generators, and 
$O=n$ referring to an operator of the Sp(4) vectors.
For later convenience, we use the following symbols
to represent the crystal momenta $\Gamma=(0,0)$,
$X=(\pi, 0)$, and $M=(\pi, \pi)$.

Previous QMC simulations show the N\'eel long-range order of the SU(4) Heisenberg 
model defined in the fundamental and anti-fundamental
representations in two sublattices~\cite{harada2003}. 
It corresponds to the SU$_B$(4) case with $\theta=90^\circ$, 
which ensures the relation of 
\bea
{S_n (\mathbf{q})=S_L (\mathbf{q} + M)},
\eea
due to the staggered definition of Sp(4) vectors 
$n_a$ in Eq.~\eqref{eq:SU4B}.
Back to the spin language, as ${\theta=90^\circ}$ the
classic energy can be minimized by arranging the two components
of ${S_z=\pm {3}/{2}}$,  or, of the other two components of
${S_z=\pm {1}/{2}}$ in a staggered way. These different
configurations are equivalent under the Sp(4) transformations.

The SU$_B$(4) relation still approximately holds when $\theta$ 
is close to $90^\circ$.
The N\'eel type correlation of $L_{ab}$ also extends to a finite
regime as $\theta$ deviates away from 90$^\circ$, and in
the same regime, $n_a$ exhibits the uniform correlations.
When ${90^\circ\ge \theta\gtrsim 60^\circ}$, the structure
factor $S_L(\mathbf{q})$ of the 10-generator channel peaks at the 
$M$-point, i.e., the nesting wavevector $(\pi,\pi)$, as shown in 
Fig.~\ref{fig:neel} ($a$). 
On the other hand, $S_n(\mathbf{q})$ of the 5-vector channel is
shown in Fig.~\ref{fig:neel}($b$), which peaks at 
the $\Gamma$-point, roughly in the same parameter range 
that $S_L(\mathbf{q})$ develops a peak at the $M$-point.

In contrast, at ${\theta\lesssim 60^\circ}$, $S_L(\mathbf{q})$ 
distributes relatively smooth over all momenta without 
dominant peaks.
The distribution of $S_n(\mathbf{q})$ also becomes smooth in a
similar way to that of $S_L(\mathbf{q})$.
Nevertheless, at small values of ${\theta<18^\circ}$, it becomes to
peak at the $M$-point, {\it i.e.}, at the momentum ${(\pi,\pi)}$.
In the case of SU(4)$_A$, {\it i.e.},
${\theta=45^{\circ}}$, the 5-vector
and the 10-generator channels become degenerate, hence,
$S_n(\mathbf{q})=S_L(\mathbf{q})$ for each $\mathbf{q}$.
As $\theta$ deviates away from $45^{\circ}$, $S_L$ and $S_n$ evolve
differently.

%-------------------------------------------------------------
\subsubsection{The columnar dimer correlations}
\label{sect:dimer}
 
To test the  possibility towards a dimerized ground state, the susceptibilities
 of translation and rotational symmetry breakings are defined
below. Two perturbations are added to the Hamiltonian Eq.~\eqref{eq:spinchain_so5}, {\it i.e.},
\bea
&&\hat{O}_{dim} (\mathbf{Q}) = \sum_{i} \cos (\mathbf{Q} \cdot \mathbf{r}_i)
H_{ex}(i,i+\hat{x}),  \nn \\
&& \hat{O}_{rot}= \sum_{i}
[H_{ex}(i,i+\hat{x})-H_{ex}(i,i+\hat{y})].
\eea
The former breaks the translation symmetry and the latter breaks
the rotation symmetry.

The ED results of susceptibilities associated with $\hat{O}_{dim}$
and $\hat{O}_{rot}$ show that both of them exhibit a peak in the interval of
${60^\circ <\theta<70^\circ}$
\cite{hung2011}.
Although no real divergences exist due to the finite size,
sharp peaks would imply the tendency
of long-range ordering.
Hence, the results imply a tendency to break both translational
and rotational symmetries, which is consistent with
a columnar dimerization in the thermodynamic limit.
In contrast, the plaquette ordering does not break the 4-fold
rotational symmetry, which is not favored in this regime.

%--------------------------------------------
\subsubsection{The plaquette form factor}
\label{sect:plaquette}
%---------------------------------------------------------
\begin{figure}[t]
\includegraphics[width=0.75\linewidth]{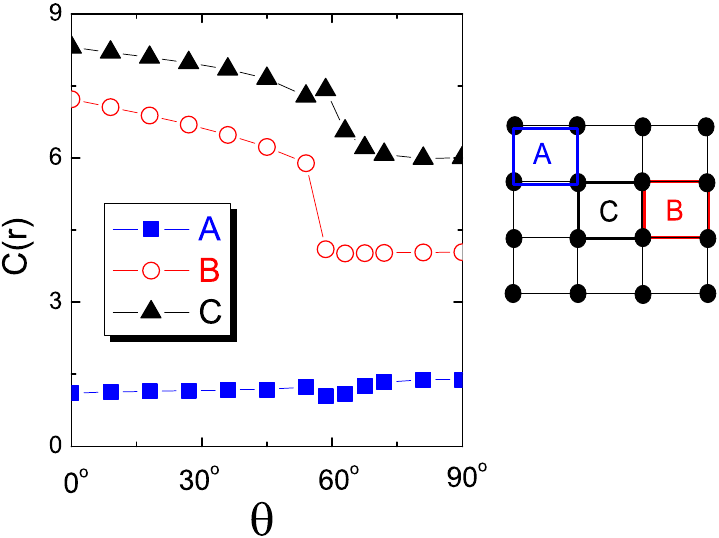}
\caption{The local Casimir $C(\mathbf{r})$  defined 
in Eq.~\eqref{eq:localspin} versus $\theta$.
The plaquettes $A$, $B$ and $C$ are depicted in the right.
}
\label{fig:Fig5}
\end{figure}
%------------------------------------------------------------

The tendency of the plaquette type ordering, {\it i.e.}, the SU(4) 
analogy to the baryon state, shows up at small
values of $\theta$, i.e., 
${\theta<50^\circ \sim 60^\circ}$.
To characterize such a state, the plaquette Sp(4) Casimir  
centered at $\mathbf{r}$ is defined as
\bea
\label{eq:localspin}
C(\mathbf{r}) =\avg{G|\sum_{1\le a<b\le 5}\big\{\sum_{i} 
L_{ab}(i)\big\}^2 |G},
\eea
where $i$ runs over the four sites of this plaquette.
A similar method was used for the SU(2) case to characterize
the competing dimer and plaquette orders \cite{richter1996}.

The relation of $C$ versus $\theta$ is calculated
under the open boundary condition.
Based on the symmetry analysis, the values of $C$ for three non-equivalent plaquettes $A$, $B$, and $C$ are shown in Fig.~\ref{fig:Fig5}.
$C(\mathbf{r_A})$ of the corner plaquette is significantly about
one order smaller than $C(\mathbf{r_B})$ and $C(\mathbf{r_C})$ of the other two plaquettes at small values of $\theta$.
This suggests that an Sp(4) plaquette tendency is pinned down by the open boundary.
As $\theta$ increases above $60^\circ$, the contrast decreases
which means that the plaquette-type pattern is weakened.
It is likely that there exists a strong plaquette-like correlation
at small values of $\theta$ 
but significantly beyond the SU(4)$_A$ line.
It covers the entire region of ${\theta<45^\circ}$ and also extends
slightly above $45^\circ$.
Nevertheless, further larger scale calculations are necessary for
a conclusion.

%%%%%%%%%%%%%%%%%%%%%%%%%%%%%%%%%%%%%%%%%%%%%%%%%%%%%%%%%%
\subsection{Quantum plaquette model for the three-dimensional SU(4) magnetism}

The SU(4)$_A$ plaquette state discussed above has been shown 
as the exact solution to the ground state of a 2-leg ladder model of spin-${3}/{2}$ fermions~\cite{chen2005}.
However, due to the geometric constraint, it cannot resonate and is a valence-bond-solid type state.

The resonating quantum plaquette model (QPM) model was constructed in three dimensions~\cite{xu2008,pankov2007}, 
which is analogous to the quantum dimer model for the SU(2) magnet~\cite{rokhsar1988,fradkin1990}.
The quantum dimer model has a gauge theory description
to the Rohksar-Kivelson (RK) Hamiltonian, which is a compact $U(1)$ gauge theory~\cite{fradkin1990}.
QPM also has a similar description as reviewed below, which is a high order gauge field theory.
Recently it has received considerable attention in the context of fracton physics~\cite{nandkishore2019}.

%--------------------------------------------------------
\subsubsection{The quantum plaquette model}

Consider a {three-dimensional (3D)} cubic lattice SU(4) model in the limit that each plaquette 
has a strong tendency to form a local SU(4) singlet.
The effective Hilbert space is spanned by all the plaquette configurations.
They are subject to the constraint that every site belongs to one and only one plaquette.

\begin{figure}
\includegraphics[width=0.6\linewidth]{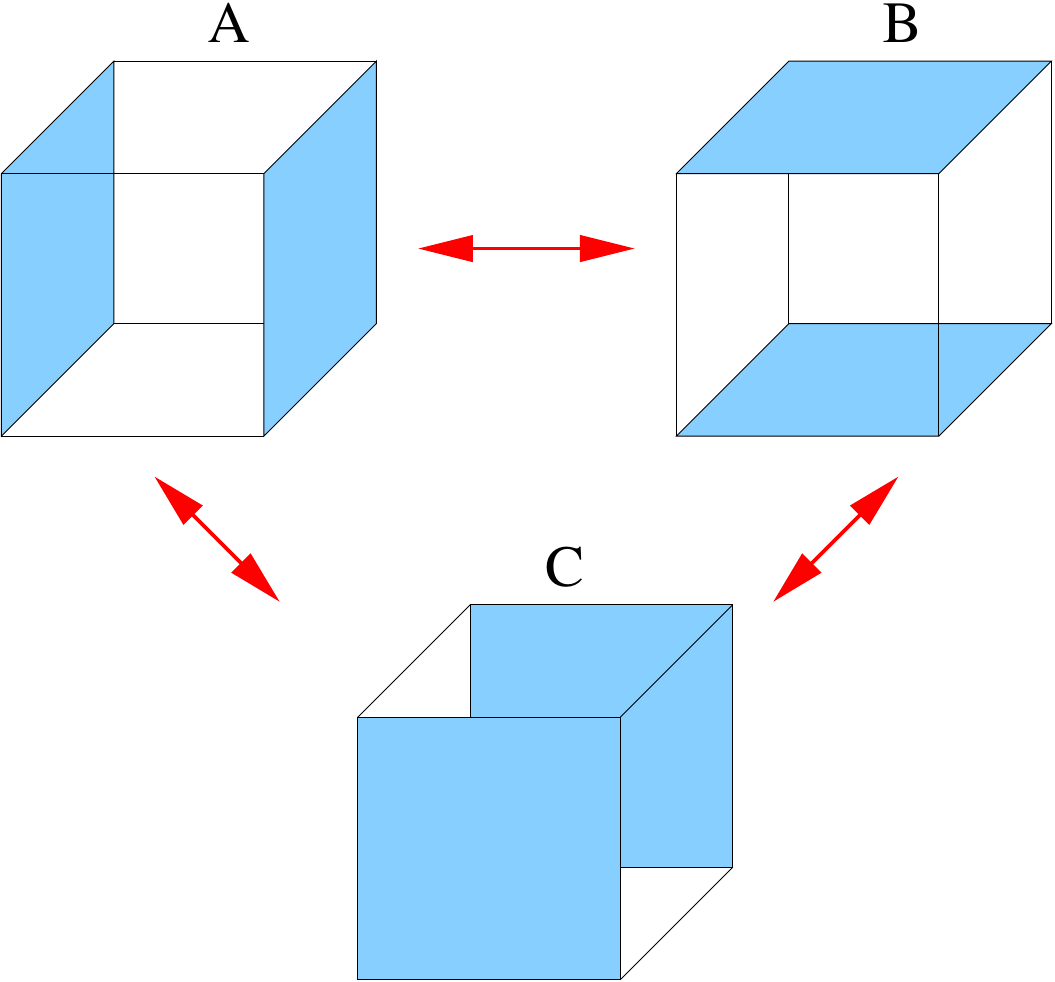}
\caption{Three flippable configurations in one cube.
Figure is adapted
from Ref.~\onlinecite{xu2008}.
}
\label{fig:resonance}
\end{figure}

Each unit cube possesses three flippable configurations:
the pairs of plaquettes of left and right, top and bottom, and
front and back denoted as $A$, $B$ and $C$ in Fig.~\ref{fig:resonance}, respectively.
A Rokhsar-Kivelson (RK) type Hamiltonian is constructed
as~\cite{rokhsar1988}:
\begin{eqnarray}
H&=&-t \sum_{\mbox{each cube}} \Big\{ |A\rangle \langle B|
+ |B\rangle \langle C|
+{|C\rangle \langle A|}+h.c.\Big\} \nonumber\\
&& + V\sum_{\mbox{each cube}} \Big\{ |A\rangle \langle A|  +
|B\rangle \langle B| + |C\rangle \langle C| \Big\},
\label{eq:RK}
\end{eqnarray}
where $t$ is assumed to be positive, and $v/t$ is arbitrary.
By defining
\begin{equation}
|Q_{1,2} \rangle=|A\rangle + e^{\pm i\frac{2}{3}\pi} |B\rangle
+e^{\mp i\frac{2}{3}\pi} {|C\rangle} ,
\end{equation}
Eq.~\eqref{eq:RK} is reformulated as
\bea H&=&
t \sum_{\mbox{each cube}} \left ( |Q_1\rangle \langle Q_1|
+|Q_2\rangle \langle Q_2| \right ) \nn \\
&+& (V-2t)
\sum_{\mbox{each cube}} \left (|A\rangle \langle A|  + |B\rangle
\langle B + |C\rangle \langle C| \right ).
\label{eq:RK2}
\eea
The RK point here corresponds to $V=2t$.

The ground state to Eq.~\eqref{eq:RK2} is the equal weight
superposition of all the plaquette configurations within the
same topological sector that can be connected by local flips.
The classical Monte Carlo simulation shows that a crystalline order
of resonating cubes at this RK point~\cite{pankov2007}.
At ${V/t<2}$, the system favors flippable cubes.
For example, as shown in Ref.~[\onlinecite{pankov2007}], 
at ${v/t\ll -1}$, the ground state exhibit the columnar ordering.
At $v>2t$, both terms in Eq.~\eqref{eq:RK2} are positive-definite.
Hence, all the configurations without flippable cubes are
the ground states.
All the transitions between different phases are of the
first order.

%------------------------------------------------
%------------------------------------------------

%------------------------------------------------
\subsubsection{A high order gauge field mapping}

It is often useful to extract the low energy physics of strong
correlated systems by mapping it to gauge theory models.
In fact, the effective gauge theory of the QPM was constructed
as a special one - a high order gauge theory.

Each square face is denoted by the location of its center:
${i+\frac{1}{2}\hat u+\frac{1}{2}\hat v}$,
where %$i$ represents the cubic lattice site and
${\hat u=\hat x, \hat y}$ and ${\hat z}$.
Each face is associated with a number $n$ and a strong local
potential ${U(n_{i+\frac{1}{2}\hat{\mu} + \frac{1}{2}\hat{\nu}} -
\frac{1}{2})^2}$ in the limit of ${U\to \infty}$, such that the
low-energy sector has only either ${n=1}$ corresponding to
an SU(4) singlet occupation, or ${n=0}$, otherwise.
The constraint is that the summation of $n$ over all
the 12 faces connecting to the same site equals to 1.
A rank-2 symmetric traceless tensor electric field
is defined as
\begin{equation}
E_{i, \mu\nu} = \eta(i)\left(n_{i+\frac{1}{2}\hat{\mu} +
\frac{1}{2}\hat{\nu}} - \frac{1}{2}\right) ,
\end{equation}
where ${\eta(i)=\pm 1}$ depends on the sublattice that $i$
belongs to.
The constraint is simply reduced to
\bea
\nabla_x \nabla_y E_{xy}+\nabla_y
\nabla_z E_{yz} + \nabla_z \nabla_x E_{zx}=5\eta(i),
\label{constraint}
\eea
where $\nabla$ is the lattice derivative.
The canonical conjugate variable of $E_{i,\mu\nu}$ is
the vector potential $A_{i,\mu\nu}$, satisfying
\begin{eqnarray}
[E_{i,\mu\nu}, A_{j,\rho\sigma} ] &=& i\delta_{ij}
(\delta_{\mu\rho}\delta_{\nu\sigma} +
\delta_{\mu\sigma}\delta_{\nu\rho} ).
\label{eq:EA}
\end{eqnarray}
$A_{i,\mu\nu}$ can be expressed by the phase variable
$\theta_{i + \frac{1}{2}\hat{\mu} + \frac{1}{2}\hat{\nu}}$,
which is the canonical conjugate variable to $n$ as
${A_{i,\mu\nu} = \eta(i) \ \theta_{i + \frac{1}{2}\hat{\mu} +
\frac{1}{2}\hat{\nu}}}$.
Because $E_{\mu\nu}$ only takes integer values,
$A_{\mu\nu}$ is an compact field with period of $2\pi$.

The plaquette flipping process changes the plaquette occupations.
For this purpose, $e^{i A_{j,\nu\sigma}}$ is employed 
which changes the eigenvalue of $E_{i,\mu\nu}$ by 1 since
\begin{equation}
[E_{i,\mu\nu}, e^{i A_{j,\nu\sigma}} ]=
(\delta_{\mu\rho}\delta_{\nu\sigma} +
\delta_{\mu\sigma}\delta_{\nu\rho} ) e^{i A_{j,\nu\sigma}}.
\end{equation}
The flipping term is represented as
\bea
H_t &=& - t [ \cos (\nabla_z
A_{xy} - \nabla_xA_{yz})
+ \cos(\nabla_x A_{yz} - \nabla_yA_{zx}) \nonumber \\
&+& 
\cos(\nabla_y A_{zx} - \nabla_zA_{xy}) ].
\label{eq:low1}
\eea
The associated gauge invariant transformation is
\bea
A_{\mu\nu} \rightarrow A_{\mu\nu}
+ \nabla_\mu\nabla_\nu f,
\eea
where $f$ an arbitrary scalar function.
The low-energy Hamiltonian of the system can be written as
\bea
H &=& H_t + \sum_{\mbox{each cube}} \Big\{
 U [E_{xy}^2+E_{yz}^2+E_{zx}^2]\nonumber\\
&+& V  [(\nabla_x E_{yz})^2
+  
(\nabla_y E_{zx})^2+ (\nabla_z E_{zx})^2 ] \Big\},
\label{low}
\eea
under the constraint of Eq.~\eqref{constraint}.

The above formulation is a compact high order gauge theory,
and the non-local topological defect excitations is crucial to determine whether its ground state is gapless or gapped.
The standard analysis is the mapping to a height model,
such that topological defects are represented by
the vertex operators \cite{xu2008}.
Due to the proliferation of topological defects, the system is generally gapped 
for the whole phase diagram of the quantum model, which corresponds to crystalline orders.
It would be interesting to further explore the possiblity
of the 3D power-law phases as the analogue to the
Kosterlitz-Thouless transition \cite{nandkishore2019}.

\section{Summary}
\label{sec6}

In summary, we have reviewed a class of Mott insulators, namely, the multiflavor Mott insulators, 
in both quantum materials and ultracold atom systems. 
In both cases, the local Hilbert states in each unit cell of the multiflavor Mott insulators consist of more 
than two flavors in contrast to the conventional spin-1/2 Mott insulators, 
and also differs fundamentally from the conventional magnets with large $S$ moments. 
They bear similarities to the orbital-active Mott-insulators but typically do not demand
the explicit orbital degeneracy. These include the breathing/cluster magnets,
the ${J=3/2}$ Mott insulators in various transition metal oxides and rare-earth systems,
twisted moir\'{e} heterostructures, the ultracold atom fermion systems 
with large hyperfine-spins, and various other physical contexts.

The study of the multiflavor Mott insulators will certainly broaden the research scope 
of quantum magnetism, and further enrich the activity of exploring exotic quantum states of matter.
As the consequence of the large local Hilbert space, this class of Mott insulators exhibit
the common feature of enhanced quantum fluctuations.  Instead of being viewed from the
large-$S$ perspective, the appropriate viewpoint should be large-$N$.
The low-energy superexchange models typically go beyond the SU(2) large-$S$ Heisenberg models,
and bear similarities to the Kugel-Khomskii models. They are expected to exhibit a variety unusual physics 
including the spin-multipolar ordering, the ``baryon-like" like physics, and even more exotic spin-liquid states.
{Quantum magnetism associated with large symmetries could exhibit certain similarities to QCD 
in spite of dramatically different energy scales. If each lattice site belongs to the fundamental 
representation of SU($N$) with ${N >2}$, the SU($N$) singlet typically involves more than 
two sites leading to the ``baryon-type" physics.}

 The review mainly focuses on the multiflavor Mott insulating states and the associated pairwise interactions 
with and without the enlarged symmetries in the strong Mott regime where the charge fluctuation
is suppressed. In many cases, the system could be proximate to the Mott transition and in the weak
Mott regime. In this regime, the charge fluctuation is strong, and the more appropriate description 
of the underlying physics would be in terms of the multiflavor Hubbard model or the exchange model 
involving the multiflavor ring exchange interactions. It is well-known that, the 
strong charge fluctuation of the weak Mott insulators or proximity to the Mott transition could stabilize various exotic quantum phases 
even for the spin-1/2 degrees of freedom~\cite{PhysRevB.72.045105,PhysRevLett.95.036403}. 
The multifavor Mott insulators actually bring a bit more interesting phenomena with their multiple favors of degrees of freedom. 
One well-known consequence is the Pomeranchuk effect or Pomeranchuk cooling
that has been observed in magic-angle graphene~\cite{Rozen_2021,Saito_2021} and been used to cool to much lower temperatures 
in the SU($N$) alkaline-earth atoms~\cite{PhysRevA.85.041604,taie2012}. 
Although both the strong charge fluctuation and the multiflavor Hilbert space could drive to exotic quantum phases
and phenomena,
the exotic phases that are favored by them, however, can be different~\cite{ASSAAD2005,PhysRevB.100.115155,PhysRevA.93.061601}. 
There has not been much systematic study
in this direction. Potentially this can be an interesting direction for both theories and experiments. In 
the ultracold atom systems, the strength of correlation can be effectively tuned experimentally to access both 
strong and weak Mott regimes. 
With the modern fabricating and controlling technology, one can effectively tune many physical properties of
 the transition metal dichalcogenide (moir\'{e}) heterostructures such as the correlation and doping,
 and drive the system through the Mott transition~\cite{Mak2022}. 
 In fact, the type of charge fluctuations in the early study of the Hubbard model is mostly concerned 
 about the physical processes above the Mott gap that involve the double occupation~\cite{PhysRevB.72.045105}. 
 The existence of large numbers of correlated moir\'{e} systems
 and the partially-filled correlated materials supports the cluster localization 
 for which the possibility of the sub-Mott-gap charge fluctuations can be relevant~\cite{PhysRevResearch.2.043424,PhysRevLett.113.197202,PhysRevB.93.245134,huang2023nonfermi}.  
 Distinct charge fluctuations would bring rather distinct physics and consequences on the 
 remaining spin and orbital degrees of freedom, especially since the charge sector often has 
 a higher energy scale than the spin and orbital and thus may be considered first. 
 The combination of distinct charge fluctuations and high symmetries is another direction to go. 
  When the charge sector is incorporated as one extra flavor of the multiflavor Mott systems, not only 
  the physics becomes richer from the interplay of more degrees of freedom, but also
 the scope of multiflavor Mott insulators can be further broadened. 
 Thus, the multiflavor Mott systems and physics are not limited to the contents and the classification
 in the current review, and many more systems may be recast into this topic. We expect
 many exciting and new results to emerge in this field in the near future.

\section*{Acknowledgments}

We are grateful from the previous collaboration and discussion
with Leon Balents, Patrick A. Lee, Fuchun Zhang, Xu-Ping Yao,
Rui Luo, Lucile Savary, Rodrigo Pereira, Fr\'{e}d\'{e}ric Mila,  
Michael Hermele, Ana Maria Rey,
Leo Radzihovsky, Jun Ye, Kaden Hazzard, Salvatore Manmana,
Arun Paramekanti, Bruce Gaulin, Yong-Baek Kim,
Michel Gingras, George Jackeli, Xiaoqun Wang, Tao Xiang, Yue Yu,
Cenke Xu, Jiang-Ping Hu, Shoucheng Zhang, Yupeng Wang,
Shu Chen and many others.

G.C. is supported by the Ministry of Science and Technology of China with Grants No.~2021YFA1400300,
by the National Science Foundation of China with
Grant No.~92065203, and by the Research Grants Council of Hong Kong with C7012-21GF. 
C.W. is supported by the National Natural Science Foundation of China under the Grants No. 12234016 and No. 12174317. 
This work has been supported by the New Cornerstone Science
Foundation.

%%%%%%%%%%%%%%%%%%%%%%%%%%%%%%%%%%%%%%%%%%%%%%%%%%%%
\appendix

\section{The SU(3) viewpoint to the spin-1 Heisenberg model with the bi-quadratic interaction}
\label{sect:append}

The two-band Hubbard model described by Eq.~\eqref{eq:twobandHubb}
exhibits different types of Mott-insulating states depending on the filling factor.
The case of quarter-filling exhibiting the orbital degree of freedom is already discussed in Sec.~\ref{sec2},
whose low-energy physics is described by a variant of the SU(4) Kugel-Khomskii model.
On the other hand, for the case of half-filling, i.e., two electrons per site, 
each orbital is occupied by one electron. Therefore, 
the orbital degree of freedom is quenched, 
and only the spin degree of freedom remains.
The Hund's rule coupling aligns two electron spins and forms the spin-1 moment.

In this case, the leading order super-exchange process across a
bond is the switching of one pair of electrons, which is at the
level of the second order perturbation theory.
It gives rise to the standard bi-linear Heisenberg superexchange
term as
\bea
H_{bl}=J_2\sum_{\avg{ij}} \mathbf{S}_i \cdot \mathbf{S}_j.
\label{eq:linear}
\eea
Moreover, the fourth order perturbation theory yields the
bi-quadratic superexchange term as
\bea
H_{bq}=J_4 \sum_{\avg{ij}} \left(\mathbf{S}_i \cdot \mathbf{S}_j
\right)^2.
\label{eq:quadratic}
\eea
Based on their perturbation orders, $J_2$ and $J_4$ are
estimated at
\bea
J_2\propto \frac{t^2}{U}, \ \ \, J_4\propto \frac{t^4}{U^3}.
\eea
Hence $J_4$ is much smaller than $J_2$.
The other microscopic mechanism for the bi-quadratic interaction is 
the spin-lattice coupling. This happens when the direct exchange 
dominates and depends sensitively on the bond length. Integrating out the 
bonds, one can obtain the bi-quadratic spin interaction~\cite{PhysRevLett.93.197203}. 
This has been applied to the Cr-based pyrochlores CdCr$_2$O$_4$ and HgCr$_2$O$_4$. 
As far as we are aware of, the sign of the bi-quadratic spin interaction is often negative and 
ferromagnetic-like. 

The dominance of the bi-linear super-exchange process applies to
even higher spin systems in quantum materials.
The onsite SU(2) multiplets are denoted as $|SS_z\rangle$ with
$S\ge S_z \ge -S$.
The bi-linear superexchange can only connect these states via
a one-dimensional weight diagram as
\bea
|S - S\rangle \longleftrightarrow
|S -S+1\rangle ... 
\longleftrightarrow
|S  S\rangle.
\eea
Hence, spin flipping between $|S -S\rangle \longleftrightarrow |SS\rangle$
takes $2S$ successive steps, which is at the $4S$-th order perturbation theory.
Therefore, no matter how large the total spin is, quantum
spin fluctuations are basically a $1/S$-effect.

The bi-quadratic term in Eq.~\eqref{eq:quadratic} and
bi-linear on in Eq.~\eqref{eq:linear} 
can be treated at equal footing as controlled by
the parameter angle $\theta$
as shown in Eq.~\eqref{eq:biquad}.
 Although for physical systems of transition metal oxides,
$0<|\theta | \ll {\pi}/{2}$, below the entire
parameter space will be considered.

Look at one bond $\langle ij \rangle$.
According to the SU(2) structure, the bond Hilbert space is
divided into the singlet, triplet and quintet states.
Their energies are denoted as $E_s$, $E_t$ and $E_q$, respectively.
It is straightforward to show that
\bea
E_s/J&=&-2\cos\theta +4 \sin\theta, \nn \\
E_t/J&=&-\cos\theta+\sin\theta, \nn \\
E_q/J&=&\cos\theta+\sin\theta.
\eea
At two special angles, degeneracy patterns appear beyond
the SU(2) multiplet: A 6-fold degeneracy appears at
$\theta= \pi/4$ and $5\pi/4$, and an
8-fold degeneracy appears at $\theta= \pi/2$
and $\theta=3\pi/2$.
They imply the hidden high symmetries of
two different types of SU(3).

To show the SU(3) symmetry explicitly, the common
representation of the Gell-mann matrix is adopted:
\bea
\lambda_{1,ij}&=&\delta_{i1}\delta_{j2}+\delta_{i2}\delta_{j1}, \ \ \,
\lambda_{2,ij}=-i\delta_{i1}\delta_{j2}+i\delta_{i2}\delta_{j1}
\nn \\
\lambda_{4,ij}&=&\delta_{i3}\delta_{j1}+\delta_{i1}\delta_{j3}, \ \ \,
\lambda_{5,ij}=-i\delta_{i3}\delta_{j1}+i\delta_{i1}\delta_{j3} \nn \\
\lambda_{6,ij}&=&\delta_{i2}\delta_{j3}+\delta_{i3}\delta_{j2}, \ \ \,
\lambda_{7,ij}=-i\delta_{i2}\delta_{j3}+i\delta_{i3}\delta_{j2}\nn \\
\lambda_{3,ij}&=&\mbox{diag}\left(1,-1,0\right), \nn \\
\lambda_{8,ij}&=&\frac{1}{\sqrt 3}\mbox{diag}\left(1,1,-2\right).
\label{eq:gellmann}
\eea
It is convenient to employ the representation that the spin-1 matrices are purely imaginary, {\it i.e.},
\bea
(S^i)_{jk}=-i\epsilon_{ijk},
\eea
which are just the three purely imaginary matrices of
Eq. (\ref{eq:gellmann}),
\bea
S^x=\lambda_7, \ \ \ S^y=-\lambda_5, \ \ \ S^z=\lambda_2.
\eea
In this representation, $S_z$ is off-diagonal, instead,
$\lambda_3$ and $\lambda_8$ are diagonal.
They are spin-quadrupole operators as
\bea
\lambda_3&=& - (S_x^2-S_y^2), \ \ \,
\lambda_8= -\frac{1}{\sqrt 3}(S_x^2+S_y^2-2S_z^2).
\eea

$\frac{1}{2}\lambda_3, \frac{1}{2}\lambda_8$ span the Catan subalgebra
of SU(3).
Their common eigenstates are just spin-1 states in the polar bases $|a\rangle$ 
with $a=x,y$ and $z$, defined as
\bea
S^a |a\rangle =0.
\eea
Each state can be represented in terms of their eigenvalues
of $(\frac{\lambda_3}{2}, \frac{\lambda_8}{2})$ as
\bea
&&|x\rangle : \left(\frac{1}{2}, \frac{1}{2\sqrt 3}\right), \ \ \,
|y\rangle : \left(-\frac{1}{2}, \frac{1}{2\sqrt 3}\right), %\nn \\ &&
~ |z\rangle : \left(0, -\frac{1}{\sqrt 3} \right).
\nn \\
\eea
They are vertices of an equilateral triangle in the plane of
$(\frac{\lambda_3}{2}, \frac{\lambda_8}{2})$, which
is just the weight diagram of the fundamental
representation of the SU(3) group.
In other words, $|x\rangle$, $|y\rangle$, and $|z\rangle$ play
the role of three colors of quarks.
They can represented as $[1]$, i.e., the single box
in terms of the Young tableau.

In this representation, Eq.~\eqref{eq:biquad} becomes
\bea
\frac{H}{J} &=&
\frac{1}{2}\sin\theta \sum_{a=1,3,4,6,8}
\sum_{\avg{ij}}  \lambda_a(i) \lambda_a(j) \nn \\
&+&\left( \cos\theta-\frac{1}{2}\sin\theta \right)
\sum_{a=2,5,7} \sum_{\avg{ij}}  \lambda_a(i) \lambda_a(j).
\label{eq:h2}
\eea
At $\theta=\frac{1}{4}\pi$ and $\frac{5}{4}\pi$,
this Hamiltonian exhibits an explicit SU(3) symmetry as
\bea
\frac{H}{J} &=&\pm\frac{\sqrt 2}{4} J \sum_{a=1}^8
\sum_{\avg{ij}} \lambda_a(i) \lambda_a(j),
\eea
with each site belonging to the fundamental representation of SU(3).
The 6-fold degeneracy comes from the decomposition of,
\bea
[1]\otimes [1]= [1]^* \oplus [2],
\eea
in terms of the Young tableau, where $[1]^*$ is the complex conjugation representation of $[1]$, 
and $[2]$ is 6D representation of one row with two boxes.

On the other hand, a staggered type of the SU(3) symmetry appears
at $\theta=\pi /2$ and $3\pi/2$.
In this case, Eq.~\eqref{eq:h2} becomes
\bea
\frac{H}{J} &=&\pm
\frac{1}{2}\sum_{\avg{ij}} \left( \sum_{a=1,3,4,6,8}
  \lambda_a(i) \lambda_a(j) -
\sum_{a=2,5,7} \lambda_a(i) \lambda_a(j)
\right), \nn
\eea
in which the terms of the real and imaginary Gell-mann matrices exhibit the opposite signs.
For a bipartite lattice, it can be cast into
\bea
H/J
&=& \mp \frac{1}{2}\sum_{i\in A, j=i\pm 1 \in B} \lambda_a(i)
\left(-\lambda_a(j)\right)^*,
\label{eq:SU3stagger}
\eea
where $A$ and $B$ represent two sublattices.
Since $-\lambda_a^*$ is the generators of $[1]^*$,
Eq.~\eqref{eq:SU3stagger} is also SU(3) symmetric
in which two sublattices lie in the fundamental and
anti-fundamental representations, respectively.
The 8-fold degeneracy arises from the decomposition of $[1]\otimes [1]^*=\bullet \otimes [2,1]$, 
where $\bullet$ means the SU(3) singlet and
$[2,1]$ is the 8D adjoint representation of the SU(3) group.

The above two different types of SU(3) symmetries lead to different
types of physics.
In 1D, the spin-1 chain Eq.~\eqref{eq:biquad} with the bilinear and bi-quadratic Heisenberg terms exhibit a variety of quantum phases.
Along the line of $\theta={\pi}/{4}$, each site lies in the
fundamental representation of SU(3).
The system is a gapless state who critical properties are described by the
SU(3)$_1$ Wess-Zumino-Witten model \cite{KNIZHNIK198483}.
On the other hand, at ${\theta=-{\pi}/{2}}$, 
the system is dimerized breaking the translation symmetry, while two sublattices belong to the fundamental and anti-fundamental
representations, respectively.
In contrast, at ${\theta=0}$, i.e., the spin-1 Heisenberg model with only the bi-linear interaction, 
the system exhibits the Haldane gap without translation symmetry breaking.
\\

%%%%%%%%%%%%%%%%%%%%%%%%%%%%%%%%%
\section{$\Gamma$-matrices as quadrupoles of spin-$3/2$}
\label{sect:gamma}

The spin quadrupole matrices are constructed as follows,
\bea
\Gamma^1&=& \frac{1}{\sqrt 3} \left( S_x S_y +S_y S_x \right),
 \nn \\
\Gamma^2&=& \frac{1}{\sqrt 3} \left( S_z S_x +S_x S_z \right),
  \nn \\
\Gamma^3&=& \frac{1}{\sqrt 3} \left( S_z S_y +S_y S_z \right),
 \nn \\
\Gamma^4&=& S_z^2-\frac{5}{4},
  \nn \\
\Gamma^5&=& \frac{1}{\sqrt 3} \left( S_x^2 -S_y^2 \right).
\label{eq:gamma}
\eea
Remarkably, when $S_{x,y,z}$ are spin-$3/2$ matrices,
the $\Gamma$-matrices defined above anticommute with each other,
$\{\Gamma^a,\Gamma^b\}=2\delta_{ab}$,
forming a representation of the Dirac $\Gamma$-matrices.
Their commutators are defined as
\bea
\Gamma^{ab}=-\frac{i}{2} [ \Gamma^a, \Gamma^b] \ \ \ (1\le a,b\le5).
\label{eq:gamma_ab}
\eea
\\
\\
\\

\noindent{\textbf{AUTHOR CONTRIBUTIONS}}\\
G.C. and C.W. organized the sections and wrote the review together. 
\\
\\
\noindent{\textbf{COMPETING INTERESTS}}\\
The authors declare no competing interests.

\bibliography{spin32}

\begin{thebibliography}{163}%
\makeatletter
\providecommand \@ifxundefined [1]{%
 \@ifx{#1\undefined}
}%
\providecommand \@ifnum [1]{%
 \ifnum #1\expandafter \@firstoftwo
 \else \expandafter \@secondoftwo
 \fi
}%
\providecommand \@ifx [1]{%
 \ifx #1\expandafter \@firstoftwo
 \else \expandafter \@secondoftwo
 \fi
}%
\providecommand \natexlab [1]{#1}%
\providecommand \enquote  [1]{``#1''}%
\providecommand \bibnamefont  [1]{#1}%
\providecommand \bibfnamefont [1]{#1}%
\providecommand \citenamefont [1]{#1}%
\providecommand \href@noop [0]{\@secondoftwo}%
\providecommand \href [0]{\begingroup \@sanitize@url \@href}%
\providecommand \@href[1]{\@@startlink{#1}\@@href}%
\providecommand \@@href[1]{\endgroup#1\@@endlink}%
\providecommand \@sanitize@url [0]{\catcode `\\12\catcode `\$12\catcode
  `\&12\catcode `\#12\catcode `\^12\catcode `\_12\catcode `\%12\relax}%
\providecommand \@@startlink[1]{}%
\providecommand \@@endlink[0]{}%
\providecommand \url  [0]{\begingroup\@sanitize@url \@url }%
\providecommand \@url [1]{\endgroup\@href {#1}{\urlprefix }}%
\providecommand \urlprefix  [0]{URL }%
\providecommand \Eprint [0]{\href }%
\providecommand \doibase [0]{http://dx.doi.org/}%
\providecommand \selectlanguage [0]{\@gobble}%
\providecommand \bibinfo  [0]{\@secondoftwo}%
\providecommand \bibfield  [0]{\@secondoftwo}%
\providecommand \translation [1]{[#1]}%
\providecommand \BibitemOpen [0]{}%
\providecommand \bibitemStop [0]{}%
\providecommand \bibitemNoStop [0]{.\EOS\space}%
\providecommand \EOS [0]{\spacefactor3000\relax}%
\providecommand \BibitemShut  [1]{\csname bibitem#1\endcsname}%
\let\auto@bib@innerbib\@empty
%</preamble>
\bibitem [{\citenamefont {Anderson}(1984)}]{Anderson1984}%
  \BibitemOpen
  \bibfield  {author} {\bibinfo {author} {\bibfnamefont {P.~W.}\ \bibnamefont
  {Anderson}},\ }\href@noop {} {\emph {\bibinfo {title} {Basic notations of
  condensed matter physics}}}\ (\bibinfo  {publisher} {The Benjamin/Cummings
  Publishing Company, Inc.},\ \bibinfo {year} {1984})\BibitemShut {NoStop}%
\bibitem [{\citenamefont {Chen}\ \emph {et~al.}(2013)\citenamefont {Chen},
  \citenamefont {Gu}, \citenamefont {Liu},\ and\ \citenamefont
  {Wen}}]{ChenXie2013}%
  \BibitemOpen
  \bibfield  {author} {\bibinfo {author} {\bibfnamefont {Xie}\ \bibnamefont
  {Chen}}, \bibinfo {author} {\bibfnamefont {Zheng-Cheng}\ \bibnamefont {Gu}},
  \bibinfo {author} {\bibfnamefont {Zheng-Xin}\ \bibnamefont {Liu}}, \ and\
  \bibinfo {author} {\bibfnamefont {Xiao-Gang}\ \bibnamefont {Wen}},\
  }\bibfield  {title} {\enquote {\bibinfo {title} {Symmetry protected
  topological orders and the group cohomology of their symmetry group},}\
  }\href {\doibase 10.1103/physrevb.87.155114} {\bibfield  {journal} {\bibinfo
  {journal} {Physical Review B}\ }\textbf {\bibinfo {volume} {87}},\ \bibinfo
  {pages} {155114} (\bibinfo {year} {2013})}\BibitemShut {NoStop}%
\bibitem [{\citenamefont {Maekawa}\ \emph {et~al.}(2004)\citenamefont
  {Maekawa}, \citenamefont {Tohyama}, \citenamefont {Barnes}, \citenamefont
  {Ishihara}, \citenamefont {Koshibae},\ and\ \citenamefont
  {Khaliullin}}]{TMO}%
  \BibitemOpen
  \bibfield  {author} {\bibinfo {author} {\bibfnamefont {Sadamichi}\
  \bibnamefont {Maekawa}}, \bibinfo {author} {\bibfnamefont {Takami}\
  \bibnamefont {Tohyama}}, \bibinfo {author} {\bibfnamefont {Stewart~E.}\
  \bibnamefont {Barnes}}, \bibinfo {author} {\bibfnamefont {Sumio}\
  \bibnamefont {Ishihara}}, \bibinfo {author} {\bibfnamefont {Wataru}\
  \bibnamefont {Koshibae}}, \ and\ \bibinfo {author} {\bibfnamefont {Giniyat}\
  \bibnamefont {Khaliullin}},\ }\href {\doibase 10.1007/978-3-662-09298-9}
  {\emph {\bibinfo {title} {Physics of Transition Metal Oxides}}}\ (\bibinfo
  {publisher} {Springer-Verlag Berlin Heidelberg},\ \bibinfo {year}
  {2004})\BibitemShut {NoStop}%
\bibitem [{\citenamefont {Khomskii}(2014)}]{khomskii2014}%
  \BibitemOpen
  \bibfield  {author} {\bibinfo {author} {\bibfnamefont {Daniel}\ \bibnamefont
  {Khomskii}},\ }\href {\doibase 10.1017/CBO9781139096782} {\emph {\bibinfo
  {title} {Transition Metal Compounds}}}\ (\bibinfo  {publisher} {Cambridge
  University Press},\ \bibinfo {year} {2014})\BibitemShut {NoStop}%
\bibitem [{\citenamefont {Nussinov}\ and\ \citenamefont {van~den
  Brink}(2015)}]{RevModPhys.87.1}%
  \BibitemOpen
  \bibfield  {author} {\bibinfo {author} {\bibfnamefont {Zohar}\ \bibnamefont
  {Nussinov}}\ and\ \bibinfo {author} {\bibfnamefont {Jeroen}\ \bibnamefont
  {van~den Brink}},\ }\bibfield  {title} {\enquote {\bibinfo {title} {Compass
  models: Theory and physical motivations},}\ }\href {\doibase
  10.1103/RevModPhys.87.1} {\bibfield  {journal} {\bibinfo  {journal} {Rev.
  Mod. Phys.}\ }\textbf {\bibinfo {volume} {87}},\ \bibinfo {pages} {1--59}
  (\bibinfo {year} {2015})}\BibitemShut {NoStop}%
\bibitem [{kit(2020)}]{kitp}%
  \BibitemOpen
  \href {www.kitp.ucsb.edu/activities/correlated20} {\enquote {\bibinfo {title}
  {{Correlated Systems with Multicomponent Local Hilbert Spaces }},}\ }\bibinfo
  {howpublished} {KITP Program} (\bibinfo {year} {2020})\BibitemShut {NoStop}%
\bibitem [{\citenamefont {Kugel}\ and\ \citenamefont
  {Khomskii}(1982)}]{kugel1982}%
  \BibitemOpen
  \bibfield  {author} {\bibinfo {author} {\bibfnamefont {K.~I.}\ \bibnamefont
  {Kugel}}\ and\ \bibinfo {author} {\bibfnamefont {D.~I.}\ \bibnamefont
  {Khomskii}},\ }\bibfield  {title} {\enquote {\bibinfo {title} {{The
  Jahn-Teller effect and magnetism: transition metal compounds}},}\ }\href
  {\doibase 10.1070/PU1982v025n04ABEH004537} {\bibfield  {journal} {\bibinfo
  {journal} {Soviet Physics Uspekhi}\ }\textbf {\bibinfo {volume} {25}},\
  \bibinfo {pages} {231} (\bibinfo {year} {1982})}\BibitemShut {NoStop}%
\bibitem [{\citenamefont {Anderson}(1950)}]{PhysRev.79.350}%
  \BibitemOpen
  \bibfield  {author} {\bibinfo {author} {\bibfnamefont {P.~W.}\ \bibnamefont
  {Anderson}},\ }\bibfield  {title} {\enquote {\bibinfo {title}
  {{Antiferromagnetism. Theory of Superexchange Interaction}},}\ }\href
  {\doibase 10.1103/PhysRev.79.350} {\bibfield  {journal} {\bibinfo  {journal}
  {Phys. Rev.}\ }\textbf {\bibinfo {volume} {79}},\ \bibinfo {pages} {350--356}
  (\bibinfo {year} {1950})}\BibitemShut {NoStop}%
\bibitem [{\citenamefont {Ole\'{s}}(2017)}]{oles2017orbital}%
  \BibitemOpen
  \bibfield  {author} {\bibinfo {author} {\bibfnamefont {Andrzej~M.}\
  \bibnamefont {Ole\'{s}}},\ }\bibfield  {title} {\enquote {\bibinfo {title}
  {{Orbital Physics}},}\ }in\ \href {https://juser.fz-juelich.de/record/837488}
  {\emph {\bibinfo {booktitle} {{T}he {P}hysics of {C}orrelated {I}nsulators,
  {M}etals, and {S}uperconductors}}},\ Vol.~\bibinfo {volume} {7},\ \bibinfo
  {editor} {edited by\ \bibinfo {editor} {\bibfnamefont {Eva}\ \bibnamefont
  {Pavarini}}, \bibinfo {editor} {\bibfnamefont {Erik}\ \bibnamefont {Koch}},
  \bibinfo {editor} {\bibfnamefont {Richard}\ \bibnamefont {Scalettar}}, \ and\
  \bibinfo {editor} {\bibfnamefont {Richard}\ \bibnamefont {Martin}}}\
  (\bibinfo  {publisher} {Forschungszentrum J\"{u}lich GmbH Zentralbibliothek,
  Verlag},\ \bibinfo {address} {J\"{u}lich},\ \bibinfo {year} {2017})\
  Chap.~\bibinfo {chapter} {7}, pp.\ \bibinfo {pages} {111--144}\BibitemShut
  {NoStop}%
\bibitem [{\citenamefont {Ole\'{s}}(2023)}]{Andrzej2023}%
  \BibitemOpen
  \bibfield  {author} {\bibinfo {author} {\bibfnamefont {Andrzej~M.}\
  \bibnamefont {Ole\'{s}}},\ }\bibfield  {title} {\enquote {\bibinfo {title}
  {{Spin-Orbital Entanglement in Mott Insulators}},}\ }in\ \href {\doibase
  10.34734/FZJ-2023-03329} {\emph {\bibinfo {booktitle} {{O}rbital {P}hysics in
  {C}orrelated {M}atter}}},\ \bibinfo {series} {Modeling and Simulation},
  Vol.~\bibinfo {volume} {13}\ (\bibinfo  {publisher} {Forschungszentrum
  J\"{u}lich GmbH Zentralbibliothek, Verlag},\ \bibinfo {address}
  {J\"{u}lich},\ \bibinfo {year} {2023})\ Chap.~\bibinfo {chapter}
  {6}\BibitemShut {NoStop}%
\bibitem [{\citenamefont {Tokura}\ and\ \citenamefont
  {Nagaosa}(2000)}]{nagaosa}%
  \BibitemOpen
  \bibfield  {author} {\bibinfo {author} {\bibfnamefont {Y.}~\bibnamefont
  {Tokura}}\ and\ \bibinfo {author} {\bibfnamefont {N.}~\bibnamefont
  {Nagaosa}},\ }\bibfield  {title} {\enquote {\bibinfo {title} {Orbital physics
  in transition-metal oxides},}\ }\href {\doibase 10.1126/science.288.5465.462}
  {\bibfield  {journal} {\bibinfo  {journal} {Science}\ }\textbf {\bibinfo
  {volume} {288}},\ \bibinfo {pages} {462--468} (\bibinfo {year}
  {2000})}\BibitemShut {NoStop}%
\bibitem [{\citenamefont {Khaliullin}(2005)}]{PTPS.160.155}%
  \BibitemOpen
  \bibfield  {author} {\bibinfo {author} {\bibfnamefont {Giniyat}\ \bibnamefont
  {Khaliullin}},\ }\bibfield  {title} {\enquote {\bibinfo {title} {{Orbital
  Order and Fluctuations in Mott Insulators}},}\ }\href {\doibase
  10.1143/PTPS.160.155} {\bibfield  {journal} {\bibinfo  {journal} {Progress of
  Theoretical Physics Supplement}\ }\textbf {\bibinfo {volume} {160}},\
  \bibinfo {pages} {155--202} (\bibinfo {year} {2005})}\BibitemShut {NoStop}%
\bibitem [{\citenamefont {Hasan}\ and\ \citenamefont
  {Kane}(2010)}]{RevModPhys.82.3045}%
  \BibitemOpen
  \bibfield  {author} {\bibinfo {author} {\bibfnamefont {M.~Z.}\ \bibnamefont
  {Hasan}}\ and\ \bibinfo {author} {\bibfnamefont {C.~L.}\ \bibnamefont
  {Kane}},\ }\bibfield  {title} {\enquote {\bibinfo {title} {Colloquium:
  Topological insulators},}\ }\href {\doibase 10.1103/RevModPhys.82.3045}
  {\bibfield  {journal} {\bibinfo  {journal} {Rev. Mod. Phys.}\ }\textbf
  {\bibinfo {volume} {82}},\ \bibinfo {pages} {3045--3067} (\bibinfo {year}
  {2010})}\BibitemShut {NoStop}%
\bibitem [{\citenamefont {Qi}\ and\ \citenamefont
  {Zhang}(2011)}]{RevModPhys.83.1057}%
  \BibitemOpen
  \bibfield  {author} {\bibinfo {author} {\bibfnamefont {Xiao-Liang}\
  \bibnamefont {Qi}}\ and\ \bibinfo {author} {\bibfnamefont {Shou-Cheng}\
  \bibnamefont {Zhang}},\ }\bibfield  {title} {\enquote {\bibinfo {title}
  {Topological insulators and superconductors},}\ }\href {\doibase
  10.1103/RevModPhys.83.1057} {\bibfield  {journal} {\bibinfo  {journal} {Rev.
  Mod. Phys.}\ }\textbf {\bibinfo {volume} {83}},\ \bibinfo {pages}
  {1057--1110} (\bibinfo {year} {2011})}\BibitemShut {NoStop}%
\bibitem [{\citenamefont {Lv}\ \emph {et~al.}(2021)\citenamefont {Lv},
  \citenamefont {Qian},\ and\ \citenamefont {Ding}}]{RevModPhys.93.025002}%
  \BibitemOpen
  \bibfield  {author} {\bibinfo {author} {\bibfnamefont {B.~Q.}\ \bibnamefont
  {Lv}}, \bibinfo {author} {\bibfnamefont {T.}~\bibnamefont {Qian}}, \ and\
  \bibinfo {author} {\bibfnamefont {H.}~\bibnamefont {Ding}},\ }\bibfield
  {title} {\enquote {\bibinfo {title} {Experimental perspective on
  three-dimensional topological semimetals},}\ }\href {\doibase
  10.1103/RevModPhys.93.025002} {\bibfield  {journal} {\bibinfo  {journal}
  {Rev. Mod. Phys.}\ }\textbf {\bibinfo {volume} {93}},\ \bibinfo {pages}
  {025002} (\bibinfo {year} {2021})}\BibitemShut {NoStop}%
\bibitem [{\citenamefont {Armitage}\ \emph {et~al.}(2018)\citenamefont
  {Armitage}, \citenamefont {Mele},\ and\ \citenamefont
  {Vishwanath}}]{RevModPhys.90.015001}%
  \BibitemOpen
  \bibfield  {author} {\bibinfo {author} {\bibfnamefont {N.~P.}\ \bibnamefont
  {Armitage}}, \bibinfo {author} {\bibfnamefont {E.~J.}\ \bibnamefont {Mele}},
  \ and\ \bibinfo {author} {\bibfnamefont {Ashvin}\ \bibnamefont
  {Vishwanath}},\ }\bibfield  {title} {\enquote {\bibinfo {title} {Weyl and
  dirac semimetals in three-dimensional solids},}\ }\href {\doibase
  10.1103/RevModPhys.90.015001} {\bibfield  {journal} {\bibinfo  {journal}
  {Rev. Mod. Phys.}\ }\textbf {\bibinfo {volume} {90}},\ \bibinfo {pages}
  {015001} (\bibinfo {year} {2018})}\BibitemShut {NoStop}%
\bibitem [{\citenamefont {Witczak-Krempa}\ \emph {et~al.}(2014)\citenamefont
  {Witczak-Krempa}, \citenamefont {Chen}, \citenamefont {Kim},\ and\
  \citenamefont {Balents}}]{annurev-conmatphys-020911-125138}%
  \BibitemOpen
  \bibfield  {author} {\bibinfo {author} {\bibfnamefont {William}\ \bibnamefont
  {Witczak-Krempa}}, \bibinfo {author} {\bibfnamefont {Gang}\ \bibnamefont
  {Chen}}, \bibinfo {author} {\bibfnamefont {Yong~Baek}\ \bibnamefont {Kim}}, \
  and\ \bibinfo {author} {\bibfnamefont {Leon}\ \bibnamefont {Balents}},\
  }\bibfield  {title} {\enquote {\bibinfo {title} {Correlated quantum phenomena
  in the strong spin-orbit regime},}\ }\href {\doibase
  10.1146/annurev-conmatphys-020911-125138} {\bibfield  {journal} {\bibinfo
  {journal} {Annual Review of Condensed Matter Physics}\ }\textbf {\bibinfo
  {volume} {5}},\ \bibinfo {pages} {57--82} (\bibinfo {year}
  {2014})}\BibitemShut {NoStop}%
\bibitem [{\citenamefont {Ament}\ \emph {et~al.}(2011)\citenamefont {Ament},
  \citenamefont {van Veenendaal}, \citenamefont {Devereaux}, \citenamefont
  {Hill},\ and\ \citenamefont {van~den Brink}}]{RevModPhys.83.705}%
  \BibitemOpen
  \bibfield  {author} {\bibinfo {author} {\bibfnamefont {Luuk J.~P.}\
  \bibnamefont {Ament}}, \bibinfo {author} {\bibfnamefont {Michel}\
  \bibnamefont {van Veenendaal}}, \bibinfo {author} {\bibfnamefont {Thomas~P.}\
  \bibnamefont {Devereaux}}, \bibinfo {author} {\bibfnamefont {John~P.}\
  \bibnamefont {Hill}}, \ and\ \bibinfo {author} {\bibfnamefont {Jeroen}\
  \bibnamefont {van~den Brink}},\ }\bibfield  {title} {\enquote {\bibinfo
  {title} {{Resonant inelastic X-ray scattering studies of elementary
  excitations}},}\ }\href {\doibase 10.1103/RevModPhys.83.705} {\bibfield
  {journal} {\bibinfo  {journal} {Rev. Mod. Phys.}\ }\textbf {\bibinfo {volume}
  {83}},\ \bibinfo {pages} {705--767} (\bibinfo {year} {2011})}\BibitemShut
  {NoStop}%
\bibitem [{\citenamefont {Schlappa}\ \emph {et~al.}(2012)\citenamefont
  {Schlappa}, \citenamefont {Wohlfeld}, \citenamefont {Zhou}, \citenamefont
  {Mourigal}, \citenamefont {Haverkort}, \citenamefont {Strocov}, \citenamefont
  {Hozoi}, \citenamefont {Monney}, \citenamefont {Nishimoto}, \citenamefont
  {Singh},\ and\ \citenamefont {et~al.}}]{RIXS_orbital}%
  \BibitemOpen
  \bibfield  {author} {\bibinfo {author} {\bibfnamefont {J.}~\bibnamefont
  {Schlappa}}, \bibinfo {author} {\bibfnamefont {K.}~\bibnamefont {Wohlfeld}},
  \bibinfo {author} {\bibfnamefont {K.~J.}\ \bibnamefont {Zhou}}, \bibinfo
  {author} {\bibfnamefont {M.}~\bibnamefont {Mourigal}}, \bibinfo {author}
  {\bibfnamefont {M.~W.}\ \bibnamefont {Haverkort}}, \bibinfo {author}
  {\bibfnamefont {V.~N.}\ \bibnamefont {Strocov}}, \bibinfo {author}
  {\bibfnamefont {L.}~\bibnamefont {Hozoi}}, \bibinfo {author} {\bibfnamefont
  {C.}~\bibnamefont {Monney}}, \bibinfo {author} {\bibfnamefont
  {S.}~\bibnamefont {Nishimoto}}, \bibinfo {author} {\bibfnamefont
  {S.}~\bibnamefont {Singh}}, \ and\ \bibinfo {author} {\bibnamefont
  {et~al.}},\ }\bibfield  {title} {\enquote {\bibinfo {title} {{Spin-orbital
  separation in the quasi-one-dimensional Mott insulator Sr$_2$CuO$_3$}},}\
  }\href {\doibase 10.1038/nature10974} {\bibfield  {journal} {\bibinfo
  {journal} {Nature}\ }\textbf {\bibinfo {volume} {485}},\ \bibinfo {pages}
  {82--85} (\bibinfo {year} {2012})}\BibitemShut {NoStop}%
\bibitem [{\citenamefont {Chen}\ \emph
  {et~al.}(2016{\natexlab{a}})\citenamefont {Chen}, \citenamefont {Kee},\ and\
  \citenamefont {Kim}}]{PhysRevB.93.245134}%
  \BibitemOpen
  \bibfield  {author} {\bibinfo {author} {\bibfnamefont {Gang}\ \bibnamefont
  {Chen}}, \bibinfo {author} {\bibfnamefont {Hae-Young}\ \bibnamefont {Kee}}, \
  and\ \bibinfo {author} {\bibfnamefont {Yong~Baek}\ \bibnamefont {Kim}},\
  }\bibfield  {title} {\enquote {\bibinfo {title} {{Cluster Mott insulators and
  two Curie-Weiss regimes on an anisotropic kagome lattice}},}\ }\href
  {\doibase 10.1103/PhysRevB.93.245134} {\bibfield  {journal} {\bibinfo
  {journal} {Phys. Rev. B}\ }\textbf {\bibinfo {volume} {93}},\ \bibinfo
  {pages} {245134} (\bibinfo {year} {2016}{\natexlab{a}})}\BibitemShut
  {NoStop}%
\bibitem [{\citenamefont {Chen}\ and\ \citenamefont
  {Lee}(2018)}]{PhysRevB.97.035124}%
  \BibitemOpen
  \bibfield  {author} {\bibinfo {author} {\bibfnamefont {Gang}\ \bibnamefont
  {Chen}}\ and\ \bibinfo {author} {\bibfnamefont {Patrick~A.}\ \bibnamefont
  {Lee}},\ }\bibfield  {title} {\enquote {\bibinfo {title} {{Emergent orbitals
  in the cluster Mott insulator on a breathing kagome lattice}},}\ }\href
  {\doibase 10.1103/PhysRevB.97.035124} {\bibfield  {journal} {\bibinfo
  {journal} {Phys. Rev. B}\ }\textbf {\bibinfo {volume} {97}},\ \bibinfo
  {pages} {035124} (\bibinfo {year} {2018})}\BibitemShut {NoStop}%
\bibitem [{\citenamefont {Chen}\ \emph {et~al.}(2014)\citenamefont {Chen},
  \citenamefont {Kee},\ and\ \citenamefont {Kim}}]{PhysRevLett.113.197202}%
  \BibitemOpen
  \bibfield  {author} {\bibinfo {author} {\bibfnamefont {Gang}\ \bibnamefont
  {Chen}}, \bibinfo {author} {\bibfnamefont {Hae-Young}\ \bibnamefont {Kee}}, \
  and\ \bibinfo {author} {\bibfnamefont {Yong~Baek}\ \bibnamefont {Kim}},\
  }\bibfield  {title} {\enquote {\bibinfo {title} {Fractionalized charge
  excitations in a spin liquid on partially filled pyrochlore lattices},}\
  }\href {\doibase 10.1103/PhysRevLett.113.197202} {\bibfield  {journal}
  {\bibinfo  {journal} {Phys. Rev. Lett.}\ }\textbf {\bibinfo {volume} {113}},\
  \bibinfo {pages} {197202} (\bibinfo {year} {2014})}\BibitemShut {NoStop}%
\bibitem [{\citenamefont {Kimura}\ \emph {et~al.}(2014)\citenamefont {Kimura},
  \citenamefont {Nakatsuji},\ and\ \citenamefont
  {Kimura}}]{PhysRevB.90.060414}%
  \BibitemOpen
  \bibfield  {author} {\bibinfo {author} {\bibfnamefont {K.}~\bibnamefont
  {Kimura}}, \bibinfo {author} {\bibfnamefont {S.}~\bibnamefont {Nakatsuji}}, \
  and\ \bibinfo {author} {\bibfnamefont {T.}~\bibnamefont {Kimura}},\
  }\bibfield  {title} {\enquote {\bibinfo {title} {Experimental realization of
  a quantum breathing pyrochlore antiferromagnet},}\ }\href {\doibase
  10.1103/PhysRevB.90.060414} {\bibfield  {journal} {\bibinfo  {journal} {Phys.
  Rev. B}\ }\textbf {\bibinfo {volume} {90}},\ \bibinfo {pages} {060414}
  (\bibinfo {year} {2014})}\BibitemShut {NoStop}%
\bibitem [{\citenamefont {Rau}\ \emph {et~al.}(2016)\citenamefont {Rau},
  \citenamefont {Wu}, \citenamefont {May}, \citenamefont {Poudel},
  \citenamefont {Winn}, \citenamefont {Garlea}, \citenamefont {Huq},
  \citenamefont {Whitfield}, \citenamefont {Taylor}, \citenamefont {Lumsden},
  \citenamefont {Gingras},\ and\ \citenamefont
  {Christianson}}]{PhysRevLett.116.257204}%
  \BibitemOpen
  \bibfield  {author} {\bibinfo {author} {\bibfnamefont {J.~G.}\ \bibnamefont
  {Rau}}, \bibinfo {author} {\bibfnamefont {L.~S.}\ \bibnamefont {Wu}},
  \bibinfo {author} {\bibfnamefont {A.~F.}\ \bibnamefont {May}}, \bibinfo
  {author} {\bibfnamefont {L.}~\bibnamefont {Poudel}}, \bibinfo {author}
  {\bibfnamefont {B.}~\bibnamefont {Winn}}, \bibinfo {author} {\bibfnamefont
  {V.~O.}\ \bibnamefont {Garlea}}, \bibinfo {author} {\bibfnamefont
  {A.}~\bibnamefont {Huq}}, \bibinfo {author} {\bibfnamefont {P.}~\bibnamefont
  {Whitfield}}, \bibinfo {author} {\bibfnamefont {A.~E.}\ \bibnamefont
  {Taylor}}, \bibinfo {author} {\bibfnamefont {M.~D.}\ \bibnamefont {Lumsden}},
  \bibinfo {author} {\bibfnamefont {M.~J.~P.}\ \bibnamefont {Gingras}}, \ and\
  \bibinfo {author} {\bibfnamefont {A.~D.}\ \bibnamefont {Christianson}},\
  }\bibfield  {title} {\enquote {\bibinfo {title} {{Anisotropic Exchange within
  Decoupled Tetrahedra in the Quantum Breathing Pyrochlore
  ${\mathrm{Ba}}_{3}{\mathrm{Yb}}_{2}{\mathrm{Zn}}_{5}{\mathrm{O}}_{11}$}},}\
  }\href {\doibase 10.1103/PhysRevLett.116.257204} {\bibfield  {journal}
  {\bibinfo  {journal} {Phys. Rev. Lett.}\ }\textbf {\bibinfo {volume} {116}},\
  \bibinfo {pages} {257204} (\bibinfo {year} {2016})}\BibitemShut {NoStop}%
\bibitem [{\citenamefont {Savary}\ \emph {et~al.}(2016)\citenamefont {Savary},
  \citenamefont {Wang}, \citenamefont {Kee}, \citenamefont {Kim}, \citenamefont
  {Yu},\ and\ \citenamefont {Chen}}]{PhysRevB.94.075146}%
  \BibitemOpen
  \bibfield  {author} {\bibinfo {author} {\bibfnamefont {Lucile}\ \bibnamefont
  {Savary}}, \bibinfo {author} {\bibfnamefont {Xiaoqun}\ \bibnamefont {Wang}},
  \bibinfo {author} {\bibfnamefont {Hae-Young}\ \bibnamefont {Kee}}, \bibinfo
  {author} {\bibfnamefont {Yong~Baek}\ \bibnamefont {Kim}}, \bibinfo {author}
  {\bibfnamefont {Yue}\ \bibnamefont {Yu}}, \ and\ \bibinfo {author}
  {\bibfnamefont {Gang}\ \bibnamefont {Chen}},\ }\bibfield  {title} {\enquote
  {\bibinfo {title} {Quantum spin ice on the breathing pyrochlore lattice},}\
  }\href {\doibase 10.1103/PhysRevB.94.075146} {\bibfield  {journal} {\bibinfo
  {journal} {Phys. Rev. B}\ }\textbf {\bibinfo {volume} {94}},\ \bibinfo
  {pages} {075146} (\bibinfo {year} {2016})}\BibitemShut {NoStop}%
\bibitem [{\citenamefont {Nikolaev}\ \emph {et~al.}(2021)\citenamefont
  {Nikolaev}, \citenamefont {Solovyev},\ and\ \citenamefont
  {Streltsov}}]{nikolaev2020quantum}%
  \BibitemOpen
  \bibfield  {author} {\bibinfo {author} {\bibfnamefont {S.~A.}\ \bibnamefont
  {Nikolaev}}, \bibinfo {author} {\bibfnamefont {I.~V.}\ \bibnamefont
  {Solovyev}}, \ and\ \bibinfo {author} {\bibfnamefont {S.~V.}\ \bibnamefont
  {Streltsov}},\ }\bibfield  {title} {\enquote {\bibinfo {title} {{Quantum spin
  liquid and cluster Mott insulator phases in the Mo$_{3}$O$_{8}$ magnets}},}\
  }\href {\doibase 10.1038/s41535-021-00316-7} {\bibfield  {journal} {\bibinfo
  {journal} {npj Quantum Mater.}\ }\textbf {\bibinfo {volume} {6}},\ \bibinfo
  {pages} {25} (\bibinfo {year} {2021})}\BibitemShut {NoStop}%
\bibitem [{\citenamefont {Sheckelton}\ \emph {et~al.}(2012)\citenamefont
  {Sheckelton}, \citenamefont {Neilson}, \citenamefont {Soltan},\ and\
  \citenamefont {McQueen}}]{Sheckelton2012}%
  \BibitemOpen
  \bibfield  {author} {\bibinfo {author} {\bibfnamefont {J.~P.}\ \bibnamefont
  {Sheckelton}}, \bibinfo {author} {\bibfnamefont {J.~R.}\ \bibnamefont
  {Neilson}}, \bibinfo {author} {\bibfnamefont {D.~G.}\ \bibnamefont {Soltan}},
  \ and\ \bibinfo {author} {\bibfnamefont {T.~M.}\ \bibnamefont {McQueen}},\
  }\bibfield  {title} {\enquote {\bibinfo {title} {{Possible valence-bond
  condensation in the frustrated cluster magnet LiZn$_2$Mo$_3$O$_8$}},}\ }\href
  {\doibase 10.1038/nmat3329} {\bibfield  {journal} {\bibinfo  {journal}
  {Nature Materials}\ }\textbf {\bibinfo {volume} {11}},\ \bibinfo {pages}
  {493?496} (\bibinfo {year} {2012})}\BibitemShut {NoStop}%
\bibitem [{\citenamefont {Kim}\ \emph {et~al.}(2014{\natexlab{a}})\citenamefont
  {Kim}, \citenamefont {Im}, \citenamefont {Han},\ and\ \citenamefont
  {Jin}}]{Heung2014}%
  \BibitemOpen
  \bibfield  {author} {\bibinfo {author} {\bibfnamefont {Heung-Sik}\
  \bibnamefont {Kim}}, \bibinfo {author} {\bibfnamefont {Jino}\ \bibnamefont
  {Im}}, \bibinfo {author} {\bibfnamefont {Myung~Joon}\ \bibnamefont {Han}}, \
  and\ \bibinfo {author} {\bibfnamefont {Hosub}\ \bibnamefont {Jin}},\
  }\bibfield  {title} {\enquote {\bibinfo {title} {{Spin-orbital entangled
  molecular $J_{eff}$ states in lacunar spinel compounds}},}\ }\href {\doibase
  10.1038/ncomms4988} {\bibfield  {journal} {\bibinfo  {journal} {Nature
  Communications}\ }\textbf {\bibinfo {volume} {5}},\ \bibinfo {pages} {3988}
  (\bibinfo {year} {2014}{\natexlab{a}})}\BibitemShut {NoStop}%
\bibitem [{\citenamefont {Yao}\ \emph {et~al.}(2020)\citenamefont {Yao},
  \citenamefont {Zhang}, \citenamefont {Kim}, \citenamefont {Wang},\ and\
  \citenamefont {Chen}}]{PhysRevResearch.2.043424}%
  \BibitemOpen
  \bibfield  {author} {\bibinfo {author} {\bibfnamefont {Xu-Ping}\ \bibnamefont
  {Yao}}, \bibinfo {author} {\bibfnamefont {Xiao-Tian}\ \bibnamefont {Zhang}},
  \bibinfo {author} {\bibfnamefont {Yong~Baek}\ \bibnamefont {Kim}}, \bibinfo
  {author} {\bibfnamefont {Xiaoqun}\ \bibnamefont {Wang}}, \ and\ \bibinfo
  {author} {\bibfnamefont {Gang}\ \bibnamefont {Chen}},\ }\bibfield  {title}
  {\enquote {\bibinfo {title} {Clusterization transition between cluster mott
  insulators on a breathing kagome lattice},}\ }\href {\doibase
  10.1103/PhysRevResearch.2.043424} {\bibfield  {journal} {\bibinfo  {journal}
  {Phys. Rev. Research}\ }\textbf {\bibinfo {volume} {2}},\ \bibinfo {pages}
  {043424} (\bibinfo {year} {2020})}\BibitemShut {NoStop}%
\bibitem [{\citenamefont {Mourigal}\ \emph {et~al.}(2014)\citenamefont
  {Mourigal}, \citenamefont {Fuhrman}, \citenamefont {Sheckelton},
  \citenamefont {Wartelle}, \citenamefont {Rodriguez-Rivera}, \citenamefont
  {Abernathy}, \citenamefont {McQueen},\ and\ \citenamefont
  {Broholm}}]{PhysRevLett.112.027202}%
  \BibitemOpen
  \bibfield  {author} {\bibinfo {author} {\bibfnamefont {M.}~\bibnamefont
  {Mourigal}}, \bibinfo {author} {\bibfnamefont {W.~T.}\ \bibnamefont
  {Fuhrman}}, \bibinfo {author} {\bibfnamefont {J.~P.}\ \bibnamefont
  {Sheckelton}}, \bibinfo {author} {\bibfnamefont {A.}~\bibnamefont
  {Wartelle}}, \bibinfo {author} {\bibfnamefont {J.~A.}\ \bibnamefont
  {Rodriguez-Rivera}}, \bibinfo {author} {\bibfnamefont {D.~L.}\ \bibnamefont
  {Abernathy}}, \bibinfo {author} {\bibfnamefont {T.~M.}\ \bibnamefont
  {McQueen}}, \ and\ \bibinfo {author} {\bibfnamefont {C.~L.}\ \bibnamefont
  {Broholm}},\ }\bibfield  {title} {\enquote {\bibinfo {title} {{Molecular
  Quantum Magnetism in
  ${\mathrm{LiZn}}_{2}{\mathrm{Mo}}_{3}{\mathrm{O}}_{8}$}},}\ }\href {\doibase
  10.1103/PhysRevLett.112.027202} {\bibfield  {journal} {\bibinfo  {journal}
  {Phys. Rev. Lett.}\ }\textbf {\bibinfo {volume} {112}},\ \bibinfo {pages}
  {027202} (\bibinfo {year} {2014})}\BibitemShut {NoStop}%
\bibitem [{\citenamefont {Dissanayake}\ \emph {et~al.}(2022)\citenamefont
  {Dissanayake}, \citenamefont {Shi}, \citenamefont {Rau}, \citenamefont {Bag},
  \citenamefont {Steinhardt}, \citenamefont {Butch}, \citenamefont {Frontzek},
  \citenamefont {Podlesnyak}, \citenamefont {Graf}, \citenamefont
  {Marjerrison}, \citenamefont {Liu}, \citenamefont {Gingras},\ and\
  \citenamefont {Haravifard}}]{dissanayake2021understanding}%
  \BibitemOpen
  \bibfield  {author} {\bibinfo {author} {\bibfnamefont {Sachith}\ \bibnamefont
  {Dissanayake}}, \bibinfo {author} {\bibfnamefont {Zhenzhong}\ \bibnamefont
  {Shi}}, \bibinfo {author} {\bibfnamefont {Jeffrey~G.}\ \bibnamefont {Rau}},
  \bibinfo {author} {\bibfnamefont {Rabindranath}\ \bibnamefont {Bag}},
  \bibinfo {author} {\bibfnamefont {William}\ \bibnamefont {Steinhardt}},
  \bibinfo {author} {\bibfnamefont {Nicholas~P.}\ \bibnamefont {Butch}},
  \bibinfo {author} {\bibfnamefont {Matthias}\ \bibnamefont {Frontzek}},
  \bibinfo {author} {\bibfnamefont {Andrey}\ \bibnamefont {Podlesnyak}},
  \bibinfo {author} {\bibfnamefont {David}\ \bibnamefont {Graf}}, \bibinfo
  {author} {\bibfnamefont {Casey}\ \bibnamefont {Marjerrison}}, \bibinfo
  {author} {\bibfnamefont {Jue}\ \bibnamefont {Liu}}, \bibinfo {author}
  {\bibfnamefont {Michel J.~P.}\ \bibnamefont {Gingras}}, \ and\ \bibinfo
  {author} {\bibfnamefont {Sara}\ \bibnamefont {Haravifard}},\ }\bibfield
  {title} {\enquote {\bibinfo {title} {{Towards understanding the magnetic
  properties of the breathing pyrochlore compound Ba$_3$Yb$_2$Zn$_5$O$_{11}$
  through single-crystal studies}},}\ }\href {\doibase
  10.1038/s41535-022-00488-w} {\bibfield  {journal} {\bibinfo  {journal} {npj
  Quantum Materials}\ }\textbf {\bibinfo {volume} {7}},\ \bibinfo {pages} {77}
  (\bibinfo {year} {2022})}\BibitemShut {NoStop}%
\bibitem [{\citenamefont {Po}\ \emph {et~al.}(2018)\citenamefont {Po},
  \citenamefont {Zou}, \citenamefont {Vishwanath},\ and\ \citenamefont
  {Senthil}}]{PhysRevX.8.031089}%
  \BibitemOpen
  \bibfield  {author} {\bibinfo {author} {\bibfnamefont {Hoi~Chun}\
  \bibnamefont {Po}}, \bibinfo {author} {\bibfnamefont {Liujun}\ \bibnamefont
  {Zou}}, \bibinfo {author} {\bibfnamefont {Ashvin}\ \bibnamefont
  {Vishwanath}}, \ and\ \bibinfo {author} {\bibfnamefont {T.}~\bibnamefont
  {Senthil}},\ }\bibfield  {title} {\enquote {\bibinfo {title} {{Origin of Mott
  Insulating Behavior and Superconductivity in Twisted Bilayer Graphene}},}\
  }\href {\doibase 10.1103/PhysRevX.8.031089} {\bibfield  {journal} {\bibinfo
  {journal} {Phys. Rev. X}\ }\textbf {\bibinfo {volume} {8}},\ \bibinfo {pages}
  {031089} (\bibinfo {year} {2018})}\BibitemShut {NoStop}%
\bibitem [{Mak(2022)}]{Mak2022}%
  \BibitemOpen
  \bibfield  {title} {\enquote {\bibinfo {title} {Semiconductor moir\'{e}
  materials},}\ }\href {\doibase 10.1038/s41565-022-01165-6} {\bibfield
  {journal} {\bibinfo  {journal} {Nat. Nanotechnol.}\ }\textbf {\bibinfo
  {volume} {17}},\ \bibinfo {pages} {686--695} (\bibinfo {year}
  {2022})}\BibitemShut {NoStop}%
\bibitem [{\citenamefont {Molavian}\ \emph {et~al.}(2007)\citenamefont
  {Molavian}, \citenamefont {Gingras},\ and\ \citenamefont
  {Canals}}]{PhysRevLett.98.157204}%
  \BibitemOpen
  \bibfield  {author} {\bibinfo {author} {\bibfnamefont {Hamid~R.}\
  \bibnamefont {Molavian}}, \bibinfo {author} {\bibfnamefont {Michel J.~P.}\
  \bibnamefont {Gingras}}, \ and\ \bibinfo {author} {\bibfnamefont {Benjamin}\
  \bibnamefont {Canals}},\ }\bibfield  {title} {\enquote {\bibinfo {title}
  {{Dynamically Induced Frustration as a Route to a Quantum Spin Ice State in
  ${\mathrm{Tb}}_{2}{\mathrm{Ti}}_{2}{\mathrm{O}}_{7}$ via Virtual Crystal
  Field Excitations and Quantum Many-Body Effects}},}\ }\href {\doibase
  10.1103/PhysRevLett.98.157204} {\bibfield  {journal} {\bibinfo  {journal}
  {Phys. Rev. Lett.}\ }\textbf {\bibinfo {volume} {98}},\ \bibinfo {pages}
  {157204} (\bibinfo {year} {2007})}\BibitemShut {NoStop}%
\bibitem [{\citenamefont {Gingras}\ \emph {et~al.}(2000)\citenamefont
  {Gingras}, \citenamefont {den Hertog}, \citenamefont {Faucher}, \citenamefont
  {Gardner}, \citenamefont {Dunsiger}, \citenamefont {Chang}, \citenamefont
  {Gaulin}, \citenamefont {Raju},\ and\ \citenamefont
  {Greedan}}]{PhysRevB.62.6496}%
  \BibitemOpen
  \bibfield  {author} {\bibinfo {author} {\bibfnamefont {M.~J.~P.}\
  \bibnamefont {Gingras}}, \bibinfo {author} {\bibfnamefont {B.~C.}\
  \bibnamefont {den Hertog}}, \bibinfo {author} {\bibfnamefont
  {M.}~\bibnamefont {Faucher}}, \bibinfo {author} {\bibfnamefont {J.~S.}\
  \bibnamefont {Gardner}}, \bibinfo {author} {\bibfnamefont {S.~R.}\
  \bibnamefont {Dunsiger}}, \bibinfo {author} {\bibfnamefont {L.~J.}\
  \bibnamefont {Chang}}, \bibinfo {author} {\bibfnamefont {B.~D.}\ \bibnamefont
  {Gaulin}}, \bibinfo {author} {\bibfnamefont {N.~P.}\ \bibnamefont {Raju}}, \
  and\ \bibinfo {author} {\bibfnamefont {J.~E.}\ \bibnamefont {Greedan}},\
  }\bibfield  {title} {\enquote {\bibinfo {title} {{Thermodynamic and
  single-ion properties of ${\mathrm{Tb}}^{3+}$ within the collective
  paramagnetic-spin liquid state of the frustrated pyrochlore antiferromagnet
  ${\mathrm{Tb}}_{2}{\mathrm{Ti}}_{2}{\mathrm{O}}_{7}$}},}\ }\href {\doibase
  10.1103/PhysRevB.62.6496} {\bibfield  {journal} {\bibinfo  {journal} {Phys.
  Rev. B}\ }\textbf {\bibinfo {volume} {62}},\ \bibinfo {pages} {6496--6511}
  (\bibinfo {year} {2000})}\BibitemShut {NoStop}%
\bibitem [{\citenamefont {Gaulin}\ \emph {et~al.}(2011)\citenamefont {Gaulin},
  \citenamefont {Gardner}, \citenamefont {McClarty},\ and\ \citenamefont
  {Gingras}}]{PhysRevB.84.140402}%
  \BibitemOpen
  \bibfield  {author} {\bibinfo {author} {\bibfnamefont {B.~D.}\ \bibnamefont
  {Gaulin}}, \bibinfo {author} {\bibfnamefont {J.~S.}\ \bibnamefont {Gardner}},
  \bibinfo {author} {\bibfnamefont {P.~A.}\ \bibnamefont {McClarty}}, \ and\
  \bibinfo {author} {\bibfnamefont {M.~J.~P.}\ \bibnamefont {Gingras}},\
  }\bibfield  {title} {\enquote {\bibinfo {title} {{Lack of evidence for a
  singlet crystal-field ground state in the magnetic pyrochlore
  Tb${}_{2}$Ti${}_{2}$O${}_{7}$}},}\ }\href {\doibase
  10.1103/PhysRevB.84.140402} {\bibfield  {journal} {\bibinfo  {journal} {Phys.
  Rev. B}\ }\textbf {\bibinfo {volume} {84}},\ \bibinfo {pages} {140402}
  (\bibinfo {year} {2011})}\BibitemShut {NoStop}%
\bibitem [{\citenamefont {Fritsch}\ \emph {et~al.}(2013)\citenamefont
  {Fritsch}, \citenamefont {Ross}, \citenamefont {Qiu}, \citenamefont {Copley},
  \citenamefont {Guidi}, \citenamefont {Bewley}, \citenamefont {Dabkowska},\
  and\ \citenamefont {Gaulin}}]{PhysRevB.87.094410}%
  \BibitemOpen
  \bibfield  {author} {\bibinfo {author} {\bibfnamefont {K.}~\bibnamefont
  {Fritsch}}, \bibinfo {author} {\bibfnamefont {K.~A.}\ \bibnamefont {Ross}},
  \bibinfo {author} {\bibfnamefont {Y.}~\bibnamefont {Qiu}}, \bibinfo {author}
  {\bibfnamefont {J.~R.~D.}\ \bibnamefont {Copley}}, \bibinfo {author}
  {\bibfnamefont {T.}~\bibnamefont {Guidi}}, \bibinfo {author} {\bibfnamefont
  {R.~I.}\ \bibnamefont {Bewley}}, \bibinfo {author} {\bibfnamefont {H.~A.}\
  \bibnamefont {Dabkowska}}, \ and\ \bibinfo {author} {\bibfnamefont {B.~D.}\
  \bibnamefont {Gaulin}},\ }\bibfield  {title} {\enquote {\bibinfo {title}
  {{Antiferromagnetic spin ice correlations at
  ($\frac{1}{2}$,$\frac{1}{2}$,$\frac{1}{2}$) in the ground state of the
  pyrochlore magnet Tb${}_{2}$Ti${}_{2}$O${}_{7}$}},}\ }\href {\doibase
  10.1103/PhysRevB.87.094410} {\bibfield  {journal} {\bibinfo  {journal} {Phys.
  Rev. B}\ }\textbf {\bibinfo {volume} {87}},\ \bibinfo {pages} {094410}
  (\bibinfo {year} {2013})}\BibitemShut {NoStop}%
\bibitem [{\citenamefont {Fritsch}\ \emph {et~al.}(2014)\citenamefont
  {Fritsch}, \citenamefont {Kermarrec}, \citenamefont {Ross}, \citenamefont
  {Qiu}, \citenamefont {Copley}, \citenamefont {Pomaranski}, \citenamefont
  {Kycia}, \citenamefont {Dabkowska},\ and\ \citenamefont
  {Gaulin}}]{PhysRevB.90.014429}%
  \BibitemOpen
  \bibfield  {author} {\bibinfo {author} {\bibfnamefont {K.}~\bibnamefont
  {Fritsch}}, \bibinfo {author} {\bibfnamefont {E.}~\bibnamefont {Kermarrec}},
  \bibinfo {author} {\bibfnamefont {K.~A.}\ \bibnamefont {Ross}}, \bibinfo
  {author} {\bibfnamefont {Y.}~\bibnamefont {Qiu}}, \bibinfo {author}
  {\bibfnamefont {J.~R.~D.}\ \bibnamefont {Copley}}, \bibinfo {author}
  {\bibfnamefont {D.}~\bibnamefont {Pomaranski}}, \bibinfo {author}
  {\bibfnamefont {J.~B.}\ \bibnamefont {Kycia}}, \bibinfo {author}
  {\bibfnamefont {H.~A.}\ \bibnamefont {Dabkowska}}, \ and\ \bibinfo {author}
  {\bibfnamefont {B.~D.}\ \bibnamefont {Gaulin}},\ }\bibfield  {title}
  {\enquote {\bibinfo {title} {{Temperature and magnetic field dependence of
  spin-ice correlations in the pyrochlore magnet
  ${\mathrm{Tb}}_{2}{\mathrm{Ti}}_{2}{\mathrm{O}}_{7}$}},}\ }\href {\doibase
  10.1103/PhysRevB.90.014429} {\bibfield  {journal} {\bibinfo  {journal} {Phys.
  Rev. B}\ }\textbf {\bibinfo {volume} {90}},\ \bibinfo {pages} {014429}
  (\bibinfo {year} {2014})}\BibitemShut {NoStop}%
\bibitem [{\citenamefont {Liu}\ \emph {et~al.}(2019)\citenamefont {Liu},
  \citenamefont {Li},\ and\ \citenamefont {Chen}}]{PhysRevB.99.224407}%
  \BibitemOpen
  \bibfield  {author} {\bibinfo {author} {\bibfnamefont {Changle}\ \bibnamefont
  {Liu}}, \bibinfo {author} {\bibfnamefont {Fei-Ye}\ \bibnamefont {Li}}, \ and\
  \bibinfo {author} {\bibfnamefont {Gang}\ \bibnamefont {Chen}},\ }\bibfield
  {title} {\enquote {\bibinfo {title} {{Upper branch magnetism in quantum
  magnets: Collapses of excited levels and emergent selection rules}},}\ }\href
  {\doibase 10.1103/PhysRevB.99.224407} {\bibfield  {journal} {\bibinfo
  {journal} {Phys. Rev. B}\ }\textbf {\bibinfo {volume} {99}},\ \bibinfo
  {pages} {224407} (\bibinfo {year} {2019})}\BibitemShut {NoStop}%
\bibitem [{\citenamefont {Chen}\ \emph {et~al.}(2010)\citenamefont {Chen},
  \citenamefont {Pereira},\ and\ \citenamefont {Balents}}]{PhysRevB.82.174440}%
  \BibitemOpen
  \bibfield  {author} {\bibinfo {author} {\bibfnamefont {Gang}\ \bibnamefont
  {Chen}}, \bibinfo {author} {\bibfnamefont {Rodrigo}\ \bibnamefont {Pereira}},
  \ and\ \bibinfo {author} {\bibfnamefont {Leon}\ \bibnamefont {Balents}},\
  }\bibfield  {title} {\enquote {\bibinfo {title} {Exotic phases induced by
  strong spin-orbit coupling in ordered double perovskites},}\ }\href {\doibase
  10.1103/PhysRevB.82.174440} {\bibfield  {journal} {\bibinfo  {journal} {Phys.
  Rev. B}\ }\textbf {\bibinfo {volume} {82}},\ \bibinfo {pages} {174440}
  (\bibinfo {year} {2010})}\BibitemShut {NoStop}%
\bibitem [{\citenamefont {Paramekanti}\ \emph {et~al.}(2020)\citenamefont
  {Paramekanti}, \citenamefont {Maharaj},\ and\ \citenamefont
  {Gaulin}}]{PhysRevB.101.054439}%
  \BibitemOpen
  \bibfield  {author} {\bibinfo {author} {\bibfnamefont {A.}~\bibnamefont
  {Paramekanti}}, \bibinfo {author} {\bibfnamefont {D.~D.}\ \bibnamefont
  {Maharaj}}, \ and\ \bibinfo {author} {\bibfnamefont {B.~D.}\ \bibnamefont
  {Gaulin}},\ }\bibfield  {title} {\enquote {\bibinfo {title} {{Octupolar order
  in $d$-orbital Mott insulators}},}\ }\href {\doibase
  10.1103/PhysRevB.101.054439} {\bibfield  {journal} {\bibinfo  {journal}
  {Phys. Rev. B}\ }\textbf {\bibinfo {volume} {101}},\ \bibinfo {pages}
  {054439} (\bibinfo {year} {2020})}\BibitemShut {NoStop}%
\bibitem [{\citenamefont {Romh\'anyi}\ \emph {et~al.}(2017)\citenamefont
  {Romh\'anyi}, \citenamefont {Balents},\ and\ \citenamefont
  {Jackeli}}]{PhysRevLett.118.217202}%
  \BibitemOpen
  \bibfield  {author} {\bibinfo {author} {\bibfnamefont {Judit}\ \bibnamefont
  {Romh\'anyi}}, \bibinfo {author} {\bibfnamefont {Leon}\ \bibnamefont
  {Balents}}, \ and\ \bibinfo {author} {\bibfnamefont {George}\ \bibnamefont
  {Jackeli}},\ }\bibfield  {title} {\enquote {\bibinfo {title} {{Spin-Orbit
  Dimers and Noncollinear Phases in ${d}^{1}$ Cubic Double Perovskites}},}\
  }\href {\doibase 10.1103/PhysRevLett.118.217202} {\bibfield  {journal}
  {\bibinfo  {journal} {Phys. Rev. Lett.}\ }\textbf {\bibinfo {volume} {118}},\
  \bibinfo {pages} {217202} (\bibinfo {year} {2017})}\BibitemShut {NoStop}%
\bibitem [{\citenamefont {Weng}\ and\ \citenamefont
  {Dong}(2021)}]{PhysRevB.104.165150}%
  \BibitemOpen
  \bibfield  {author} {\bibinfo {author} {\bibfnamefont {Yakui}\ \bibnamefont
  {Weng}}\ and\ \bibinfo {author} {\bibfnamefont {Shuai}\ \bibnamefont
  {Dong}},\ }\bibfield  {title} {\enquote {\bibinfo {title} {{Manipulation of
  ${J}_{\mathrm{eff}}\phantom{\rule{4pt}{0ex}}=\phantom{\rule{4pt}{0ex}}\frac{3}{2}$
  states by tuning the tetragonal distortion}},}\ }\href {\doibase
  10.1103/PhysRevB.104.165150} {\bibfield  {journal} {\bibinfo  {journal}
  {Phys. Rev. B}\ }\textbf {\bibinfo {volume} {104}},\ \bibinfo {pages}
  {165150} (\bibinfo {year} {2021})}\BibitemShut {NoStop}%
\bibitem [{\citenamefont {Weng}\ \emph {et~al.}(2020)\citenamefont {Weng},
  \citenamefont {Li},\ and\ \citenamefont {Dong}}]{PhysRevB.102.180401}%
  \BibitemOpen
  \bibfield  {author} {\bibinfo {author} {\bibfnamefont {Yakui}\ \bibnamefont
  {Weng}}, \bibinfo {author} {\bibfnamefont {Xing'ao}\ \bibnamefont {Li}}, \
  and\ \bibinfo {author} {\bibfnamefont {Shuai}\ \bibnamefont {Dong}},\
  }\bibfield  {title} {\enquote {\bibinfo {title} {Strong tuning of magnetism
  and electronic structure by spin orientation},}\ }\href {\doibase
  10.1103/PhysRevB.102.180401} {\bibfield  {journal} {\bibinfo  {journal}
  {Phys. Rev. B}\ }\textbf {\bibinfo {volume} {102}},\ \bibinfo {pages}
  {180401} (\bibinfo {year} {2020})}\BibitemShut {NoStop}%
\bibitem [{\citenamefont {Yamada}\ \emph {et~al.}(2018)\citenamefont {Yamada},
  \citenamefont {Oshikawa},\ and\ \citenamefont
  {Jackeli}}]{PhysRevLett.121.097201}%
  \BibitemOpen
  \bibfield  {author} {\bibinfo {author} {\bibfnamefont {Masahiko~G.}\
  \bibnamefont {Yamada}}, \bibinfo {author} {\bibfnamefont {Masaki}\
  \bibnamefont {Oshikawa}}, \ and\ \bibinfo {author} {\bibfnamefont {George}\
  \bibnamefont {Jackeli}},\ }\bibfield  {title} {\enquote {\bibinfo {title}
  {{Emergent $\mathrm{SU}(4)$ Symmetry in
  $\ensuremath{\alpha}\text{\ensuremath{-}}{\mathrm{ZrCl}}_{3}$ and Crystalline
  Spin-Orbital Liquids}},}\ }\href {\doibase 10.1103/PhysRevLett.121.097201}
  {\bibfield  {journal} {\bibinfo  {journal} {Phys. Rev. Lett.}\ }\textbf
  {\bibinfo {volume} {121}},\ \bibinfo {pages} {097201} (\bibinfo {year}
  {2018})}\BibitemShut {NoStop}%
\bibitem [{\citenamefont {Ole{\'{s}}}(2009)}]{oles2009spinorbital}%
  \BibitemOpen
  \bibfield  {author} {\bibinfo {author} {\bibfnamefont {Andrzej~M.}\
  \bibnamefont {Ole{\'{s}}}},\ }\bibfield  {title} {\enquote {\bibinfo {title}
  {{Spin-Orbital Physics in Transition Metal Oxides}},}\ }\href {\doibase
  10.12693/aphyspola.115.36} {\bibfield  {journal} {\bibinfo  {journal} {Acta
  Physica Polonica A}\ }\textbf {\bibinfo {volume} {115}},\ \bibinfo {pages}
  {36--46} (\bibinfo {year} {2009})}\BibitemShut {NoStop}%
\bibitem [{\citenamefont {Wu}\ \emph {et~al.}(2003)\citenamefont {Wu},
  \citenamefont {Hu},\ and\ \citenamefont {Zhang}}]{wu2003}%
  \BibitemOpen
  \bibfield  {author} {\bibinfo {author} {\bibfnamefont {Congjun}\ \bibnamefont
  {Wu}}, \bibinfo {author} {\bibfnamefont {Jiangping}\ \bibnamefont {Hu}}, \
  and\ \bibinfo {author} {\bibfnamefont {Shoucheng}\ \bibnamefont {Zhang}},\
  }\bibfield  {title} {\enquote {\bibinfo {title} {{Exact SO(5) symmetry in the
  spin-3/2 fermionic system}},}\ }\href {\doibase
  10.1103/PhysRevLett.91.186402} {\bibfield  {journal} {\bibinfo  {journal}
  {Phys. Rev. Lett.}\ }\textbf {\bibinfo {volume} {91}},\ \bibinfo {pages}
  {186402} (\bibinfo {year} {2003})}\BibitemShut {NoStop}%
\bibitem [{\citenamefont {Wu}(2006{\natexlab{a}})}]{wu2006}%
  \BibitemOpen
  \bibfield  {author} {\bibinfo {author} {\bibfnamefont {Congjun}\ \bibnamefont
  {Wu}},\ }\bibfield  {title} {\enquote {\bibinfo {title} {Hidden symmetry and
  quantum phases in spin-3/2 cold atomic systems},}\ }\href {\doibase
  10.1142/s0217984906012213} {\bibfield  {journal} {\bibinfo  {journal} {Modern
  Physics Letters B}\ }\textbf {\bibinfo {volume} {20}},\ \bibinfo {pages}
  {1707--1738} (\bibinfo {year} {2006}{\natexlab{a}})}\BibitemShut {NoStop}%
\bibitem [{\citenamefont {Gorshkov}\ \emph {et~al.}(2010)\citenamefont
  {Gorshkov}, \citenamefont {Hermele}, \citenamefont {Gurarie}, \citenamefont
  {Xu}, \citenamefont {Julienne}, \citenamefont {Ye}, \citenamefont {Zoller},
  \citenamefont {Demler}, \citenamefont {Lukin},\ and\ \citenamefont
  {Rey}}]{Gorshkov2010}%
  \BibitemOpen
  \bibfield  {author} {\bibinfo {author} {\bibfnamefont {A.~V.}\ \bibnamefont
  {Gorshkov}}, \bibinfo {author} {\bibfnamefont {M.}~\bibnamefont {Hermele}},
  \bibinfo {author} {\bibfnamefont {V.}~\bibnamefont {Gurarie}}, \bibinfo
  {author} {\bibfnamefont {C.}~\bibnamefont {Xu}}, \bibinfo {author}
  {\bibfnamefont {P.~S.}\ \bibnamefont {Julienne}}, \bibinfo {author}
  {\bibfnamefont {J.}~\bibnamefont {Ye}}, \bibinfo {author} {\bibfnamefont
  {P.}~\bibnamefont {Zoller}}, \bibinfo {author} {\bibfnamefont
  {E.}~\bibnamefont {Demler}}, \bibinfo {author} {\bibfnamefont {M.~D.}\
  \bibnamefont {Lukin}}, \ and\ \bibinfo {author} {\bibfnamefont {A.~M.}\
  \bibnamefont {Rey}},\ }\bibfield  {title} {\enquote {\bibinfo {title}
  {{Two-orbital SU(N) magnetism with ultracold alkaline-earth atoms}},}\ }\href
  {\doibase 10.1038/nphys1535} {\bibfield  {journal} {\bibinfo  {journal}
  {Nature Physics}\ }\textbf {\bibinfo {volume} {6}},\ \bibinfo {pages} {289}
  (\bibinfo {year} {2010})}\BibitemShut {NoStop}%
\bibitem [{\citenamefont {Controzzi}\ and\ \citenamefont
  {Tsvelik}(2006)}]{controzzi2006}%
  \BibitemOpen
  \bibfield  {author} {\bibinfo {author} {\bibfnamefont {D.}~\bibnamefont
  {Controzzi}}\ and\ \bibinfo {author} {\bibfnamefont {A.~M.}\ \bibnamefont
  {Tsvelik}},\ }\bibfield  {title} {\enquote {\bibinfo {title} {{Exactly
  Solvable Model for Isospin $S=3/2$ Fermionic Atoms on an Optical Lattice}},}\
  }\href {\doibase 10.1103/PhysRevLett.96.097205} {\bibfield  {journal}
  {\bibinfo  {journal} {Phys. Rev. Lett.}\ }\textbf {\bibinfo {volume} {96}},\
  \bibinfo {pages} {097205} (\bibinfo {year} {2006})}\BibitemShut {NoStop}%
\bibitem [{\citenamefont {Wu}(2010)}]{wu2010}%
  \BibitemOpen
  \bibfield  {author} {\bibinfo {author} {\bibfnamefont {Congjun}\ \bibnamefont
  {Wu}},\ }\bibfield  {title} {\enquote {\bibinfo {title} {Exotic many-body
  physics with large-spin fermi gases},}\ }\href {\doibase
  10.1103/Physics.3.92} {\bibfield  {journal} {\bibinfo  {journal} {Physics}\
  }\textbf {\bibinfo {volume} {3}},\ \bibinfo {pages} {92} (\bibinfo {year}
  {2010})}\BibitemShut {NoStop}%
\bibitem [{\citenamefont {Wu}(2012)}]{wu2012}%
  \BibitemOpen
  \bibfield  {author} {\bibinfo {author} {\bibfnamefont {Congjun}\ \bibnamefont
  {Wu}},\ }\bibfield  {title} {\enquote {\bibinfo {title} {Mott made easy},}\
  }\href {\doibase 10.1038/nphys2432} {\bibfield  {journal} {\bibinfo
  {journal} {Nat. Phys., New and Views}\ }\textbf {\bibinfo {volume} {8}},\
  \bibinfo {pages} {784} (\bibinfo {year} {2012})}\BibitemShut {NoStop}%
\bibitem [{\citenamefont {DeSalvo}\ \emph {et~al.}(2010)\citenamefont
  {DeSalvo}, \citenamefont {Yan}, \citenamefont {Mickelson}, \citenamefont
  {Martinez~de Escobar},\ and\ \citenamefont {Killian}}]{killian2010}%
  \BibitemOpen
  \bibfield  {author} {\bibinfo {author} {\bibfnamefont {B.~J.}\ \bibnamefont
  {DeSalvo}}, \bibinfo {author} {\bibfnamefont {M.}~\bibnamefont {Yan}},
  \bibinfo {author} {\bibfnamefont {P.~G.}\ \bibnamefont {Mickelson}}, \bibinfo
  {author} {\bibfnamefont {Y.~N.}\ \bibnamefont {Martinez~de Escobar}}, \ and\
  \bibinfo {author} {\bibfnamefont {T.~C.}\ \bibnamefont {Killian}},\
  }\bibfield  {title} {\enquote {\bibinfo {title} {{Degenerate Fermi Gas of
  $^{87}\mathrm{Sr}$}},}\ }\href {\doibase 10.1103/PhysRevLett.105.030402}
  {\bibfield  {journal} {\bibinfo  {journal} {Phys. Rev. Lett.}\ }\textbf
  {\bibinfo {volume} {105}},\ \bibinfo {pages} {030402} (\bibinfo {year}
  {2010})}\BibitemShut {NoStop}%
\bibitem [{\citenamefont {Taie}\ \emph {et~al.}(2012)\citenamefont {Taie},
  \citenamefont {Yamazaki}, \citenamefont {Sugawa},\ and\ \citenamefont
  {Takahashi}}]{taie2012}%
  \BibitemOpen
  \bibfield  {author} {\bibinfo {author} {\bibfnamefont {Shintaro}\
  \bibnamefont {Taie}}, \bibinfo {author} {\bibfnamefont {Rekishu}\
  \bibnamefont {Yamazaki}}, \bibinfo {author} {\bibfnamefont {Seiji}\
  \bibnamefont {Sugawa}}, \ and\ \bibinfo {author} {\bibfnamefont {Yoshiro}\
  \bibnamefont {Takahashi}},\ }\bibfield  {title} {\enquote {\bibinfo {title}
  {{An SU(6) Mott insulator of an atomic Fermi gas realized by large-spin
  Pomeranchuk cooling}},}\ }\href {\doibase 10.1038/nphys2430} {\bibfield
  {journal} {\bibinfo  {journal} {Nature Physics}\ }\textbf {\bibinfo {volume}
  {8}},\ \bibinfo {pages} {825--830} (\bibinfo {year} {2012})}\BibitemShut
  {NoStop}%
\bibitem [{\citenamefont {Cazalilla}\ \emph {et~al.}(2009)\citenamefont
  {Cazalilla}, \citenamefont {Ho},\ and\ \citenamefont {Ueda}}]{cazalilla2009}%
  \BibitemOpen
  \bibfield  {author} {\bibinfo {author} {\bibfnamefont {M.~A.}\ \bibnamefont
  {Cazalilla}}, \bibinfo {author} {\bibfnamefont {A.~F.}\ \bibnamefont {Ho}}, \
  and\ \bibinfo {author} {\bibfnamefont {M.}~\bibnamefont {Ueda}},\ }\bibfield
  {title} {\enquote {\bibinfo {title} {{Ultracold gases of ytterbium:
  ferromagnetism and Mott states in an SU(6) Fermi system}},}\ }\href {\doibase
  10.1088/1367-2630/11/10/103033} {\bibfield  {journal} {\bibinfo  {journal}
  {New J. Phys.}\ }\textbf {\bibinfo {volume} {11}},\ \bibinfo {pages} {103033}
  (\bibinfo {year} {2009})}\BibitemShut {NoStop}%
\bibitem [{\citenamefont {Hermele}\ \emph {et~al.}(2009)\citenamefont
  {Hermele}, \citenamefont {Gurarie},\ and\ \citenamefont {Rey}}]{Hermele2009}%
  \BibitemOpen
  \bibfield  {author} {\bibinfo {author} {\bibfnamefont {Michael}\ \bibnamefont
  {Hermele}}, \bibinfo {author} {\bibfnamefont {Victor}\ \bibnamefont
  {Gurarie}}, \ and\ \bibinfo {author} {\bibfnamefont {Ana~Maria}\ \bibnamefont
  {Rey}},\ }\bibfield  {title} {\enquote {\bibinfo {title} {{Mott Insulators of
  Ultracold Fermionic Alkaline Earth Atoms: Underconstrained Magnetism and
  Chiral Spin Liquid}},}\ }\href {\doibase 10.1103/PhysRevLett.103.135301}
  {\bibfield  {journal} {\bibinfo  {journal} {Phys. Rev. Lett.}\ }\textbf
  {\bibinfo {volume} {103}},\ \bibinfo {pages} {135301} (\bibinfo {year}
  {2009})}\BibitemShut {NoStop}%
\bibitem [{\citenamefont {Goodenough}(1963)}]{goo63}%
  \BibitemOpen
  \bibfield  {author} {\bibinfo {author} {\bibfnamefont {J.~B.}\ \bibnamefont
  {Goodenough}},\ }\href@noop {} {\emph {\bibinfo {title} {Magnetism and
  Chemical Bond}}}\ (\bibinfo  {publisher} {Interscience},\ \bibinfo {address}
  {New York/London},\ \bibinfo {year} {1963})\BibitemShut {NoStop}%
\bibitem [{\citenamefont {{Kanamori}}(1957)}]{1957PThPh..17..177K}%
  \BibitemOpen
  \bibfield  {author} {\bibinfo {author} {\bibfnamefont {J.}~\bibnamefont
  {{Kanamori}}},\ }\bibfield  {title} {\enquote {\bibinfo {title} {{Theory of
  the Magnetic Properties of Ferrous and Cobaltous Oxides, I}},}\ }\href
  {\doibase 10.1143/PTP.17.177} {\bibfield  {journal} {\bibinfo  {journal}
  {Progress of Theoretical Physics}\ }\textbf {\bibinfo {volume} {17}},\
  \bibinfo {pages} {177--196} (\bibinfo {year} {1957})}\BibitemShut {NoStop}%
\bibitem [{\citenamefont {Kugel}\ and\ \citenamefont
  {Khomskii}(1973)}]{kugel1973crystal}%
  \BibitemOpen
  \bibfield  {author} {\bibinfo {author} {\bibfnamefont {KI}~\bibnamefont
  {Kugel}}\ and\ \bibinfo {author} {\bibfnamefont {DI}~\bibnamefont
  {Khomskii}},\ }\bibfield  {title} {\enquote {\bibinfo {title}
  {{Crystal-structure and magnetic properties of substances with orbital
  degeneracy}},}\ }\href@noop {} {\bibfield  {journal} {\bibinfo  {journal}
  {Zh. Eksp. Teor. Fiz}\ }\textbf {\bibinfo {volume} {64}},\ \bibinfo {pages}
  {1429--1439} (\bibinfo {year} {1973})}\BibitemShut {NoStop}%
\bibitem [{\citenamefont {Joshi}\ \emph {et~al.}(1999)\citenamefont {Joshi},
  \citenamefont {Ma}, \citenamefont {Mila}, \citenamefont {Shi},\ and\
  \citenamefont {Zhang}}]{PhysRevB.60.6584}%
  \BibitemOpen
  \bibfield  {author} {\bibinfo {author} {\bibfnamefont {A.}~\bibnamefont
  {Joshi}}, \bibinfo {author} {\bibfnamefont {M.}~\bibnamefont {Ma}}, \bibinfo
  {author} {\bibfnamefont {F.}~\bibnamefont {Mila}}, \bibinfo {author}
  {\bibfnamefont {D.~N.}\ \bibnamefont {Shi}}, \ and\ \bibinfo {author}
  {\bibfnamefont {F.~C.}\ \bibnamefont {Zhang}},\ }\bibfield  {title} {\enquote
  {\bibinfo {title} {Elementary excitations in magnetically ordered systems
  with orbital degeneracy},}\ }\href {\doibase 10.1103/PhysRevB.60.6584}
  {\bibfield  {journal} {\bibinfo  {journal} {Phys. Rev. B}\ }\textbf {\bibinfo
  {volume} {60}},\ \bibinfo {pages} {6584--6587} (\bibinfo {year}
  {1999})}\BibitemShut {NoStop}%
\bibitem [{\citenamefont {Goodenough}(1968)}]{PhysRev.171.466}%
  \BibitemOpen
  \bibfield  {author} {\bibinfo {author} {\bibfnamefont {John~B.}\ \bibnamefont
  {Goodenough}},\ }\bibfield  {title} {\enquote {\bibinfo {title}
  {{Spin-Orbit-Coupling Effects in Transition-Metal Compounds}},}\ }\href
  {\doibase 10.1103/PhysRev.171.466} {\bibfield  {journal} {\bibinfo  {journal}
  {Phys. Rev.}\ }\textbf {\bibinfo {volume} {171}},\ \bibinfo {pages}
  {466--479} (\bibinfo {year} {1968})}\BibitemShut {NoStop}%
\bibitem [{\citenamefont {Chen}\ and\ \citenamefont
  {Balents}(2008)}]{PhysRevB.78.094403}%
  \BibitemOpen
  \bibfield  {author} {\bibinfo {author} {\bibfnamefont {Gang}\ \bibnamefont
  {Chen}}\ and\ \bibinfo {author} {\bibfnamefont {Leon}\ \bibnamefont
  {Balents}},\ }\bibfield  {title} {\enquote {\bibinfo {title} {{Spin-orbit
  effects in ${\text{Na}}_{4}{\text{Ir}}_{3}{\text{O}}_{8}$: A hyper-kagome
  lattice antiferromagnet}},}\ }\href {\doibase 10.1103/PhysRevB.78.094403}
  {\bibfield  {journal} {\bibinfo  {journal} {Phys. Rev. B}\ }\textbf {\bibinfo
  {volume} {78}},\ \bibinfo {pages} {094403} (\bibinfo {year}
  {2008})}\BibitemShut {NoStop}%
\bibitem [{\citenamefont {Feiner}\ \emph {et~al.}(1997)\citenamefont {Feiner},
  \citenamefont {Ole\ifmmode~\acute{s}\else \'{s}\fi{}},\ and\ \citenamefont
  {Zaanen}}]{PhysRevLett.78.2799}%
  \BibitemOpen
  \bibfield  {author} {\bibinfo {author} {\bibfnamefont {Louis~Felix}\
  \bibnamefont {Feiner}}, \bibinfo {author} {\bibfnamefont {Andrzej~M.}\
  \bibnamefont {Ole\ifmmode~\acute{s}\else \'{s}\fi{}}}, \ and\ \bibinfo
  {author} {\bibfnamefont {Jan}\ \bibnamefont {Zaanen}},\ }\bibfield  {title}
  {\enquote {\bibinfo {title} {{Quantum Melting of Magnetic Order due to
  Orbital Fluctuations}},}\ }\href {\doibase 10.1103/PhysRevLett.78.2799}
  {\bibfield  {journal} {\bibinfo  {journal} {Phys. Rev. Lett.}\ }\textbf
  {\bibinfo {volume} {78}},\ \bibinfo {pages} {2799--2802} (\bibinfo {year}
  {1997})}\BibitemShut {NoStop}%
\bibitem [{\citenamefont {Coldea}\ and\ \citenamefont
  {Watson}(2018)}]{annurev-conmatphys-033117-054137}%
  \BibitemOpen
  \bibfield  {author} {\bibinfo {author} {\bibfnamefont {Amalia~I.}\
  \bibnamefont {Coldea}}\ and\ \bibinfo {author} {\bibfnamefont {Matthew~D.}\
  \bibnamefont {Watson}},\ }\bibfield  {title} {\enquote {\bibinfo {title}
  {{The Key Ingredients of the Electronic Structure of FeSe}},}\ }\href
  {\doibase 10.1146/annurev-conmatphys-033117-054137} {\bibfield  {journal}
  {\bibinfo  {journal} {Annual Review of Condensed Matter Physics}\ }\textbf
  {\bibinfo {volume} {9}},\ \bibinfo {pages} {125--146} (\bibinfo {year}
  {2018})}\BibitemShut {NoStop}%
\bibitem [{\citenamefont {Chubukov}(2012)}]{annurev-conmatphys-020911-125055}%
  \BibitemOpen
  \bibfield  {author} {\bibinfo {author} {\bibfnamefont {Andrey}\ \bibnamefont
  {Chubukov}},\ }\bibfield  {title} {\enquote {\bibinfo {title} {{Pairing
  Mechanism in Fe-Based Superconductors}},}\ }\href {\doibase
  10.1146/annurev-conmatphys-020911-125055} {\bibfield  {journal} {\bibinfo
  {journal} {Annual Review of Condensed Matter Physics}\ }\textbf {\bibinfo
  {volume} {3}},\ \bibinfo {pages} {57--92} (\bibinfo {year}
  {2012})}\BibitemShut {NoStop}%
\bibitem [{\citenamefont {Canfield}\ and\ \citenamefont
  {Bud'ko}(2010)}]{annurev-conmatphys-070909-104041}%
  \BibitemOpen
  \bibfield  {author} {\bibinfo {author} {\bibfnamefont {Paul~C.}\ \bibnamefont
  {Canfield}}\ and\ \bibinfo {author} {\bibfnamefont {Sergey~L.}\ \bibnamefont
  {Bud'ko}},\ }\bibfield  {title} {\enquote {\bibinfo {title} {{FeAs-Based
  Superconductivity: A Case Study of the Effects of Transition Metal Doping on
  BaFe$_2$As$_2$}},}\ }\href {\doibase
  10.1146/annurev-conmatphys-070909-104041} {\bibfield  {journal} {\bibinfo
  {journal} {Annual Review of Condensed Matter Physics}\ }\textbf {\bibinfo
  {volume} {1}},\ \bibinfo {pages} {27--50} (\bibinfo {year}
  {2010})}\BibitemShut {NoStop}%
\bibitem [{\citenamefont {Bohmer}\ and\ \citenamefont
  {Kreisel}(2017)}]{Bohmer2017}%
  \BibitemOpen
  \bibfield  {author} {\bibinfo {author} {\bibfnamefont {Anna~E}\ \bibnamefont
  {Bohmer}}\ and\ \bibinfo {author} {\bibfnamefont {Andreas}\ \bibnamefont
  {Kreisel}},\ }\bibfield  {title} {\enquote {\bibinfo {title} {{Nematicity,
  magnetism and superconductivity in FeSe}},}\ }\href {\doibase
  10.1088/1361-648x/aa9caa} {\bibfield  {journal} {\bibinfo  {journal} {Journal
  of Physics: Condensed Matter}\ }\textbf {\bibinfo {volume} {30}},\ \bibinfo
  {pages} {023001} (\bibinfo {year} {2017})}\BibitemShut {NoStop}%
\bibitem [{\citenamefont {Kreisel}\ \emph {et~al.}(2020)\citenamefont
  {Kreisel}, \citenamefont {Hirschfeld},\ and\ \citenamefont
  {Andersen}}]{Kreisel2020}%
  \BibitemOpen
  \bibfield  {author} {\bibinfo {author} {\bibfnamefont {Andreas}\ \bibnamefont
  {Kreisel}}, \bibinfo {author} {\bibfnamefont {Peter}\ \bibnamefont
  {Hirschfeld}}, \ and\ \bibinfo {author} {\bibfnamefont {Brian}\ \bibnamefont
  {Andersen}},\ }\bibfield  {title} {\enquote {\bibinfo {title} {{On the
  Remarkable Superconductivity of FeSe and Its Close Cousins}},}\ }\href
  {\doibase 10.3390/sym12091402} {\bibfield  {journal} {\bibinfo  {journal}
  {Symmetry}\ }\textbf {\bibinfo {volume} {12}},\ \bibinfo {pages} {1402}
  (\bibinfo {year} {2020})}\BibitemShut {NoStop}%
\bibitem [{\citenamefont {Wang}\ \emph
  {et~al.}(2015{\natexlab{a}})\citenamefont {Wang}, \citenamefont {Kivelson},\
  and\ \citenamefont {Lee}}]{WangFa2015}%
  \BibitemOpen
  \bibfield  {author} {\bibinfo {author} {\bibfnamefont {Fa}~\bibnamefont
  {Wang}}, \bibinfo {author} {\bibfnamefont {Steven~A.}\ \bibnamefont
  {Kivelson}}, \ and\ \bibinfo {author} {\bibfnamefont {Dung-Hai}\ \bibnamefont
  {Lee}},\ }\bibfield  {title} {\enquote {\bibinfo {title} {{Nematicity and
  quantum paramagnetism in FeSe}},}\ }\href {\doibase 10.1038/nphys3456}
  {\bibfield  {journal} {\bibinfo  {journal} {Nature Physics}\ }\textbf
  {\bibinfo {volume} {11}},\ \bibinfo {pages} {959--963} (\bibinfo {year}
  {2015}{\natexlab{a}})}\BibitemShut {NoStop}%
\bibitem [{\citenamefont {Hung}\ \emph {et~al.}(2012)\citenamefont {Hung},
  \citenamefont {Song}, \citenamefont {Chen}, \citenamefont {Ma}, \citenamefont
  {Xue},\ and\ \citenamefont {Wu}}]{PhysRevB.85.104510}%
  \BibitemOpen
  \bibfield  {author} {\bibinfo {author} {\bibfnamefont {Hsiang-Hsuan}\
  \bibnamefont {Hung}}, \bibinfo {author} {\bibfnamefont {Can-Li}\ \bibnamefont
  {Song}}, \bibinfo {author} {\bibfnamefont {Xi}~\bibnamefont {Chen}}, \bibinfo
  {author} {\bibfnamefont {Xucun}\ \bibnamefont {Ma}}, \bibinfo {author}
  {\bibfnamefont {Qi-kun}\ \bibnamefont {Xue}}, \ and\ \bibinfo {author}
  {\bibfnamefont {Congjun}\ \bibnamefont {Wu}},\ }\bibfield  {title} {\enquote
  {\bibinfo {title} {{Anisotropic vortex lattice structures in the FeSe
  superconductor}},}\ }\href {\doibase 10.1103/PhysRevB.85.104510} {\bibfield
  {journal} {\bibinfo  {journal} {Phys. Rev. B}\ }\textbf {\bibinfo {volume}
  {85}},\ \bibinfo {pages} {104510} (\bibinfo {year} {2012})}\BibitemShut
  {NoStop}%
\bibitem [{\citenamefont {Liu}\ \emph {et~al.}(2018{\natexlab{a}})\citenamefont
  {Liu}, \citenamefont {Li}, \citenamefont {Huang}, \citenamefont {Lei},
  \citenamefont {Wang}, \citenamefont {Wu}, \citenamefont {Shen}, \citenamefont
  {Gao}, \citenamefont {Zhang}, \citenamefont {Liu}, \citenamefont {Hu},
  \citenamefont {Xu}, \citenamefont {Liang}, \citenamefont {Liu}, \citenamefont
  {Ai}, \citenamefont {Zhao}, \citenamefont {He}, \citenamefont {Yu},
  \citenamefont {Liu}, \citenamefont {Mao}, \citenamefont {Dong}, \citenamefont
  {Jia}, \citenamefont {Zhang}, \citenamefont {Zhang}, \citenamefont {Yang},
  \citenamefont {Wang}, \citenamefont {Peng}, \citenamefont {Shi},
  \citenamefont {Hu}, \citenamefont {Xiang}, \citenamefont {Chen},
  \citenamefont {Xu}, \citenamefont {Chen},\ and\ \citenamefont
  {Zhou}}]{PhysRevX.8.031033}%
  \BibitemOpen
  \bibfield  {author} {\bibinfo {author} {\bibfnamefont {Defa}\ \bibnamefont
  {Liu}}, \bibinfo {author} {\bibfnamefont {Cong}\ \bibnamefont {Li}}, \bibinfo
  {author} {\bibfnamefont {Jianwei}\ \bibnamefont {Huang}}, \bibinfo {author}
  {\bibfnamefont {Bin}\ \bibnamefont {Lei}}, \bibinfo {author} {\bibfnamefont
  {Le}~\bibnamefont {Wang}}, \bibinfo {author} {\bibfnamefont {Xianxin}\
  \bibnamefont {Wu}}, \bibinfo {author} {\bibfnamefont {Bing}\ \bibnamefont
  {Shen}}, \bibinfo {author} {\bibfnamefont {Qiang}\ \bibnamefont {Gao}},
  \bibinfo {author} {\bibfnamefont {Yuxiao}\ \bibnamefont {Zhang}}, \bibinfo
  {author} {\bibfnamefont {Xu}~\bibnamefont {Liu}}, \bibinfo {author}
  {\bibfnamefont {Yong}\ \bibnamefont {Hu}}, \bibinfo {author} {\bibfnamefont
  {Yu}~\bibnamefont {Xu}}, \bibinfo {author} {\bibfnamefont {Aiji}\
  \bibnamefont {Liang}}, \bibinfo {author} {\bibfnamefont {Jing}\ \bibnamefont
  {Liu}}, \bibinfo {author} {\bibfnamefont {Ping}\ \bibnamefont {Ai}}, \bibinfo
  {author} {\bibfnamefont {Lin}\ \bibnamefont {Zhao}}, \bibinfo {author}
  {\bibfnamefont {Shaolong}\ \bibnamefont {He}}, \bibinfo {author}
  {\bibfnamefont {Li}~\bibnamefont {Yu}}, \bibinfo {author} {\bibfnamefont
  {Guodong}\ \bibnamefont {Liu}}, \bibinfo {author} {\bibfnamefont {Yiyuan}\
  \bibnamefont {Mao}}, \bibinfo {author} {\bibfnamefont {Xiaoli}\ \bibnamefont
  {Dong}}, \bibinfo {author} {\bibfnamefont {Xiaowen}\ \bibnamefont {Jia}},
  \bibinfo {author} {\bibfnamefont {Fengfeng}\ \bibnamefont {Zhang}}, \bibinfo
  {author} {\bibfnamefont {Shenjin}\ \bibnamefont {Zhang}}, \bibinfo {author}
  {\bibfnamefont {Feng}\ \bibnamefont {Yang}}, \bibinfo {author} {\bibfnamefont
  {Zhimin}\ \bibnamefont {Wang}}, \bibinfo {author} {\bibfnamefont {Qinjun}\
  \bibnamefont {Peng}}, \bibinfo {author} {\bibfnamefont {Youguo}\ \bibnamefont
  {Shi}}, \bibinfo {author} {\bibfnamefont {Jiangping}\ \bibnamefont {Hu}},
  \bibinfo {author} {\bibfnamefont {Tao}\ \bibnamefont {Xiang}}, \bibinfo
  {author} {\bibfnamefont {Xianhui}\ \bibnamefont {Chen}}, \bibinfo {author}
  {\bibfnamefont {Zuyan}\ \bibnamefont {Xu}}, \bibinfo {author} {\bibfnamefont
  {Chuangtian}\ \bibnamefont {Chen}}, \ and\ \bibinfo {author} {\bibfnamefont
  {X.~J.}\ \bibnamefont {Zhou}},\ }\bibfield  {title} {\enquote {\bibinfo
  {title} {{Orbital Origin of Extremely Anisotropic Superconducting Gap in
  Nematic Phase of FeSe Superconductor}},}\ }\href {\doibase
  10.1103/PhysRevX.8.031033} {\bibfield  {journal} {\bibinfo  {journal} {Phys.
  Rev. X}\ }\textbf {\bibinfo {volume} {8}},\ \bibinfo {pages} {031033}
  (\bibinfo {year} {2018}{\natexlab{a}})}\BibitemShut {NoStop}%
\bibitem [{\citenamefont {Song}\ \emph {et~al.}(2011)\citenamefont {Song},
  \citenamefont {Wang}, \citenamefont {Cheng}, \citenamefont {Jiang},
  \citenamefont {Li}, \citenamefont {T}, \citenamefont {Li}, \citenamefont
  {He}, \citenamefont {Wang}, \citenamefont {Jia}, \citenamefont {Hung},
  \citenamefont {Wu}, \citenamefont {Ma}, \citenamefont {Chen},\ and\
  \citenamefont {QK}}]{science1202226}%
  \BibitemOpen
  \bibfield  {author} {\bibinfo {author} {\bibfnamefont {CL}~\bibnamefont
  {Song}}, \bibinfo {author} {\bibfnamefont {YL}~\bibnamefont {Wang}}, \bibinfo
  {author} {\bibfnamefont {P}~\bibnamefont {Cheng}}, \bibinfo {author}
  {\bibfnamefont {YP}~\bibnamefont {Jiang}}, \bibinfo {author} {\bibfnamefont
  {W}~\bibnamefont {Li}}, \bibinfo {author} {\bibfnamefont {Zhang}\
  \bibnamefont {T}}, \bibinfo {author} {\bibfnamefont {Z}~\bibnamefont {Li}},
  \bibinfo {author} {\bibfnamefont {K}~\bibnamefont {He}}, \bibinfo {author}
  {\bibfnamefont {L}~\bibnamefont {Wang}}, \bibinfo {author} {\bibfnamefont
  {JF}~\bibnamefont {Jia}}, \bibinfo {author} {\bibfnamefont {HH}~\bibnamefont
  {Hung}}, \bibinfo {author} {\bibfnamefont {C}~\bibnamefont {Wu}}, \bibinfo
  {author} {\bibfnamefont {X}~\bibnamefont {Ma}}, \bibinfo {author}
  {\bibfnamefont {X}~\bibnamefont {Chen}}, \ and\ \bibinfo {author}
  {\bibfnamefont {Xue}\ \bibnamefont {QK}},\ }\bibfield  {title} {\enquote
  {\bibinfo {title} {{Direct observation of nodes and twofold symmetry in FeSe
  superconductor}},}\ }\href {\doibase 10.1126/science.1202226} {\bibfield
  {journal} {\bibinfo  {journal} {Science}\ }\textbf {\bibinfo {volume}
  {332}},\ \bibinfo {pages} {1410} (\bibinfo {year} {2011})}\BibitemShut
  {NoStop}%
\bibitem [{\citenamefont {Miller}\ and\ \citenamefont
  {Gatteschi}(2011)}]{C1CS90019F}%
  \BibitemOpen
  \bibfield  {author} {\bibinfo {author} {\bibfnamefont {Joel~S.}\ \bibnamefont
  {Miller}}\ and\ \bibinfo {author} {\bibfnamefont {Dante}\ \bibnamefont
  {Gatteschi}},\ }\bibfield  {title} {\enquote {\bibinfo {title}
  {Molecule-based magnets},}\ }\href {\doibase 10.1039/C1CS90019F} {\bibfield
  {journal} {\bibinfo  {journal} {Chem. Soc. Rev.}\ }\textbf {\bibinfo {volume}
  {40}},\ \bibinfo {pages} {3065--3066} (\bibinfo {year} {2011})}\BibitemShut
  {NoStop}%
\bibitem [{\citenamefont {Gatteschi}\ and\ \citenamefont
  {Sessoli}(1992)}]{GATTESCHI19922092}%
  \BibitemOpen
  \bibfield  {author} {\bibinfo {author} {\bibfnamefont {Dante}\ \bibnamefont
  {Gatteschi}}\ and\ \bibinfo {author} {\bibfnamefont {Roberta}\ \bibnamefont
  {Sessoli}},\ }\bibfield  {title} {\enquote {\bibinfo {title} {Molecular based
  magnetic materials},}\ }\href {\doibase
  https://doi.org/10.1016/0304-8853(92)91683-K} {\bibfield  {journal} {\bibinfo
   {journal} {Journal of Magnetism and Magnetic Materials}\ }\textbf {\bibinfo
  {volume} {104-107}},\ \bibinfo {pages} {2092--2095} (\bibinfo {year}
  {1992})}\BibitemShut {NoStop}%
\bibitem [{\citenamefont {{Maniaki, Diamantoula and Pilichos, Evangelos and
  Perlepes, Spyros P.}}(2018)}]{fchem.2018.00461}%
  \BibitemOpen
  \bibfield  {author} {\bibinfo {author} {\bibnamefont {{Maniaki, Diamantoula
  and Pilichos, Evangelos and Perlepes, Spyros P.}}},\ }\bibfield  {title}
  {\enquote {\bibinfo {title} {{Coordination Clusters of 3d-Metals That Behave
  as Single-Molecule Magnets (SMMs): Synthetic Routes and Strategies}},}\
  }\href {\doibase 10.3389/fchem.2018.00461} {\bibfield  {journal} {\bibinfo
  {journal} {Frontiers in Chemistry}\ }\textbf {\bibinfo {volume} {6}},\
  \bibinfo {pages} {461} (\bibinfo {year} {2018})}\BibitemShut {NoStop}%
\bibitem [{\citenamefont {Shimizu}\ \emph {et~al.}(2003)\citenamefont
  {Shimizu}, \citenamefont {Miyagawa}, \citenamefont {Kanoda}, \citenamefont
  {Maesato},\ and\ \citenamefont {Saito}}]{PhysRevLett.91.107001}%
  \BibitemOpen
  \bibfield  {author} {\bibinfo {author} {\bibfnamefont {Y.}~\bibnamefont
  {Shimizu}}, \bibinfo {author} {\bibfnamefont {K.}~\bibnamefont {Miyagawa}},
  \bibinfo {author} {\bibfnamefont {K.}~\bibnamefont {Kanoda}}, \bibinfo
  {author} {\bibfnamefont {M.}~\bibnamefont {Maesato}}, \ and\ \bibinfo
  {author} {\bibfnamefont {G.}~\bibnamefont {Saito}},\ }\bibfield  {title}
  {\enquote {\bibinfo {title} {{Spin Liquid State in an Organic Mott Insulator
  with a Triangular Lattice}},}\ }\href {\doibase
  10.1103/PhysRevLett.91.107001} {\bibfield  {journal} {\bibinfo  {journal}
  {Phys. Rev. Lett.}\ }\textbf {\bibinfo {volume} {91}},\ \bibinfo {pages}
  {107001} (\bibinfo {year} {2003})}\BibitemShut {NoStop}%
\bibitem [{\citenamefont {Nakamura}\ \emph {et~al.}(2009)\citenamefont
  {Nakamura}, \citenamefont {Yoshimoto}, \citenamefont {Kosugi}, \citenamefont
  {Arita},\ and\ \citenamefont {Imada}}]{JPSJ.78.083710}%
  \BibitemOpen
  \bibfield  {author} {\bibinfo {author} {\bibfnamefont {Kazuma}\ \bibnamefont
  {Nakamura}}, \bibinfo {author} {\bibfnamefont {Yoshihide}\ \bibnamefont
  {Yoshimoto}}, \bibinfo {author} {\bibfnamefont {Taichi}\ \bibnamefont
  {Kosugi}}, \bibinfo {author} {\bibfnamefont {Ryotaro}\ \bibnamefont {Arita}},
  \ and\ \bibinfo {author} {\bibfnamefont {Masatoshi}\ \bibnamefont {Imada}},\
  }\bibfield  {title} {\enquote {\bibinfo {title} {{Ab initio Derivation of
  Low-Energy Model for $\kappa$-ET Type Organic Conductors}},}\ }\href
  {\doibase 10.1143/JPSJ.78.083710} {\bibfield  {journal} {\bibinfo  {journal}
  {Journal of the Physical Society of Japan}\ }\textbf {\bibinfo {volume}
  {78}},\ \bibinfo {pages} {083710} (\bibinfo {year} {2009})}\BibitemShut
  {NoStop}%
\bibitem [{\citenamefont {Okamoto}\ \emph {et~al.}(2015)\citenamefont
  {Okamoto}, \citenamefont {Nilsen}, \citenamefont {Nakazono},\ and\
  \citenamefont {Hiroi}}]{JPSJ.84.043707}%
  \BibitemOpen
  \bibfield  {author} {\bibinfo {author} {\bibfnamefont {Yoshihiko}\
  \bibnamefont {Okamoto}}, \bibinfo {author} {\bibfnamefont {Goran~J.}\
  \bibnamefont {Nilsen}}, \bibinfo {author} {\bibfnamefont {Taishi}\
  \bibnamefont {Nakazono}}, \ and\ \bibinfo {author} {\bibfnamefont {Zenji}\
  \bibnamefont {Hiroi}},\ }\bibfield  {title} {\enquote {\bibinfo {title}
  {{Magnetic Phase Diagram of the Breathing Pyrochlore Antiferromagnet
  LiGa$_{1-x}$In$_x$Cr$_4$O$_8$}},}\ }\href {\doibase 10.7566/JPSJ.84.043707}
  {\bibfield  {journal} {\bibinfo  {journal} {Journal of the Physical Society
  of Japan}\ }\textbf {\bibinfo {volume} {84}},\ \bibinfo {pages} {043707}
  (\bibinfo {year} {2015})}\BibitemShut {NoStop}%
\bibitem [{\citenamefont {Mila}(1998)}]{PhysRevLett.81.2356}%
  \BibitemOpen
  \bibfield  {author} {\bibinfo {author} {\bibfnamefont {F.}~\bibnamefont
  {Mila}},\ }\bibfield  {title} {\enquote {\bibinfo {title} {{Low-Energy Sector
  of the
  $\mathit{S}\mathit{}\phantom{\rule{0ex}{0ex}}=\phantom{\rule{0ex}{0ex}}1/2$
  Kagome Antiferromagnet}},}\ }\href {\doibase 10.1103/PhysRevLett.81.2356}
  {\bibfield  {journal} {\bibinfo  {journal} {Phys. Rev. Lett.}\ }\textbf
  {\bibinfo {volume} {81}},\ \bibinfo {pages} {2356--2359} (\bibinfo {year}
  {1998})}\BibitemShut {NoStop}%
\bibitem [{\citenamefont {Kim}\ \emph {et~al.}(2014{\natexlab{b}})\citenamefont
  {Kim}, \citenamefont {Im}, \citenamefont {Han},\ and\ \citenamefont
  {Jin}}]{Kim_2014}%
  \BibitemOpen
  \bibfield  {author} {\bibinfo {author} {\bibfnamefont {Heung-Sik}\
  \bibnamefont {Kim}}, \bibinfo {author} {\bibfnamefont {Jino}\ \bibnamefont
  {Im}}, \bibinfo {author} {\bibfnamefont {Myung~Joon}\ \bibnamefont {Han}}, \
  and\ \bibinfo {author} {\bibfnamefont {Hosub}\ \bibnamefont {Jin}},\
  }\bibfield  {title} {\enquote {\bibinfo {title} {Spin-orbital entangled
  molecular $j_{eff}$ states in lacunar spinel compounds},}\ }\href {\doibase
  10.1038/ncomms4988} {\bibfield  {journal} {\bibinfo  {journal} {Nature
  Communications}\ }\textbf {\bibinfo {volume} {5}} (\bibinfo {year}
  {2014}{\natexlab{b}}),\ 10.1038/ncomms4988}\BibitemShut {NoStop}%
\bibitem [{\citenamefont {Pokharel}\ \emph {et~al.}(2021)\citenamefont
  {Pokharel}, \citenamefont {Arachchige}, \citenamefont {Gao}, \citenamefont
  {Do}, \citenamefont {Fishman}, \citenamefont {Ehlers}, \citenamefont {Qiu},
  \citenamefont {Rodriguez-Rivera}, \citenamefont {Stone}, \citenamefont
  {Zhang}, \citenamefont {Wilson}, \citenamefont {Mandrus},\ and\ \citenamefont
  {Christianson}}]{PhysRevB.104.224425}%
  \BibitemOpen
  \bibfield  {author} {\bibinfo {author} {\bibfnamefont {G.}~\bibnamefont
  {Pokharel}}, \bibinfo {author} {\bibfnamefont {H.~Suriya}\ \bibnamefont
  {Arachchige}}, \bibinfo {author} {\bibfnamefont {S.}~\bibnamefont {Gao}},
  \bibinfo {author} {\bibfnamefont {S.-H.}\ \bibnamefont {Do}}, \bibinfo
  {author} {\bibfnamefont {R.~S.}\ \bibnamefont {Fishman}}, \bibinfo {author}
  {\bibfnamefont {G.}~\bibnamefont {Ehlers}}, \bibinfo {author} {\bibfnamefont
  {Y.}~\bibnamefont {Qiu}}, \bibinfo {author} {\bibfnamefont {J.~A.}\
  \bibnamefont {Rodriguez-Rivera}}, \bibinfo {author} {\bibfnamefont {M.~B.}\
  \bibnamefont {Stone}}, \bibinfo {author} {\bibfnamefont {H.}~\bibnamefont
  {Zhang}}, \bibinfo {author} {\bibfnamefont {S.~D.}\ \bibnamefont {Wilson}},
  \bibinfo {author} {\bibfnamefont {D.}~\bibnamefont {Mandrus}}, \ and\
  \bibinfo {author} {\bibfnamefont {A.~D.}\ \bibnamefont {Christianson}},\
  }\bibfield  {title} {\enquote {\bibinfo {title} {{Spin dynamics in the
  skyrmion-host lacunar spinel ${\mathrm{GaV}}_{4}{\mathrm{S}}_{8}$}},}\ }\href
  {\doibase 10.1103/PhysRevB.104.224425} {\bibfield  {journal} {\bibinfo
  {journal} {Phys. Rev. B}\ }\textbf {\bibinfo {volume} {104}},\ \bibinfo
  {pages} {224425} (\bibinfo {year} {2021})}\BibitemShut {NoStop}%
\bibitem [{\citenamefont {Curnoe}(2008)}]{PhysRevB.78.094418}%
  \BibitemOpen
  \bibfield  {author} {\bibinfo {author} {\bibfnamefont {S.~H.}\ \bibnamefont
  {Curnoe}},\ }\bibfield  {title} {\enquote {\bibinfo {title} {{Structural
  distortion and the spin liquid state in
  ${\text{Tb}}_{2}{\text{Ti}}_{2}{\text{O}}_{7}$}},}\ }\href {\doibase
  10.1103/PhysRevB.78.094418} {\bibfield  {journal} {\bibinfo  {journal} {Phys.
  Rev. B}\ }\textbf {\bibinfo {volume} {78}},\ \bibinfo {pages} {094418}
  (\bibinfo {year} {2008})}\BibitemShut {NoStop}%
\bibitem [{\citenamefont {Li}\ \emph {et~al.}(2016)\citenamefont {Li},
  \citenamefont {Wang},\ and\ \citenamefont {Chen}}]{PhysRevB.94.035107}%
  \BibitemOpen
  \bibfield  {author} {\bibinfo {author} {\bibfnamefont {Yao-Dong}\
  \bibnamefont {Li}}, \bibinfo {author} {\bibfnamefont {Xiaoqun}\ \bibnamefont
  {Wang}}, \ and\ \bibinfo {author} {\bibfnamefont {Gang}\ \bibnamefont
  {Chen}},\ }\bibfield  {title} {\enquote {\bibinfo {title} {Anisotropic spin
  model of strong spin-orbit-coupled triangular antiferromagnets},}\ }\href
  {\doibase 10.1103/PhysRevB.94.035107} {\bibfield  {journal} {\bibinfo
  {journal} {Phys. Rev. B}\ }\textbf {\bibinfo {volume} {94}},\ \bibinfo
  {pages} {035107} (\bibinfo {year} {2016})}\BibitemShut {NoStop}%
\bibitem [{\citenamefont {Huang}\ \emph {et~al.}(2014)\citenamefont {Huang},
  \citenamefont {Chen},\ and\ \citenamefont
  {Hermele}}]{PhysRevLett.112.167203}%
  \BibitemOpen
  \bibfield  {author} {\bibinfo {author} {\bibfnamefont {Yi-Ping}\ \bibnamefont
  {Huang}}, \bibinfo {author} {\bibfnamefont {Gang}\ \bibnamefont {Chen}}, \
  and\ \bibinfo {author} {\bibfnamefont {Michael}\ \bibnamefont {Hermele}},\
  }\bibfield  {title} {\enquote {\bibinfo {title} {Quantum spin ices and
  topological phases from dipolar-octupolar doublets on the pyrochlore
  lattice},}\ }\href {\doibase 10.1103/PhysRevLett.112.167203} {\bibfield
  {journal} {\bibinfo  {journal} {Phys. Rev. Lett.}\ }\textbf {\bibinfo
  {volume} {112}},\ \bibinfo {pages} {167203} (\bibinfo {year}
  {2014})}\BibitemShut {NoStop}%
\bibitem [{\citenamefont {Onoda}\ and\ \citenamefont
  {Tanaka}(2011)}]{PhysRevB.83.094411}%
  \BibitemOpen
  \bibfield  {author} {\bibinfo {author} {\bibfnamefont {Shigeki}\ \bibnamefont
  {Onoda}}\ and\ \bibinfo {author} {\bibfnamefont {Yoichi}\ \bibnamefont
  {Tanaka}},\ }\bibfield  {title} {\enquote {\bibinfo {title} {Quantum
  fluctuations in the effective pseudospin-$\frac{1}{2}$ model for magnetic
  pyrochlore oxides},}\ }\href {\doibase 10.1103/PhysRevB.83.094411} {\bibfield
   {journal} {\bibinfo  {journal} {Phys. Rev. B}\ }\textbf {\bibinfo {volume}
  {83}},\ \bibinfo {pages} {094411} (\bibinfo {year} {2011})}\BibitemShut
  {NoStop}%
\bibitem [{\citenamefont {Savary}\ \emph {et~al.}(2012)\citenamefont {Savary},
  \citenamefont {Ross}, \citenamefont {Gaulin}, \citenamefont {Ruff},\ and\
  \citenamefont {Balents}}]{PhysRevLett.109.167201}%
  \BibitemOpen
  \bibfield  {author} {\bibinfo {author} {\bibfnamefont {Lucile}\ \bibnamefont
  {Savary}}, \bibinfo {author} {\bibfnamefont {Kate~A.}\ \bibnamefont {Ross}},
  \bibinfo {author} {\bibfnamefont {Bruce~D.}\ \bibnamefont {Gaulin}}, \bibinfo
  {author} {\bibfnamefont {Jacob P.~C.}\ \bibnamefont {Ruff}}, \ and\ \bibinfo
  {author} {\bibfnamefont {Leon}\ \bibnamefont {Balents}},\ }\bibfield  {title}
  {\enquote {\bibinfo {title} {{Order by Quantum Disorder in
  ${\mathrm{Er}}_{2}{\mathrm{Ti}}_{2}{\mathbf{O}}_{7}$}},}\ }\href {\doibase
  10.1103/PhysRevLett.109.167201} {\bibfield  {journal} {\bibinfo  {journal}
  {Phys. Rev. Lett.}\ }\textbf {\bibinfo {volume} {109}},\ \bibinfo {pages}
  {167201} (\bibinfo {year} {2012})}\BibitemShut {NoStop}%
\bibitem [{\citenamefont {Zhang}\ \emph {et~al.}(2014)\citenamefont {Zhang},
  \citenamefont {Fritsch}, \citenamefont {Hao}, \citenamefont {Bagheri},
  \citenamefont {Gingras}, \citenamefont {Granroth}, \citenamefont
  {Jiramongkolchai}, \citenamefont {Cava},\ and\ \citenamefont
  {Gaulin}}]{PhysRevB.89.134410}%
  \BibitemOpen
  \bibfield  {author} {\bibinfo {author} {\bibfnamefont {J.}~\bibnamefont
  {Zhang}}, \bibinfo {author} {\bibfnamefont {K.}~\bibnamefont {Fritsch}},
  \bibinfo {author} {\bibfnamefont {Z.}~\bibnamefont {Hao}}, \bibinfo {author}
  {\bibfnamefont {B.~V.}\ \bibnamefont {Bagheri}}, \bibinfo {author}
  {\bibfnamefont {M.~J.~P.}\ \bibnamefont {Gingras}}, \bibinfo {author}
  {\bibfnamefont {G.~E.}\ \bibnamefont {Granroth}}, \bibinfo {author}
  {\bibfnamefont {P.}~\bibnamefont {Jiramongkolchai}}, \bibinfo {author}
  {\bibfnamefont {R.~J.}\ \bibnamefont {Cava}}, \ and\ \bibinfo {author}
  {\bibfnamefont {B.~D.}\ \bibnamefont {Gaulin}},\ }\bibfield  {title}
  {\enquote {\bibinfo {title} {{Neutron spectroscopic study of crystal field
  excitations in ${\mathrm{Tb}}_{2}{\mathrm{Ti}}_{2}{\mathrm{O}}_{7}$ and
  ${\mathrm{Tb}}_{2}{\mathrm{Sn}}_{2}{\mathrm{O}}_{7}$}},}\ }\href {\doibase
  10.1103/PhysRevB.89.134410} {\bibfield  {journal} {\bibinfo  {journal} {Phys.
  Rev. B}\ }\textbf {\bibinfo {volume} {89}},\ \bibinfo {pages} {134410}
  (\bibinfo {year} {2014})}\BibitemShut {NoStop}%
\bibitem [{\citenamefont {Petit}\ \emph {et~al.}(2012)\citenamefont {Petit},
  \citenamefont {Bonville}, \citenamefont {Mirebeau}, \citenamefont {Mutka},\
  and\ \citenamefont {Robert}}]{PhysRevB.85.054428}%
  \BibitemOpen
  \bibfield  {author} {\bibinfo {author} {\bibfnamefont {Sylvain}\ \bibnamefont
  {Petit}}, \bibinfo {author} {\bibfnamefont {Pierre}\ \bibnamefont
  {Bonville}}, \bibinfo {author} {\bibfnamefont {Isabelle}\ \bibnamefont
  {Mirebeau}}, \bibinfo {author} {\bibfnamefont {Hannu}\ \bibnamefont {Mutka}},
  \ and\ \bibinfo {author} {\bibfnamefont {Julien}\ \bibnamefont {Robert}},\
  }\bibfield  {title} {\enquote {\bibinfo {title} {{Spin dynamics in the
  ordered spin ice Tb${}_{2}$Sn${}_{2}$O${}_{7}$}},}\ }\href {\doibase
  10.1103/PhysRevB.85.054428} {\bibfield  {journal} {\bibinfo  {journal} {Phys.
  Rev. B}\ }\textbf {\bibinfo {volume} {85}},\ \bibinfo {pages} {054428}
  (\bibinfo {year} {2012})}\BibitemShut {NoStop}%
\bibitem [{\citenamefont {Mirebeau}\ \emph {et~al.}(2007)\citenamefont
  {Mirebeau}, \citenamefont {Bonville},\ and\ \citenamefont
  {Hennion}}]{PhysRevB.76.184436}%
  \BibitemOpen
  \bibfield  {author} {\bibinfo {author} {\bibfnamefont {I.}~\bibnamefont
  {Mirebeau}}, \bibinfo {author} {\bibfnamefont {P.}~\bibnamefont {Bonville}},
  \ and\ \bibinfo {author} {\bibfnamefont {M.}~\bibnamefont {Hennion}},\
  }\bibfield  {title} {\enquote {\bibinfo {title} {{Magnetic excitations in
  ${\mathrm{Tb}}_{2}{\mathrm{Sn}}_{2}{\mathrm{O}}_{7}$ and
  ${\mathrm{Tb}}_{2}{\mathrm{Ti}}_{2}{\mathrm{O}}_{7}$ as measured by inelastic
  neutron scattering}},}\ }\href {\doibase 10.1103/PhysRevB.76.184436}
  {\bibfield  {journal} {\bibinfo  {journal} {Phys. Rev. B}\ }\textbf {\bibinfo
  {volume} {76}},\ \bibinfo {pages} {184436} (\bibinfo {year}
  {2007})}\BibitemShut {NoStop}%
\bibitem [{\citenamefont {Chaloupka}\ \emph {et~al.}(2010)\citenamefont
  {Chaloupka}, \citenamefont {Jackeli},\ and\ \citenamefont
  {Khaliullin}}]{PhysRevLett.105.027204}%
  \BibitemOpen
  \bibfield  {author} {\bibinfo {author} {\bibfnamefont {Ji\ifmmode
  \check{r}\else~\v{r}\fi{}\'{\i}}\ \bibnamefont {Chaloupka}}, \bibinfo
  {author} {\bibfnamefont {George}\ \bibnamefont {Jackeli}}, \ and\ \bibinfo
  {author} {\bibfnamefont {Giniyat}\ \bibnamefont {Khaliullin}},\ }\bibfield
  {title} {\enquote {\bibinfo {title} {{Kitaev-Heisenberg Model on a Honeycomb
  Lattice: Possible Exotic Phases in Iridium Oxides
  ${A}_{2}{\mathrm{IrO}}_{3}$}},}\ }\href {\doibase
  10.1103/PhysRevLett.105.027204} {\bibfield  {journal} {\bibinfo  {journal}
  {Phys. Rev. Lett.}\ }\textbf {\bibinfo {volume} {105}},\ \bibinfo {pages}
  {027204} (\bibinfo {year} {2010})}\BibitemShut {NoStop}%
\bibitem [{\citenamefont {Jackeli}\ and\ \citenamefont
  {Khaliullin}(2009)}]{PhysRevLett.102.017205}%
  \BibitemOpen
  \bibfield  {author} {\bibinfo {author} {\bibfnamefont {G.}~\bibnamefont
  {Jackeli}}\ and\ \bibinfo {author} {\bibfnamefont {G.}~\bibnamefont
  {Khaliullin}},\ }\bibfield  {title} {\enquote {\bibinfo {title} {{Mott
  Insulators in the Strong Spin-Orbit Coupling Limit: From Heisenberg to a
  Quantum Compass and Kitaev Models}},}\ }\href {\doibase
  10.1103/PhysRevLett.102.017205} {\bibfield  {journal} {\bibinfo  {journal}
  {Phys. Rev. Lett.}\ }\textbf {\bibinfo {volume} {102}},\ \bibinfo {pages}
  {017205} (\bibinfo {year} {2009})}\BibitemShut {NoStop}%
\bibitem [{\citenamefont {Kim}\ \emph {et~al.}(2008)\citenamefont {Kim},
  \citenamefont {Jin}, \citenamefont {Moon}, \citenamefont {Kim}, \citenamefont
  {Park}, \citenamefont {Leem}, \citenamefont {Yu}, \citenamefont {Noh},
  \citenamefont {Kim}, \citenamefont {Oh}, \citenamefont {Park}, \citenamefont
  {Durairaj}, \citenamefont {Cao},\ and\ \citenamefont
  {Rotenberg}}]{PhysRevLett.101.076402}%
  \BibitemOpen
  \bibfield  {author} {\bibinfo {author} {\bibfnamefont {B.~J.}\ \bibnamefont
  {Kim}}, \bibinfo {author} {\bibfnamefont {Hosub}\ \bibnamefont {Jin}},
  \bibinfo {author} {\bibfnamefont {S.~J.}\ \bibnamefont {Moon}}, \bibinfo
  {author} {\bibfnamefont {J.-Y.}\ \bibnamefont {Kim}}, \bibinfo {author}
  {\bibfnamefont {B.-G.}\ \bibnamefont {Park}}, \bibinfo {author}
  {\bibfnamefont {C.~S.}\ \bibnamefont {Leem}}, \bibinfo {author}
  {\bibfnamefont {Jaejun}\ \bibnamefont {Yu}}, \bibinfo {author} {\bibfnamefont
  {T.~W.}\ \bibnamefont {Noh}}, \bibinfo {author} {\bibfnamefont
  {C.}~\bibnamefont {Kim}}, \bibinfo {author} {\bibfnamefont {S.-J.}\
  \bibnamefont {Oh}}, \bibinfo {author} {\bibfnamefont {J.-H.}\ \bibnamefont
  {Park}}, \bibinfo {author} {\bibfnamefont {V.}~\bibnamefont {Durairaj}},
  \bibinfo {author} {\bibfnamefont {G.}~\bibnamefont {Cao}}, \ and\ \bibinfo
  {author} {\bibfnamefont {E.}~\bibnamefont {Rotenberg}},\ }\bibfield  {title}
  {\enquote {\bibinfo {title} {{Novel ${J}_{\mathrm{eff}}=1/2$ Mott State
  Induced by Relativistic Spin-Orbit Coupling in
  ${\mathrm{Sr}}_{2}{\mathrm{IrO}}_{4}$}},}\ }\href {\doibase
  10.1103/PhysRevLett.101.076402} {\bibfield  {journal} {\bibinfo  {journal}
  {Phys. Rev. Lett.}\ }\textbf {\bibinfo {volume} {101}},\ \bibinfo {pages}
  {076402} (\bibinfo {year} {2008})}\BibitemShut {NoStop}%
\bibitem [{\citenamefont {Plumb}\ \emph {et~al.}(2014)\citenamefont {Plumb},
  \citenamefont {Clancy}, \citenamefont {Sandilands}, \citenamefont {Shankar},
  \citenamefont {Hu}, \citenamefont {Burch}, \citenamefont {Kee},\ and\
  \citenamefont {Kim}}]{PhysRevB.90.041112}%
  \BibitemOpen
  \bibfield  {author} {\bibinfo {author} {\bibfnamefont {K.~W.}\ \bibnamefont
  {Plumb}}, \bibinfo {author} {\bibfnamefont {J.~P.}\ \bibnamefont {Clancy}},
  \bibinfo {author} {\bibfnamefont {L.~J.}\ \bibnamefont {Sandilands}},
  \bibinfo {author} {\bibfnamefont {V.~Vijay}\ \bibnamefont {Shankar}},
  \bibinfo {author} {\bibfnamefont {Y.~F.}\ \bibnamefont {Hu}}, \bibinfo
  {author} {\bibfnamefont {K.~S.}\ \bibnamefont {Burch}}, \bibinfo {author}
  {\bibfnamefont {Hae-Young}\ \bibnamefont {Kee}}, \ and\ \bibinfo {author}
  {\bibfnamefont {Young-June}\ \bibnamefont {Kim}},\ }\bibfield  {title}
  {\enquote {\bibinfo {title}
  {{$\ensuremath{\alpha}\ensuremath{-}{\mathrm{RuCl}}_{3}$: A spin-orbit
  assisted Mott insulator on a honeycomb lattice}},}\ }\href {\doibase
  10.1103/PhysRevB.90.041112} {\bibfield  {journal} {\bibinfo  {journal} {Phys.
  Rev. B}\ }\textbf {\bibinfo {volume} {90}},\ \bibinfo {pages} {041112}
  (\bibinfo {year} {2014})}\BibitemShut {NoStop}%
\bibitem [{\citenamefont {Liu}\ \emph {et~al.}(2020)\citenamefont {Liu},
  \citenamefont {Chaloupka},\ and\ \citenamefont
  {Khaliullin}}]{PhysRevLett.125.047201}%
  \BibitemOpen
  \bibfield  {author} {\bibinfo {author} {\bibfnamefont {Huimei}\ \bibnamefont
  {Liu}}, \bibinfo {author} {\bibfnamefont {Ji\ifmmode
  \check{r}\else~\v{r}\fi{}\'{\i}}\ \bibnamefont {Chaloupka}}, \ and\ \bibinfo
  {author} {\bibfnamefont {Giniyat}\ \bibnamefont {Khaliullin}},\ }\bibfield
  {title} {\enquote {\bibinfo {title} {{Kitaev Spin Liquid in $3d$ Transition
  Metal Compounds}},}\ }\href {\doibase 10.1103/PhysRevLett.125.047201}
  {\bibfield  {journal} {\bibinfo  {journal} {Phys. Rev. Lett.}\ }\textbf
  {\bibinfo {volume} {125}},\ \bibinfo {pages} {047201} (\bibinfo {year}
  {2020})}\BibitemShut {NoStop}%
\bibitem [{\citenamefont {Liu}\ and\ \citenamefont
  {Khaliullin}(2018)}]{PhysRevB.97.014407}%
  \BibitemOpen
  \bibfield  {author} {\bibinfo {author} {\bibfnamefont {Huimei}\ \bibnamefont
  {Liu}}\ and\ \bibinfo {author} {\bibfnamefont {Giniyat}\ \bibnamefont
  {Khaliullin}},\ }\bibfield  {title} {\enquote {\bibinfo {title} {{Pseudospin
  exchange interactions in ${d}^{7}$ cobalt compounds: Possible realization of
  the Kitaev model}},}\ }\href {\doibase 10.1103/PhysRevB.97.014407} {\bibfield
   {journal} {\bibinfo  {journal} {Phys. Rev. B}\ }\textbf {\bibinfo {volume}
  {97}},\ \bibinfo {pages} {014407} (\bibinfo {year} {2018})}\BibitemShut
  {NoStop}%
\bibitem [{\citenamefont {Motome}\ \emph {et~al.}(2020)\citenamefont {Motome},
  \citenamefont {Sano}, \citenamefont {Jang}, \citenamefont {Sugita},\ and\
  \citenamefont {Kato}}]{Motome2020}%
  \BibitemOpen
  \bibfield  {author} {\bibinfo {author} {\bibfnamefont {Yukitoshi}\
  \bibnamefont {Motome}}, \bibinfo {author} {\bibfnamefont {Ryoya}\
  \bibnamefont {Sano}}, \bibinfo {author} {\bibfnamefont {Seonghoon}\
  \bibnamefont {Jang}}, \bibinfo {author} {\bibfnamefont {Yusuke}\ \bibnamefont
  {Sugita}}, \ and\ \bibinfo {author} {\bibfnamefont {Yasuyuki}\ \bibnamefont
  {Kato}},\ }\bibfield  {title} {\enquote {\bibinfo {title} {{Materials design
  of Kitaev spin liquids beyond the Jackeli-Khaliullin mechanism}},}\ }\href
  {\doibase 10.1088/1361-648x/ab8525} {\bibfield  {journal} {\bibinfo
  {journal} {Journal of Physics: Condensed Matter}\ }\textbf {\bibinfo {volume}
  {32}},\ \bibinfo {pages} {404001} (\bibinfo {year} {2020})}\BibitemShut
  {NoStop}%
\bibitem [{\citenamefont {Sano}\ \emph {et~al.}(2018)\citenamefont {Sano},
  \citenamefont {Kato},\ and\ \citenamefont {Motome}}]{PhysRevB.97.014408}%
  \BibitemOpen
  \bibfield  {author} {\bibinfo {author} {\bibfnamefont {Ryoya}\ \bibnamefont
  {Sano}}, \bibinfo {author} {\bibfnamefont {Yasuyuki}\ \bibnamefont {Kato}}, \
  and\ \bibinfo {author} {\bibfnamefont {Yukitoshi}\ \bibnamefont {Motome}},\
  }\bibfield  {title} {\enquote {\bibinfo {title} {{Kitaev-Heisenberg
  Hamiltonian for high-spin ${d}^{7}$ Mott insulators}},}\ }\href {\doibase
  10.1103/PhysRevB.97.014408} {\bibfield  {journal} {\bibinfo  {journal} {Phys.
  Rev. B}\ }\textbf {\bibinfo {volume} {97}},\ \bibinfo {pages} {014408}
  (\bibinfo {year} {2018})}\BibitemShut {NoStop}%
\bibitem [{\citenamefont {Elliot}\ \emph {et~al.}(2021)\citenamefont {Elliot},
  \citenamefont {McClarty}, \citenamefont {Prabhakaran}, \citenamefont
  {Johnson}, \citenamefont {Walker}, \citenamefont {Manuel},\ and\
  \citenamefont {Coldea}}]{Coldea2021}%
  \BibitemOpen
  \bibfield  {author} {\bibinfo {author} {\bibfnamefont {M.}~\bibnamefont
  {Elliot}}, \bibinfo {author} {\bibfnamefont {P.~A.}\ \bibnamefont
  {McClarty}}, \bibinfo {author} {\bibfnamefont {D.}~\bibnamefont
  {Prabhakaran}}, \bibinfo {author} {\bibfnamefont {R.~D.}\ \bibnamefont
  {Johnson}}, \bibinfo {author} {\bibfnamefont {H.~C.}\ \bibnamefont {Walker}},
  \bibinfo {author} {\bibfnamefont {P.}~\bibnamefont {Manuel}}, \ and\ \bibinfo
  {author} {\bibfnamefont {R.}~\bibnamefont {Coldea}},\ }\bibfield  {title}
  {\enquote {\bibinfo {title} {Order-by-disorder from bond-dependent exchange
  and intensity signature of nodal quasiparticles in a honeycomb cobaltate},}\
  }\href {\doibase 10.1038/s41467-021-23851-0} {\bibfield  {journal} {\bibinfo
  {journal} {Nature Communications}\ }\textbf {\bibinfo {volume} {12}},\
  \bibinfo {pages} {3936} (\bibinfo {year} {2021})}\BibitemShut {NoStop}%
\bibitem [{\citenamefont {Moriya}(1960)}]{PhysRev.120.91}%
  \BibitemOpen
  \bibfield  {author} {\bibinfo {author} {\bibfnamefont {T\^oru}\ \bibnamefont
  {Moriya}},\ }\bibfield  {title} {\enquote {\bibinfo {title} {Anisotropic
  superexchange interaction and weak ferromagnetism},}\ }\href {\doibase
  10.1103/PhysRev.120.91} {\bibfield  {journal} {\bibinfo  {journal} {Phys.
  Rev.}\ }\textbf {\bibinfo {volume} {120}},\ \bibinfo {pages} {91--98}
  (\bibinfo {year} {1960})}\BibitemShut {NoStop}%
\bibitem [{\citenamefont {Luttinger}\ and\ \citenamefont
  {Kohn}(1955)}]{PhysRev.97.869}%
  \BibitemOpen
  \bibfield  {author} {\bibinfo {author} {\bibfnamefont {J.~M.}\ \bibnamefont
  {Luttinger}}\ and\ \bibinfo {author} {\bibfnamefont {W.}~\bibnamefont
  {Kohn}},\ }\bibfield  {title} {\enquote {\bibinfo {title} {{Motion of
  Electrons and Holes in Perturbed Periodic Fields}},}\ }\href {\doibase
  10.1103/PhysRev.97.869} {\bibfield  {journal} {\bibinfo  {journal} {Phys.
  Rev.}\ }\textbf {\bibinfo {volume} {97}},\ \bibinfo {pages} {869--883}
  (\bibinfo {year} {1955})}\BibitemShut {NoStop}%
\bibitem [{\citenamefont {Szab\'o}\ \emph {et~al.}(2021)\citenamefont
  {Szab\'o}, \citenamefont {Moessner},\ and\ \citenamefont
  {Roy}}]{PhysRevB.103.165139}%
  \BibitemOpen
  \bibfield  {author} {\bibinfo {author} {\bibfnamefont {Andr\'as~L.}\
  \bibnamefont {Szab\'o}}, \bibinfo {author} {\bibfnamefont {Roderich}\
  \bibnamefont {Moessner}}, \ and\ \bibinfo {author} {\bibfnamefont {Bitan}\
  \bibnamefont {Roy}},\ }\bibfield  {title} {\enquote {\bibinfo {title}
  {{Interacting spin-$\frac{3}{2}$ fermions in a Luttinger semimetal: Competing
  phases and their selection in the global phase diagram}},}\ }\href {\doibase
  10.1103/PhysRevB.103.165139} {\bibfield  {journal} {\bibinfo  {journal}
  {Phys. Rev. B}\ }\textbf {\bibinfo {volume} {103}},\ \bibinfo {pages}
  {165139} (\bibinfo {year} {2021})}\BibitemShut {NoStop}%
\bibitem [{\citenamefont {Yao}\ and\ \citenamefont
  {Chen}(2018)}]{PhysRevX.8.041039}%
  \BibitemOpen
  \bibfield  {author} {\bibinfo {author} {\bibfnamefont {Xu-Ping}\ \bibnamefont
  {Yao}}\ and\ \bibinfo {author} {\bibfnamefont {Gang}\ \bibnamefont {Chen}},\
  }\bibfield  {title} {\enquote {\bibinfo {title}
  {{${\mathrm{Pr}}_{2}{\mathrm{Ir}}_{2}{\mathrm{O}}_{7}$: When Luttinger
  Semimetal Meets Melko-Hertog-Gingras Spin Ice State}},}\ }\href {\doibase
  10.1103/PhysRevX.8.041039} {\bibfield  {journal} {\bibinfo  {journal} {Phys.
  Rev. X}\ }\textbf {\bibinfo {volume} {8}},\ \bibinfo {pages} {041039}
  (\bibinfo {year} {2018})}\BibitemShut {NoStop}%
\bibitem [{\citenamefont {Sim}\ \emph {et~al.}(2020)\citenamefont {Sim},
  \citenamefont {Mishra}, \citenamefont {Park}, \citenamefont {Kim},
  \citenamefont {Cho},\ and\ \citenamefont {Lee}}]{PhysRevResearch.2.023416}%
  \BibitemOpen
  \bibfield  {author} {\bibinfo {author} {\bibfnamefont {GiBaik}\ \bibnamefont
  {Sim}}, \bibinfo {author} {\bibfnamefont {Archana}\ \bibnamefont {Mishra}},
  \bibinfo {author} {\bibfnamefont {Moon~Jip}\ \bibnamefont {Park}}, \bibinfo
  {author} {\bibfnamefont {Yong~Baek}\ \bibnamefont {Kim}}, \bibinfo {author}
  {\bibfnamefont {Gil~Young}\ \bibnamefont {Cho}}, \ and\ \bibinfo {author}
  {\bibfnamefont {SungBin}\ \bibnamefont {Lee}},\ }\bibfield  {title} {\enquote
  {\bibinfo {title} {{Multipolar superconductivity in Luttinger semimetals}},}\
  }\href {\doibase 10.1103/PhysRevResearch.2.023416} {\bibfield  {journal}
  {\bibinfo  {journal} {Phys. Rev. Research}\ }\textbf {\bibinfo {volume}
  {2}},\ \bibinfo {pages} {023416} (\bibinfo {year} {2020})}\BibitemShut
  {NoStop}%
\bibitem [{\citenamefont {Moon}\ \emph {et~al.}(2013)\citenamefont {Moon},
  \citenamefont {Xu}, \citenamefont {Kim},\ and\ \citenamefont
  {Balents}}]{PhysRevLett.111.206401}%
  \BibitemOpen
  \bibfield  {author} {\bibinfo {author} {\bibfnamefont {Eun-Gook}\
  \bibnamefont {Moon}}, \bibinfo {author} {\bibfnamefont {Cenke}\ \bibnamefont
  {Xu}}, \bibinfo {author} {\bibfnamefont {Yong~Baek}\ \bibnamefont {Kim}}, \
  and\ \bibinfo {author} {\bibfnamefont {Leon}\ \bibnamefont {Balents}},\
  }\bibfield  {title} {\enquote {\bibinfo {title} {{Non-Fermi-Liquid and
  Topological States with Strong Spin-Orbit Coupling}},}\ }\href {\doibase
  10.1103/PhysRevLett.111.206401} {\bibfield  {journal} {\bibinfo  {journal}
  {Phys. Rev. Lett.}\ }\textbf {\bibinfo {volume} {111}},\ \bibinfo {pages}
  {206401} (\bibinfo {year} {2013})}\BibitemShut {NoStop}%
\bibitem [{\citenamefont {Kharitonov}\ \emph {et~al.}(2017)\citenamefont
  {Kharitonov}, \citenamefont {Mayer},\ and\ \citenamefont
  {Hankiewicz}}]{PhysRevLett.119.266402}%
  \BibitemOpen
  \bibfield  {author} {\bibinfo {author} {\bibfnamefont {Maxim}\ \bibnamefont
  {Kharitonov}}, \bibinfo {author} {\bibfnamefont {Julian-Benedikt}\
  \bibnamefont {Mayer}}, \ and\ \bibinfo {author} {\bibfnamefont {Ewelina~M.}\
  \bibnamefont {Hankiewicz}},\ }\bibfield  {title} {\enquote {\bibinfo {title}
  {{Universality and Stability of the Edge States of Chiral-Symmetric
  Topological Semimetals and Surface States of the Luttinger Semimetal}},}\
  }\href {\doibase 10.1103/PhysRevLett.119.266402} {\bibfield  {journal}
  {\bibinfo  {journal} {Phys. Rev. Lett.}\ }\textbf {\bibinfo {volume} {119}},\
  \bibinfo {pages} {266402} (\bibinfo {year} {2017})}\BibitemShut {NoStop}%
\bibitem [{\citenamefont {Roy}\ \emph {et~al.}(2019)\citenamefont {Roy},
  \citenamefont {Ghorashi}, \citenamefont {Foster},\ and\ \citenamefont
  {Nevidomskyy}}]{PhysRevB.99.054505}%
  \BibitemOpen
  \bibfield  {author} {\bibinfo {author} {\bibfnamefont {Bitan}\ \bibnamefont
  {Roy}}, \bibinfo {author} {\bibfnamefont {Sayed Ali~Akbar}\ \bibnamefont
  {Ghorashi}}, \bibinfo {author} {\bibfnamefont {Matthew~S.}\ \bibnamefont
  {Foster}}, \ and\ \bibinfo {author} {\bibfnamefont {Andriy~H.}\ \bibnamefont
  {Nevidomskyy}},\ }\bibfield  {title} {\enquote {\bibinfo {title}
  {{Topological superconductivity of spin-$3/2$ carriers in a three-dimensional
  doped Luttinger semimetal}},}\ }\href {\doibase 10.1103/PhysRevB.99.054505}
  {\bibfield  {journal} {\bibinfo  {journal} {Phys. Rev. B}\ }\textbf {\bibinfo
  {volume} {99}},\ \bibinfo {pages} {054505} (\bibinfo {year}
  {2019})}\BibitemShut {NoStop}%
\bibitem [{\citenamefont {Boettcher}\ and\ \citenamefont
  {Herbut}(2018)}]{PhysRevLett.120.057002}%
  \BibitemOpen
  \bibfield  {author} {\bibinfo {author} {\bibfnamefont {Igor}\ \bibnamefont
  {Boettcher}}\ and\ \bibinfo {author} {\bibfnamefont {Igor~F.}\ \bibnamefont
  {Herbut}},\ }\bibfield  {title} {\enquote {\bibinfo {title} {{Unconventional
  Superconductivity in Luttinger Semimetals: Theory of Complex Tensor Order and
  the Emergence of the Uniaxial Nematic State}},}\ }\href {\doibase
  10.1103/PhysRevLett.120.057002} {\bibfield  {journal} {\bibinfo  {journal}
  {Phys. Rev. Lett.}\ }\textbf {\bibinfo {volume} {120}},\ \bibinfo {pages}
  {057002} (\bibinfo {year} {2018})}\BibitemShut {NoStop}%
\bibitem [{\citenamefont {Mostovoy}\ and\ \citenamefont
  {Khomskii}(2004)}]{PhysRevLett.92.167201}%
  \BibitemOpen
  \bibfield  {author} {\bibinfo {author} {\bibfnamefont {M.~V.}\ \bibnamefont
  {Mostovoy}}\ and\ \bibinfo {author} {\bibfnamefont {D.~I.}\ \bibnamefont
  {Khomskii}},\ }\bibfield  {title} {\enquote {\bibinfo {title} {Orbital
  ordering in charge transfer insulators},}\ }\href {\doibase
  10.1103/PhysRevLett.92.167201} {\bibfield  {journal} {\bibinfo  {journal}
  {Phys. Rev. Lett.}\ }\textbf {\bibinfo {volume} {92}},\ \bibinfo {pages}
  {167201} (\bibinfo {year} {2004})}\BibitemShut {NoStop}%
\bibitem [{\citenamefont {Mostovoy}\ and\ \citenamefont
  {Khomskii}(2002)}]{PhysRevLett.89.227203}%
  \BibitemOpen
  \bibfield  {author} {\bibinfo {author} {\bibfnamefont {M.~V.}\ \bibnamefont
  {Mostovoy}}\ and\ \bibinfo {author} {\bibfnamefont {D.~I.}\ \bibnamefont
  {Khomskii}},\ }\bibfield  {title} {\enquote {\bibinfo {title} {{Orbital
  Ordering in Frustrated Jahn-Teller Systems with
  90\ifmmode^\circ\else\textdegree\fi{} Exchange}},}\ }\href {\doibase
  10.1103/PhysRevLett.89.227203} {\bibfield  {journal} {\bibinfo  {journal}
  {Phys. Rev. Lett.}\ }\textbf {\bibinfo {volume} {89}},\ \bibinfo {pages}
  {227203} (\bibinfo {year} {2002})}\BibitemShut {NoStop}%
\bibitem [{\citenamefont {Khomskii}\ and\ \citenamefont
  {Mostovoy}(2003)}]{Khomskii_2003}%
  \BibitemOpen
  \bibfield  {author} {\bibinfo {author} {\bibfnamefont {D~I}\ \bibnamefont
  {Khomskii}}\ and\ \bibinfo {author} {\bibfnamefont {M~V}\ \bibnamefont
  {Mostovoy}},\ }\bibfield  {title} {\enquote {\bibinfo {title} {Orbital
  ordering and frustrations},}\ }\href {\doibase 10.1088/0305-4470/36/35/307}
  {\bibfield  {journal} {\bibinfo  {journal} {Journal of Physics A:
  Mathematical and General}\ }\textbf {\bibinfo {volume} {36}},\ \bibinfo
  {pages} {9197} (\bibinfo {year} {2003})}\BibitemShut {NoStop}%
\bibitem [{\citenamefont {Pavarini}\ \emph {et~al.}(2008)\citenamefont
  {Pavarini}, \citenamefont {Koch},\ and\ \citenamefont
  {Lichtenstein}}]{PhysRevLett.101.266405}%
  \BibitemOpen
  \bibfield  {author} {\bibinfo {author} {\bibfnamefont {E.}~\bibnamefont
  {Pavarini}}, \bibinfo {author} {\bibfnamefont {E.}~\bibnamefont {Koch}}, \
  and\ \bibinfo {author} {\bibfnamefont {A.~I.}\ \bibnamefont {Lichtenstein}},\
  }\bibfield  {title} {\enquote {\bibinfo {title} {{Mechanism for Orbital
  Ordering in ${\mathrm{KCuF}}_{3}$}},}\ }\href {\doibase
  10.1103/PhysRevLett.101.266405} {\bibfield  {journal} {\bibinfo  {journal}
  {Phys. Rev. Lett.}\ }\textbf {\bibinfo {volume} {101}},\ \bibinfo {pages}
  {266405} (\bibinfo {year} {2008})}\BibitemShut {NoStop}%
\bibitem [{\citenamefont {Li}\ \emph {et~al.}(2021)\citenamefont {Li},
  \citenamefont {Xu}, \citenamefont {Garcia-Fernandez}, \citenamefont {Nag},
  \citenamefont {Robarts}, \citenamefont {Walters}, \citenamefont {Liu},
  \citenamefont {Zhou}, \citenamefont {Wohlfeld}, \citenamefont {van~den
  Brink}, \citenamefont {Ding},\ and\ \citenamefont
  {Zhou}}]{PhysRevLett.126.106401}%
  \BibitemOpen
  \bibfield  {author} {\bibinfo {author} {\bibfnamefont {Jiemin}\ \bibnamefont
  {Li}}, \bibinfo {author} {\bibfnamefont {Lei}\ \bibnamefont {Xu}}, \bibinfo
  {author} {\bibfnamefont {Mirian}\ \bibnamefont {Garcia-Fernandez}}, \bibinfo
  {author} {\bibfnamefont {Abhishek}\ \bibnamefont {Nag}}, \bibinfo {author}
  {\bibfnamefont {H.~C.}\ \bibnamefont {Robarts}}, \bibinfo {author}
  {\bibfnamefont {A.~C.}\ \bibnamefont {Walters}}, \bibinfo {author}
  {\bibfnamefont {X.}~\bibnamefont {Liu}}, \bibinfo {author} {\bibfnamefont
  {Jianshi}\ \bibnamefont {Zhou}}, \bibinfo {author} {\bibfnamefont
  {Krzysztof}\ \bibnamefont {Wohlfeld}}, \bibinfo {author} {\bibfnamefont
  {Jeroen}\ \bibnamefont {van~den Brink}}, \bibinfo {author} {\bibfnamefont
  {Hong}\ \bibnamefont {Ding}}, \ and\ \bibinfo {author} {\bibfnamefont
  {Ke-Jin}\ \bibnamefont {Zhou}},\ }\bibfield  {title} {\enquote {\bibinfo
  {title} {{Unraveling the Orbital Physics in a Canonical Orbital System
  ${\mathrm{KCuF}}_{3}$}},}\ }\href {\doibase 10.1103/PhysRevLett.126.106401}
  {\bibfield  {journal} {\bibinfo  {journal} {Phys. Rev. Lett.}\ }\textbf
  {\bibinfo {volume} {126}},\ \bibinfo {pages} {106401} (\bibinfo {year}
  {2021})}\BibitemShut {NoStop}%
\bibitem [{\citenamefont {Lee}\ \emph {et~al.}(2011)\citenamefont {Lee},
  \citenamefont {Yuan}, \citenamefont {Lal}, \citenamefont {Joe}, \citenamefont
  {Gan}, \citenamefont {Smadici}, \citenamefont {Finkelstein}, \citenamefont
  {Feng}, \citenamefont {Rusydi}, \citenamefont {Goldbart}, \citenamefont
  {Cooper},\ and\ \citenamefont {Abbamonte}}]{Lee_2011}%
  \BibitemOpen
  \bibfield  {author} {\bibinfo {author} {\bibfnamefont {James C.~T.}\
  \bibnamefont {Lee}}, \bibinfo {author} {\bibfnamefont {Shi}\ \bibnamefont
  {Yuan}}, \bibinfo {author} {\bibfnamefont {Siddhartha}\ \bibnamefont {Lal}},
  \bibinfo {author} {\bibfnamefont {Young~Il}\ \bibnamefont {Joe}}, \bibinfo
  {author} {\bibfnamefont {Yu}~\bibnamefont {Gan}}, \bibinfo {author}
  {\bibfnamefont {Serban}\ \bibnamefont {Smadici}}, \bibinfo {author}
  {\bibfnamefont {Ken}\ \bibnamefont {Finkelstein}}, \bibinfo {author}
  {\bibfnamefont {Yejun}\ \bibnamefont {Feng}}, \bibinfo {author}
  {\bibfnamefont {Andrivo}\ \bibnamefont {Rusydi}}, \bibinfo {author}
  {\bibfnamefont {Paul~M.}\ \bibnamefont {Goldbart}}, \bibinfo {author}
  {\bibfnamefont {S.~Lance}\ \bibnamefont {Cooper}}, \ and\ \bibinfo {author}
  {\bibfnamefont {Peter}\ \bibnamefont {Abbamonte}},\ }\bibfield  {title}
  {\enquote {\bibinfo {title} {{Two-stage orbital order and dynamical spin
  frustration in {KCuF}$_3$}},}\ }\href {\doibase 10.1038/nphys2117} {\bibfield
   {journal} {\bibinfo  {journal} {Nature Physics}\ }\textbf {\bibinfo {volume}
  {8}},\ \bibinfo {pages} {63--66} (\bibinfo {year} {2011})}\BibitemShut
  {NoStop}%
\bibitem [{\citenamefont {Wang}\ \emph
  {et~al.}(2015{\natexlab{b}})\citenamefont {Wang}, \citenamefont {Kivelson},\
  and\ \citenamefont {Lee}}]{Wang_2015}%
  \BibitemOpen
  \bibfield  {author} {\bibinfo {author} {\bibfnamefont {Fa}~\bibnamefont
  {Wang}}, \bibinfo {author} {\bibfnamefont {Steven~A.}\ \bibnamefont
  {Kivelson}}, \ and\ \bibinfo {author} {\bibfnamefont {Dung-Hai}\ \bibnamefont
  {Lee}},\ }\bibfield  {title} {\enquote {\bibinfo {title} {Nematicity and
  quantum paramagnetism in {FeSe}},}\ }\href {\doibase 10.1038/nphys3456}
  {\bibfield  {journal} {\bibinfo  {journal} {Nature Physics}\ }\textbf
  {\bibinfo {volume} {11}},\ \bibinfo {pages} {959--963} (\bibinfo {year}
  {2015}{\natexlab{b}})}\BibitemShut {NoStop}%
\bibitem [{\citenamefont {Savary}(2021)}]{Savarync}%
  \BibitemOpen
  \bibfield  {author} {\bibinfo {author} {\bibfnamefont {L.}~\bibnamefont
  {Savary}},\ }\bibfield  {title} {\enquote {\bibinfo {title} {Quantum loop
  states in spin-orbital models on the honeycomb lattice},}\ }\href {\doibase
  10.1038/s41467-021-23033-y} {\bibfield  {journal} {\bibinfo  {journal}
  {Nature Communications}\ }\textbf {\bibinfo {volume} {12}},\ \bibinfo {pages}
  {3004} (\bibinfo {year} {2021})}\BibitemShut {NoStop}%
\bibitem [{\citenamefont {T\'oth}\ \emph {et~al.}(2010)\citenamefont {T\'oth},
  \citenamefont {L\"auchli}, \citenamefont {Mila},\ and\ \citenamefont
  {Penc}}]{PhysRevLett.105.265301}%
  \BibitemOpen
  \bibfield  {author} {\bibinfo {author} {\bibfnamefont {Tam\'as~A.}\
  \bibnamefont {T\'oth}}, \bibinfo {author} {\bibfnamefont {Andreas~M.}\
  \bibnamefont {L\"auchli}}, \bibinfo {author} {\bibfnamefont {Fr\'ed\'eric}\
  \bibnamefont {Mila}}, \ and\ \bibinfo {author} {\bibfnamefont {Karlo}\
  \bibnamefont {Penc}},\ }\bibfield  {title} {\enquote {\bibinfo {title}
  {{Three-Sublattice Ordering of the SU(3) Heisenberg Model of Three-Flavor
  Fermions on the Square and Cubic Lattices}},}\ }\href {\doibase
  10.1103/PhysRevLett.105.265301} {\bibfield  {journal} {\bibinfo  {journal}
  {Phys. Rev. Lett.}\ }\textbf {\bibinfo {volume} {105}},\ \bibinfo {pages}
  {265301} (\bibinfo {year} {2010})}\BibitemShut {NoStop}%
\bibitem [{\citenamefont {Hermele}\ and\ \citenamefont
  {Gurarie}(2011)}]{PhysRevB.84.174441}%
  \BibitemOpen
  \bibfield  {author} {\bibinfo {author} {\bibfnamefont {Michael}\ \bibnamefont
  {Hermele}}\ and\ \bibinfo {author} {\bibfnamefont {Victor}\ \bibnamefont
  {Gurarie}},\ }\bibfield  {title} {\enquote {\bibinfo {title} {{Topological
  liquids and valence cluster states in two-dimensional SU$(N)$ magnets}},}\
  }\href {\doibase 10.1103/PhysRevB.84.174441} {\bibfield  {journal} {\bibinfo
  {journal} {Phys. Rev. B}\ }\textbf {\bibinfo {volume} {84}},\ \bibinfo
  {pages} {174441} (\bibinfo {year} {2011})}\BibitemShut {NoStop}%
\bibitem [{\citenamefont {Nataf}\ and\ \citenamefont
  {Mila}(2014)}]{PhysRevLett.113.127204}%
  \BibitemOpen
  \bibfield  {author} {\bibinfo {author} {\bibfnamefont {Pierre}\ \bibnamefont
  {Nataf}}\ and\ \bibinfo {author} {\bibfnamefont {Fr\'ed\'eric}\ \bibnamefont
  {Mila}},\ }\bibfield  {title} {\enquote {\bibinfo {title} {{Exact
  Diagonalization of Heisenberg $\mathrm{SU}(N)$ Models}},}\ }\href {\doibase
  10.1103/PhysRevLett.113.127204} {\bibfield  {journal} {\bibinfo  {journal}
  {Phys. Rev. Lett.}\ }\textbf {\bibinfo {volume} {113}},\ \bibinfo {pages}
  {127204} (\bibinfo {year} {2014})}\BibitemShut {NoStop}%
\bibitem [{\citenamefont {Corboz}\ \emph {et~al.}(2011)\citenamefont {Corboz},
  \citenamefont {L\"auchli}, \citenamefont {Penc}, \citenamefont {Troyer},\
  and\ \citenamefont {Mila}}]{PhysRevLett.107.215301}%
  \BibitemOpen
  \bibfield  {author} {\bibinfo {author} {\bibfnamefont {Philippe}\
  \bibnamefont {Corboz}}, \bibinfo {author} {\bibfnamefont {Andreas~M.}\
  \bibnamefont {L\"auchli}}, \bibinfo {author} {\bibfnamefont {Karlo}\
  \bibnamefont {Penc}}, \bibinfo {author} {\bibfnamefont {Matthias}\
  \bibnamefont {Troyer}}, \ and\ \bibinfo {author} {\bibfnamefont
  {Fr\'ed\'eric}\ \bibnamefont {Mila}},\ }\bibfield  {title} {\enquote
  {\bibinfo {title} {{Simultaneous Dimerization and SU(4) Symmetry Breaking of
  4-Color Fermions on the Square Lattice}},}\ }\href {\doibase
  10.1103/PhysRevLett.107.215301} {\bibfield  {journal} {\bibinfo  {journal}
  {Phys. Rev. Lett.}\ }\textbf {\bibinfo {volume} {107}},\ \bibinfo {pages}
  {215301} (\bibinfo {year} {2011})}\BibitemShut {NoStop}%
\bibitem [{\citenamefont {Yao}\ \emph {et~al.}(2022)\citenamefont {Yao},
  \citenamefont {Luo},\ and\ \citenamefont {Chen}}]{yao2021intertwining}%
  \BibitemOpen
  \bibfield  {author} {\bibinfo {author} {\bibfnamefont {Xu-Ping}\ \bibnamefont
  {Yao}}, \bibinfo {author} {\bibfnamefont {Rui~Leonard}\ \bibnamefont {Luo}},
  \ and\ \bibinfo {author} {\bibfnamefont {Gang}\ \bibnamefont {Chen}},\
  }\bibfield  {title} {\enquote {\bibinfo {title} {{Intertwining SU($N$)
  symmetry and frustration on a honeycomb lattice}},}\ }\href {\doibase
  10.1103/PhysRevB.105.024401} {\bibfield  {journal} {\bibinfo  {journal}
  {Phys. Rev. B}\ }\textbf {\bibinfo {volume} {105}},\ \bibinfo {pages}
  {024401} (\bibinfo {year} {2022})}\BibitemShut {NoStop}%
\bibitem [{\citenamefont {Chen}\ \emph
  {et~al.}(2016{\natexlab{b}})\citenamefont {Chen}, \citenamefont {Hazzard},
  \citenamefont {Rey},\ and\ \citenamefont {Hermele}}]{PhysRevA.93.061601}%
  \BibitemOpen
  \bibfield  {author} {\bibinfo {author} {\bibfnamefont {Gang}\ \bibnamefont
  {Chen}}, \bibinfo {author} {\bibfnamefont {Kaden R.~A.}\ \bibnamefont
  {Hazzard}}, \bibinfo {author} {\bibfnamefont {Ana~Maria}\ \bibnamefont
  {Rey}}, \ and\ \bibinfo {author} {\bibfnamefont {Michael}\ \bibnamefont
  {Hermele}},\ }\bibfield  {title} {\enquote {\bibinfo {title}
  {Synthetic-gauge-field stabilization of the chiral-spin-liquid phase},}\
  }\href {\doibase 10.1103/PhysRevA.93.061601} {\bibfield  {journal} {\bibinfo
  {journal} {Phys. Rev. A}\ }\textbf {\bibinfo {volume} {93}},\ \bibinfo
  {pages} {061601} (\bibinfo {year} {2016}{\natexlab{b}})}\BibitemShut
  {NoStop}%
\bibitem [{\citenamefont {Yao}\ \emph {et~al.}(2021)\citenamefont {Yao},
  \citenamefont {Gao},\ and\ \citenamefont {Chen}}]{PhysRevResearch.3.023138}%
  \BibitemOpen
  \bibfield  {author} {\bibinfo {author} {\bibfnamefont {Xu-Ping}\ \bibnamefont
  {Yao}}, \bibinfo {author} {\bibfnamefont {Yonghao}\ \bibnamefont {Gao}}, \
  and\ \bibinfo {author} {\bibfnamefont {Gang}\ \bibnamefont {Chen}},\
  }\bibfield  {title} {\enquote {\bibinfo {title} {{Topological chiral spin
  liquids and competing states in triangular lattice $\mathrm{SU}(N)$ Mott
  insulators}},}\ }\href {\doibase 10.1103/PhysRevResearch.3.023138} {\bibfield
   {journal} {\bibinfo  {journal} {Phys. Rev. Research}\ }\textbf {\bibinfo
  {volume} {3}},\ \bibinfo {pages} {023138} (\bibinfo {year}
  {2021})}\BibitemShut {NoStop}%
\bibitem [{\citenamefont {Smerald}\ and\ \citenamefont
  {Mila}(2014)}]{PhysRevB.90.094422}%
  \BibitemOpen
  \bibfield  {author} {\bibinfo {author} {\bibfnamefont {Andrew}\ \bibnamefont
  {Smerald}}\ and\ \bibinfo {author} {\bibfnamefont {Fr\'ed\'eric}\
  \bibnamefont {Mila}},\ }\bibfield  {title} {\enquote {\bibinfo {title}
  {{Exploring the spin-orbital ground state of
  ${\mathrm{Ba}}_{3}{\mathrm{CuSb}}_{2}{\mathrm{O}}_{9}$}},}\ }\href {\doibase
  10.1103/PhysRevB.90.094422} {\bibfield  {journal} {\bibinfo  {journal} {Phys.
  Rev. B}\ }\textbf {\bibinfo {volume} {90}},\ \bibinfo {pages} {094422}
  (\bibinfo {year} {2014})}\BibitemShut {NoStop}%
\bibitem [{\citenamefont {Katayama}\ \emph {et~al.}(2015)\citenamefont
  {Katayama}, \citenamefont {Kimura}, \citenamefont {Han}, \citenamefont
  {Nasu}, \citenamefont {Drichko}, \citenamefont {Nakanishi}, \citenamefont
  {Halim}, \citenamefont {Ishiguro}, \citenamefont {Satake}, \citenamefont
  {Nishibori}, \citenamefont {Yoshizawa}, \citenamefont {Nakano}, \citenamefont
  {Nozue}, \citenamefont {Wakabayashi}, \citenamefont {Ishihara}, \citenamefont
  {Hagiwara}, \citenamefont {Sawa},\ and\ \citenamefont
  {Nakatsuji}}]{Katayama_2015}%
  \BibitemOpen
  \bibfield  {author} {\bibinfo {author} {\bibfnamefont {Naoyuki}\ \bibnamefont
  {Katayama}}, \bibinfo {author} {\bibfnamefont {Kenta}\ \bibnamefont
  {Kimura}}, \bibinfo {author} {\bibfnamefont {Yibo}\ \bibnamefont {Han}},
  \bibinfo {author} {\bibfnamefont {Joji}\ \bibnamefont {Nasu}}, \bibinfo
  {author} {\bibfnamefont {Natalia}\ \bibnamefont {Drichko}}, \bibinfo {author}
  {\bibfnamefont {Yoshiki}\ \bibnamefont {Nakanishi}}, \bibinfo {author}
  {\bibfnamefont {Mario}\ \bibnamefont {Halim}}, \bibinfo {author}
  {\bibfnamefont {Yuki}\ \bibnamefont {Ishiguro}}, \bibinfo {author}
  {\bibfnamefont {Ryuta}\ \bibnamefont {Satake}}, \bibinfo {author}
  {\bibfnamefont {Eiji}\ \bibnamefont {Nishibori}}, \bibinfo {author}
  {\bibfnamefont {Masahito}\ \bibnamefont {Yoshizawa}}, \bibinfo {author}
  {\bibfnamefont {Takehito}\ \bibnamefont {Nakano}}, \bibinfo {author}
  {\bibfnamefont {Yasuo}\ \bibnamefont {Nozue}}, \bibinfo {author}
  {\bibfnamefont {Yusuke}\ \bibnamefont {Wakabayashi}}, \bibinfo {author}
  {\bibfnamefont {Sumio}\ \bibnamefont {Ishihara}}, \bibinfo {author}
  {\bibfnamefont {Masayuki}\ \bibnamefont {Hagiwara}}, \bibinfo {author}
  {\bibfnamefont {Hiroshi}\ \bibnamefont {Sawa}}, \ and\ \bibinfo {author}
  {\bibfnamefont {Satoru}\ \bibnamefont {Nakatsuji}},\ }\bibfield  {title}
  {\enquote {\bibinfo {title} {{Absence of Jahn-Teller transition in the
  hexagonal ${\mathrm{Ba}}_{3}{\mathrm{CuSb}}_{2}{\mathrm{O}}_{9}$ single
  crystal}},}\ }\href {\doibase 10.1073/pnas.1508941112} {\bibfield  {journal}
  {\bibinfo  {journal} {Proceedings of the National Academy of Sciences}\
  }\textbf {\bibinfo {volume} {112}},\ \bibinfo {pages} {9305--9309} (\bibinfo
  {year} {2015})}\BibitemShut {NoStop}%
\bibitem [{\citenamefont {Altmeyer}\ \emph {et~al.}(2017)\citenamefont
  {Altmeyer}, \citenamefont {Mila}, \citenamefont {Smerald},\ and\
  \citenamefont {Valent\'{\i}}}]{PhysRevB.96.115116}%
  \BibitemOpen
  \bibfield  {author} {\bibinfo {author} {\bibfnamefont {Michaela}\
  \bibnamefont {Altmeyer}}, \bibinfo {author} {\bibfnamefont {Frederic}\
  \bibnamefont {Mila}}, \bibinfo {author} {\bibfnamefont {Andrew}\ \bibnamefont
  {Smerald}}, \ and\ \bibinfo {author} {\bibfnamefont {Roser}\ \bibnamefont
  {Valent\'{\i}}},\ }\bibfield  {title} {\enquote {\bibinfo {title} {{Cu-Sb
  dumbbell arrangement in the spin-orbital liquid candidate
  ${\mathrm{Ba}}_{3}{\mathrm{CuSb}}_{2}{\mathrm{O}}_{9}$}},}\ }\href {\doibase
  10.1103/PhysRevB.96.115116} {\bibfield  {journal} {\bibinfo  {journal} {Phys.
  Rev. B}\ }\textbf {\bibinfo {volume} {96}},\ \bibinfo {pages} {115116}
  (\bibinfo {year} {2017})}\BibitemShut {NoStop}%
\bibitem [{\citenamefont {Yamada}\ \emph {et~al.}(2021)\citenamefont {Yamada},
  \citenamefont {Oshikawa},\ and\ \citenamefont
  {Jackeli}}]{PhysRevB.104.224436}%
  \BibitemOpen
  \bibfield  {author} {\bibinfo {author} {\bibfnamefont {Masahiko~G.}\
  \bibnamefont {Yamada}}, \bibinfo {author} {\bibfnamefont {Masaki}\
  \bibnamefont {Oshikawa}}, \ and\ \bibinfo {author} {\bibfnamefont {George}\
  \bibnamefont {Jackeli}},\ }\bibfield  {title} {\enquote {\bibinfo {title}
  {{SU(4)-symmetric quantum spin-orbital liquids on various lattices}},}\
  }\href {\doibase 10.1103/PhysRevB.104.224436} {\bibfield  {journal} {\bibinfo
   {journal} {Phys. Rev. B}\ }\textbf {\bibinfo {volume} {104}},\ \bibinfo
  {pages} {224436} (\bibinfo {year} {2021})}\BibitemShut {NoStop}%
\bibitem [{\citenamefont {Zhang}\ \emph {et~al.}(2021)\citenamefont {Zhang},
  \citenamefont {Sheng},\ and\ \citenamefont {Vishwanath}}]{Zhang_2021}%
  \BibitemOpen
  \bibfield  {author} {\bibinfo {author} {\bibfnamefont {Ya-Hui}\ \bibnamefont
  {Zhang}}, \bibinfo {author} {\bibfnamefont {D.-N.}\ \bibnamefont {Sheng}}, \
  and\ \bibinfo {author} {\bibfnamefont {Ashvin}\ \bibnamefont {Vishwanath}},\
  }\bibfield  {title} {\enquote {\bibinfo {title} {{SU}(4) chiral spin liquid,
  exciton supersolid, and electric detection in moir{\'{e}} bilayers},}\ }\href
  {\doibase 10.1103/physrevlett.127.247701} {\bibfield  {journal} {\bibinfo
  {journal} {Physical Review Letters}\ }\textbf {\bibinfo {volume} {127}},\
  \bibinfo {pages} {247701} (\bibinfo {year} {2021})}\BibitemShut {NoStop}%
\bibitem [{\citenamefont {Zhang}\ and\ \citenamefont
  {Vishwanath}(2020)}]{zhang2020electrical}%
  \BibitemOpen
  \bibfield  {author} {\bibinfo {author} {\bibfnamefont {Ya-Hui}\ \bibnamefont
  {Zhang}}\ and\ \bibinfo {author} {\bibfnamefont {Ashvin}\ \bibnamefont
  {Vishwanath}},\ }\href@noop {} {\enquote {\bibinfo {title} {Electrical
  detection of spin liquids in double moir\'e layers},}\ } (\bibinfo {year}
  {2020}),\ \Eprint {http://arxiv.org/abs/2005.12925} {arXiv:2005.12925
  [cond-mat.str-el]} \BibitemShut {NoStop}%
\bibitem [{\citenamefont {Liu}\ \emph {et~al.}(2018{\natexlab{b}})\citenamefont
  {Liu}, \citenamefont {Li},\ and\ \citenamefont {Chen}}]{PhysRevB.98.045119}%
  \BibitemOpen
  \bibfield  {author} {\bibinfo {author} {\bibfnamefont {Changle}\ \bibnamefont
  {Liu}}, \bibinfo {author} {\bibfnamefont {Yao-Dong}\ \bibnamefont {Li}}, \
  and\ \bibinfo {author} {\bibfnamefont {Gang}\ \bibnamefont {Chen}},\
  }\bibfield  {title} {\enquote {\bibinfo {title} {Selective measurements of
  intertwined multipolar orders: Non-kramers doublets on a triangular
  lattice},}\ }\href {\doibase 10.1103/PhysRevB.98.045119} {\bibfield
  {journal} {\bibinfo  {journal} {Phys. Rev. B}\ }\textbf {\bibinfo {volume}
  {98}},\ \bibinfo {pages} {045119} (\bibinfo {year}
  {2018}{\natexlab{b}})}\BibitemShut {NoStop}%
\bibitem [{\citenamefont {Harris}\ \emph {et~al.}(2003)\citenamefont {Harris},
  \citenamefont {Yildirim}, \citenamefont {Aharony}, \citenamefont
  {Entin-Wohlman},\ and\ \citenamefont {Korenblit}}]{PhysRevLett.91.087206}%
  \BibitemOpen
  \bibfield  {author} {\bibinfo {author} {\bibfnamefont {A.~B.}\ \bibnamefont
  {Harris}}, \bibinfo {author} {\bibfnamefont {Taner}\ \bibnamefont
  {Yildirim}}, \bibinfo {author} {\bibfnamefont {Amnon}\ \bibnamefont
  {Aharony}}, \bibinfo {author} {\bibfnamefont {Ora}\ \bibnamefont
  {Entin-Wohlman}}, \ and\ \bibinfo {author} {\bibfnamefont {I.~Ya.}\
  \bibnamefont {Korenblit}},\ }\bibfield  {title} {\enquote {\bibinfo {title}
  {{Unusual Symmetries in the Kugel-Khomskii Hamiltonian}},}\ }\href {\doibase
  10.1103/PhysRevLett.91.087206} {\bibfield  {journal} {\bibinfo  {journal}
  {Phys. Rev. Lett.}\ }\textbf {\bibinfo {volume} {91}},\ \bibinfo {pages}
  {087206} (\bibinfo {year} {2003})}\BibitemShut {NoStop}%
\bibitem [{\citenamefont {Khaliullin}\ and\ \citenamefont
  {Oudovenko}(1997)}]{PhysRevB.56.R14243}%
  \BibitemOpen
  \bibfield  {author} {\bibinfo {author} {\bibfnamefont {G.}~\bibnamefont
  {Khaliullin}}\ and\ \bibinfo {author} {\bibfnamefont {V.}~\bibnamefont
  {Oudovenko}},\ }\bibfield  {title} {\enquote {\bibinfo {title} {{Spin and
  orbital excitation spectrum in the Kugel-Khomskii model}},}\ }\href {\doibase
  10.1103/PhysRevB.56.R14243} {\bibfield  {journal} {\bibinfo  {journal} {Phys.
  Rev. B}\ }\textbf {\bibinfo {volume} {56}},\ \bibinfo {pages}
  {R14243--R14246} (\bibinfo {year} {1997})}\BibitemShut {NoStop}%
\bibitem [{\citenamefont {Di~Matteo}\ \emph {et~al.}(2004)\citenamefont
  {Di~Matteo}, \citenamefont {Jackeli}, \citenamefont {Lacroix},\ and\
  \citenamefont {Perkins}}]{PhysRevLett.93.077208}%
  \BibitemOpen
  \bibfield  {author} {\bibinfo {author} {\bibfnamefont {S.}~\bibnamefont
  {Di~Matteo}}, \bibinfo {author} {\bibfnamefont {G.}~\bibnamefont {Jackeli}},
  \bibinfo {author} {\bibfnamefont {C.}~\bibnamefont {Lacroix}}, \ and\
  \bibinfo {author} {\bibfnamefont {N.~B.}\ \bibnamefont {Perkins}},\
  }\bibfield  {title} {\enquote {\bibinfo {title} {Valence-bond crystal in a
  pyrochlore antiferromagnet with orbital degeneracy},}\ }\href {\doibase
  10.1103/PhysRevLett.93.077208} {\bibfield  {journal} {\bibinfo  {journal}
  {Phys. Rev. Lett.}\ }\textbf {\bibinfo {volume} {93}},\ \bibinfo {pages}
  {077208} (\bibinfo {year} {2004})}\BibitemShut {NoStop}%
\bibitem [{\citenamefont {Wu}(2006{\natexlab{b}})}]{wu2006a}%
  \BibitemOpen
  \bibfield  {author} {\bibinfo {author} {\bibfnamefont {Congjun}\ \bibnamefont
  {Wu}},\ }\bibfield  {title} {\enquote {\bibinfo {title} {Hidden symmetry and
  quantum phases in spin-3/2 cold atomic systems},}\ }\href {\doibase
  10.1142/S0217984906012213} {\bibfield  {journal} {\bibinfo  {journal} {Mod.
  Phys. Lett. B}\ }\textbf {\bibinfo {volume} {20}},\ \bibinfo {pages} {1707}
  (\bibinfo {year} {2006}{\natexlab{b}})}\BibitemShut {NoStop}%
\bibitem [{\citenamefont {Chen}\ \emph {et~al.}(2005)\citenamefont {Chen},
  \citenamefont {Wu}, \citenamefont {Wang},\ and\ \citenamefont
  {Zhang}}]{chen2005}%
  \BibitemOpen
  \bibfield  {author} {\bibinfo {author} {\bibfnamefont {S.}~\bibnamefont
  {Chen}}, \bibinfo {author} {\bibfnamefont {C.}~\bibnamefont {Wu}}, \bibinfo
  {author} {\bibfnamefont {Y.~P.}\ \bibnamefont {Wang}}, \ and\ \bibinfo
  {author} {\bibfnamefont {S.~C.}\ \bibnamefont {Zhang}},\ }\bibfield  {title}
  {\enquote {\bibinfo {title} {Exact spontaneous plaquette ground states for
  high-spin ladder models},}\ }\href {\doibase 10.1103/PhysRevB.72.214428}
  {\bibfield  {journal} {\bibinfo  {journal} {Phys. Rev. B}\ }\textbf {\bibinfo
  {volume} {72}},\ \bibinfo {pages} {214428} (\bibinfo {year}
  {2005})}\BibitemShut {NoStop}%
\bibitem [{\citenamefont {Wu}(2005)}]{wu2005a}%
  \BibitemOpen
  \bibfield  {author} {\bibinfo {author} {\bibfnamefont {Congjun}\ \bibnamefont
  {Wu}},\ }\bibfield  {title} {\enquote {\bibinfo {title} {Competing orders in
  one dimensional spin 3/2 fermionic systems},}\ }\href {\doibase
  10.1103/PhysRevLett.95.266404} {\bibfield  {journal} {\bibinfo  {journal}
  {Phys. Rev. Lett.}\ }\textbf {\bibinfo {volume} {95}},\ \bibinfo {pages}
  {266404} (\bibinfo {year} {2005})}\BibitemShut {NoStop}%
\bibitem [{\citenamefont {Xu}\ and\ \citenamefont {Wu}(2008)}]{xu2008}%
  \BibitemOpen
  \bibfield  {author} {\bibinfo {author} {\bibfnamefont {Cenke}\ \bibnamefont
  {Xu}}\ and\ \bibinfo {author} {\bibfnamefont {Congjun}\ \bibnamefont {Wu}},\
  }\bibfield  {title} {\enquote {\bibinfo {title} {{Resonating plaquette phases
  in SU(4) Heisenberg antiferromagnet}},}\ }\href {\doibase
  10.1103/PhysRevB.77.134449} {\bibfield  {journal} {\bibinfo  {journal} {Phys.
  Rev. B}\ }\textbf {\bibinfo {volume} {77}},\ \bibinfo {pages} {134449}
  (\bibinfo {year} {2008})}\BibitemShut {NoStop}%
\bibitem [{\citenamefont {Hung}\ \emph {et~al.}(2011)\citenamefont {Hung},
  \citenamefont {Wang},\ and\ \citenamefont {Wu}}]{hung2011}%
  \BibitemOpen
  \bibfield  {author} {\bibinfo {author} {\bibfnamefont {Hsiang-Hsuan}\
  \bibnamefont {Hung}}, \bibinfo {author} {\bibfnamefont {Yupeng}\ \bibnamefont
  {Wang}}, \ and\ \bibinfo {author} {\bibfnamefont {Congjun}\ \bibnamefont
  {Wu}},\ }\bibfield  {title} {\enquote {\bibinfo {title} {Quantum magnetism in
  ultracold alkali and alkaline-earth fermion systems with symplectic
  symmetry},}\ }\href {\doibase 10.1103/PhysRevB.84.054406} {\bibfield
  {journal} {\bibinfo  {journal} {Phys. Rev. B}\ }\textbf {\bibinfo {volume}
  {84}},\ \bibinfo {pages} {054406} (\bibinfo {year} {2011})}\BibitemShut
  {NoStop}%
\bibitem [{\citenamefont {Rapp}\ \emph {et~al.}(2007)\citenamefont {Rapp},
  \citenamefont {Zarand}, \citenamefont {Honerkamp},\ and\ \citenamefont
  {Hofstetter}}]{rapp2007}%
  \BibitemOpen
  \bibfield  {author} {\bibinfo {author} {\bibfnamefont {A.}~\bibnamefont
  {Rapp}}, \bibinfo {author} {\bibfnamefont {G.}~\bibnamefont {Zarand}},
  \bibinfo {author} {\bibfnamefont {C.}~\bibnamefont {Honerkamp}}, \ and\
  \bibinfo {author} {\bibfnamefont {W.}~\bibnamefont {Hofstetter}},\ }\bibfield
   {title} {\enquote {\bibinfo {title} {Color superfluidity and "baryon"
  formation in ultracold fermions},}\ }\href {\doibase
  10.1103/PhysRevLett.98.160405} {\bibfield  {journal} {\bibinfo  {journal}
  {Phys. Rev. Lett.}\ }\textbf {\bibinfo {volume} {98}},\ \bibinfo {pages}
  {160405} (\bibinfo {year} {2007})}\BibitemShut {NoStop}%
\bibitem [{\citenamefont {Lecheminant}\ \emph {et~al.}(2008)\citenamefont
  {Lecheminant}, \citenamefont {Azaria},\ and\ \citenamefont
  {Boulat}}]{lecheminant2008}%
  \BibitemOpen
  \bibfield  {author} {\bibinfo {author} {\bibfnamefont {P.}~\bibnamefont
  {Lecheminant}}, \bibinfo {author} {\bibfnamefont {P.}~\bibnamefont {Azaria}},
  \ and\ \bibinfo {author} {\bibfnamefont {E.}~\bibnamefont {Boulat}},\
  }\bibfield  {title} {\enquote {\bibinfo {title} {Competing orders in
  one-dimensional half-integer fermionic cold atoms: A conformal field theory
  approach},}\ }\href {\doibase 10.1016/j.nuclphysb.2007.12.034} {\bibfield
  {journal} {\bibinfo  {journal} {Nucl. Phys. B}\ }\textbf {\bibinfo {volume}
  {798}},\ \bibinfo {pages} {443} (\bibinfo {year} {2008})}\BibitemShut
  {NoStop}%
\bibitem [{\citenamefont {Bossche}\ \emph {et~al.}(2000)\citenamefont
  {Bossche}, \citenamefont {Zhang},\ and\ \citenamefont {Mila}}]{bossche2000}%
  \BibitemOpen
  \bibfield  {author} {\bibinfo {author} {\bibfnamefont {M.~V.~D.}\
  \bibnamefont {Bossche}}, \bibinfo {author} {\bibfnamefont {F.~C.}\
  \bibnamefont {Zhang}}, \ and\ \bibinfo {author} {\bibfnamefont
  {F.}~\bibnamefont {Mila}},\ }\bibfield  {title} {\enquote {\bibinfo {title}
  {{Plaquette Ground State in the Two-dimensional SU(4) Spin-Orbital Model}},}\
  }\href {\doibase 10.1007/PL00011085} {\bibfield  {journal} {\bibinfo
  {journal} {Eur. Phys. J. B}\ }\textbf {\bibinfo {volume} {17}},\ \bibinfo
  {pages} {367} (\bibinfo {year} {2000})}\BibitemShut {NoStop}%
\bibitem [{\citenamefont {Mishra}\ \emph {et~al.}(2002)\citenamefont {Mishra},
  \citenamefont {Ma},\ and\ \citenamefont {Zhang}}]{mishra2002}%
  \BibitemOpen
  \bibfield  {author} {\bibinfo {author} {\bibfnamefont {A.}~\bibnamefont
  {Mishra}}, \bibinfo {author} {\bibfnamefont {M.}~\bibnamefont {Ma}}, \ and\
  \bibinfo {author} {\bibfnamefont {F.~C.}\ \bibnamefont {Zhang}},\ }\bibfield
  {title} {\enquote {\bibinfo {title} {{Plaquette ordering in SU(4)
  antiferromagnets }},}\ }\href {\doibase 10.1103/PhysRevB.65.214411}
  {\bibfield  {journal} {\bibinfo  {journal} {Phys. Rev. B}\ }\textbf {\bibinfo
  {volume} {65}},\ \bibinfo {pages} {214411} (\bibinfo {year}
  {2002})}\BibitemShut {NoStop}%
\bibitem [{\citenamefont {Sutherland}(1975)}]{sutherland1975}%
  \BibitemOpen
  \bibfield  {author} {\bibinfo {author} {\bibfnamefont {B.}~\bibnamefont
  {Sutherland}},\ }\bibfield  {title} {\enquote {\bibinfo {title} {Model for a
  multicomponent quantum system},}\ }\href {\doibase 10.1103/PhysRevB.12.3795}
  {\bibfield  {journal} {\bibinfo  {journal} {Phys. Rev. B}\ }\textbf {\bibinfo
  {volume} {12}},\ \bibinfo {pages} {3795--3805} (\bibinfo {year}
  {1975})}\BibitemShut {NoStop}%
\bibitem [{\citenamefont {Li}\ \emph {et~al.}(1998)\citenamefont {Li},
  \citenamefont {Ma}, \citenamefont {Shi},\ and\ \citenamefont
  {Zhang}}]{li1998}%
  \BibitemOpen
  \bibfield  {author} {\bibinfo {author} {\bibfnamefont {Y.~Q.}\ \bibnamefont
  {Li}}, \bibinfo {author} {\bibfnamefont {Michael}\ \bibnamefont {Ma}},
  \bibinfo {author} {\bibfnamefont {D.~N.}\ \bibnamefont {Shi}}, \ and\
  \bibinfo {author} {\bibfnamefont {F.~C.}\ \bibnamefont {Zhang}},\ }\bibfield
  {title} {\enquote {\bibinfo {title} {{SU(4) Theory for Spin Systems with
  Orbital Degeneracy}},}\ }\href {\doibase 10.1103/PhysRevLett.81.3527}
  {\bibfield  {journal} {\bibinfo  {journal} {Phys. Rev. Lett.}\ }\textbf
  {\bibinfo {volume} {81}},\ \bibinfo {pages} {3527} (\bibinfo {year}
  {1998})}\BibitemShut {NoStop}%
\bibitem [{\citenamefont {van~den Bossche}\ \emph {et~al.}(2001)\citenamefont
  {van~den Bossche}, \citenamefont {Azaria}, \citenamefont {Lecheminant},\ and\
  \citenamefont {Mila}}]{bossche2001}%
  \BibitemOpen
  \bibfield  {author} {\bibinfo {author} {\bibfnamefont {Mathias}\ \bibnamefont
  {van~den Bossche}}, \bibinfo {author} {\bibfnamefont {P.}~\bibnamefont
  {Azaria}}, \bibinfo {author} {\bibfnamefont {P.}~\bibnamefont {Lecheminant}},
  \ and\ \bibinfo {author} {\bibfnamefont {F.}~\bibnamefont {Mila}},\
  }\bibfield  {title} {\enquote {\bibinfo {title} {{Spontaneous Plaquette
  Formation in the SU(4) Spin-Orbital Ladder}},}\ }\href {\doibase
  10.1103/PhysRevLett.86.4124} {\bibfield  {journal} {\bibinfo  {journal}
  {Phys. Rev. Lett.}\ }\textbf {\bibinfo {volume} {86}},\ \bibinfo {pages}
  {4124} (\bibinfo {year} {2001})}\BibitemShut {NoStop}%
\bibitem [{\citenamefont {Yamashita}\ \emph {et~al.}(1998)\citenamefont
  {Yamashita}, \citenamefont {Shibta},\ and\ \citenamefont
  {Ueda}}]{yamashita1998}%
  \BibitemOpen
  \bibfield  {author} {\bibinfo {author} {\bibfnamefont {Y.}~\bibnamefont
  {Yamashita}}, \bibinfo {author} {\bibfnamefont {N.}~\bibnamefont {Shibta}}, \
  and\ \bibinfo {author} {\bibfnamefont {K.}~\bibnamefont {Ueda}},\ }\bibfield
  {title} {\enquote {\bibinfo {title} {{SU(4) spin-orbit critical state in one
  dimension}},}\ }\href {\doibase 10.1103/PhysRevB.58.9114} {\bibfield
  {journal} {\bibinfo  {journal} {Phys. Rev. B}\ }\textbf {\bibinfo {volume}
  {58}},\ \bibinfo {pages} {9114} (\bibinfo {year} {1998})}\BibitemShut
  {NoStop}%
\bibitem [{\citenamefont {Azaria}\ \emph {et~al.}(1999)\citenamefont {Azaria},
  \citenamefont {Gogolin}, \citenamefont {Lecheminant},\ and\ \citenamefont
  {Nersesyan}}]{azaria1999}%
  \BibitemOpen
  \bibfield  {author} {\bibinfo {author} {\bibfnamefont {P.}~\bibnamefont
  {Azaria}}, \bibinfo {author} {\bibfnamefont {A.~O.}\ \bibnamefont {Gogolin}},
  \bibinfo {author} {\bibfnamefont {P.}~\bibnamefont {Lecheminant}}, \ and\
  \bibinfo {author} {\bibfnamefont {A.~A.}\ \bibnamefont {Nersesyan}},\
  }\bibfield  {title} {\enquote {\bibinfo {title} {{One-Dimensional SU(4)
  Spin-Orbital Model: A Low-Energy Effective Theory}},}\ }\href {\doibase
  10.1103/PhysRevLett.83.624} {\bibfield  {journal} {\bibinfo  {journal} {Phys.
  Rev. Lett.}\ }\textbf {\bibinfo {volume} {83}},\ \bibinfo {pages} {624}
  (\bibinfo {year} {1999})}\BibitemShut {NoStop}%
\bibitem [{\citenamefont {Harada}\ \emph {et~al.}(2003)\citenamefont {Harada},
  \citenamefont {Kawashima},\ and\ \citenamefont {Troyer}}]{harada2003}%
  \BibitemOpen
  \bibfield  {author} {\bibinfo {author} {\bibfnamefont {K.}~\bibnamefont
  {Harada}}, \bibinfo {author} {\bibfnamefont {N.}~\bibnamefont {Kawashima}}, \
  and\ \bibinfo {author} {\bibfnamefont {M.}~\bibnamefont {Troyer}},\
  }\bibfield  {title} {\enquote {\bibinfo {title} {{N\'eel and Spin-Peierls
  Ground States of Two-Dimensional SU(N) Quantum Antiferromagnets}},}\ }\href
  {\doibase 10.1103/PhysRevLett.90.117203} {\bibfield  {journal} {\bibinfo
  {journal} {Phys. Rev. Lett.}\ }\textbf {\bibinfo {volume} {90}},\ \bibinfo
  {pages} {117203} (\bibinfo {year} {2003})}\BibitemShut {NoStop}%
\bibitem [{\citenamefont {Schulz}\ and\ \citenamefont
  {Ziman}(1992)}]{schulz1992}%
  \BibitemOpen
  \bibfield  {author} {\bibinfo {author} {\bibfnamefont {H.~J.}\ \bibnamefont
  {Schulz}}\ and\ \bibinfo {author} {\bibfnamefont {T.~A.~L.}\ \bibnamefont
  {Ziman}},\ }\bibfield  {title} {\enquote {\bibinfo {title} {Finite-size
  scaling for the two-dimensional frustrated quantum heisenberg
  antiferromagnet},}\ }\href {\doibase 10.1209/0295-5075/18/4/013} {\bibfield
  {journal} {\bibinfo  {journal} {Europhys. Lett.}\ }\textbf {\bibinfo {volume}
  {18}},\ \bibinfo {pages} {355} (\bibinfo {year} {1992})}\BibitemShut
  {NoStop}%
\bibitem [{\citenamefont {Richter}\ and\ \citenamefont
  {Ivanove}(1996)}]{richter1996}%
  \BibitemOpen
  \bibfield  {author} {\bibinfo {author} {\bibfnamefont {J.}~\bibnamefont
  {Richter}}\ and\ \bibinfo {author} {\bibfnamefont {N.~B.}\ \bibnamefont
  {Ivanove}},\ }\bibfield  {title} {\enquote {\bibinfo {title}
  {Zero-temperature quantum disorder in spin systems by competition between
  dimer and plaquette bonds},}\ }\href {\doibase 10.1007/BF02570951} {\bibfield
   {journal} {\bibinfo  {journal} {Czechoslovak J. Phys.}\ }\textbf {\bibinfo
  {volume} {46}},\ \bibinfo {pages} {1919} (\bibinfo {year}
  {1996})}\BibitemShut {NoStop}%
\bibitem [{\citenamefont {Pankov}\ \emph {et~al.}(2007)\citenamefont {Pankov},
  \citenamefont {Moessner},\ and\ \citenamefont {Sondhi}}]{pankov2007}%
  \BibitemOpen
  \bibfield  {author} {\bibinfo {author} {\bibfnamefont {S.}~\bibnamefont
  {Pankov}}, \bibinfo {author} {\bibfnamefont {R.}~\bibnamefont {Moessner}}, \
  and\ \bibinfo {author} {\bibfnamefont {S.~L.}\ \bibnamefont {Sondhi}},\
  }\bibfield  {title} {\enquote {\bibinfo {title} {Resonating singlet valence
  plaquettes},}\ }\href {doi:10.1103/PhysRevB.76.104436} {\bibfield  {journal}
  {\bibinfo  {journal} {Phys. Rev. B}\ }\textbf {\bibinfo {volume} {76}},\
  \bibinfo {pages} {104436} (\bibinfo {year} {2007})}\BibitemShut {NoStop}%
\bibitem [{\citenamefont {Rokhsar}\ and\ \citenamefont
  {Kivelson}(1988)}]{rokhsar1988}%
  \BibitemOpen
  \bibfield  {author} {\bibinfo {author} {\bibfnamefont {Daniel~S.}\
  \bibnamefont {Rokhsar}}\ and\ \bibinfo {author} {\bibfnamefont {Steven~A.}\
  \bibnamefont {Kivelson}},\ }\bibfield  {title} {\enquote {\bibinfo {title}
  {Superconductivity and the quantum hard-core dimer gas},}\ }\href {\doibase
  10.1103/PhysRevLett.61.2376} {\bibfield  {journal} {\bibinfo  {journal}
  {Phys. Rev. Lett.}\ }\textbf {\bibinfo {volume} {61}},\ \bibinfo {pages}
  {2376--2379} (\bibinfo {year} {1988})}\BibitemShut {NoStop}%
\bibitem [{\citenamefont {Fradkin}\ and\ \citenamefont
  {Kivelson}(1990)}]{fradkin1990}%
  \BibitemOpen
  \bibfield  {author} {\bibinfo {author} {\bibfnamefont {Eduardo~H.}\
  \bibnamefont {Fradkin}}\ and\ \bibinfo {author} {\bibfnamefont {Steven}\
  \bibnamefont {Kivelson}},\ }\bibfield  {title} {\enquote {\bibinfo {title}
  {{Short Range Resonating Valence Bond Theories and Superconductivity}},}\
  }\href {\doibase 10.1142/S0217984990000295} {\bibfield  {journal} {\bibinfo
  {journal} {Mod. Phys. Lett. B}\ }\textbf {\bibinfo {volume} {4}},\ \bibinfo
  {pages} {225} (\bibinfo {year} {1990})}\BibitemShut {NoStop}%
\bibitem [{\citenamefont {Nandkishore}\ and\ \citenamefont
  {Hermele}(2019)}]{nandkishore2019}%
  \BibitemOpen
  \bibfield  {author} {\bibinfo {author} {\bibfnamefont {Rahul~M.}\
  \bibnamefont {Nandkishore}}\ and\ \bibinfo {author} {\bibfnamefont {Michael}\
  \bibnamefont {Hermele}},\ }\bibfield  {title} {\enquote {\bibinfo {title}
  {Fractons},}\ }\href {\doibase 10.1146/annurev-conmatphys-031218-013604}
  {\bibfield  {journal} {\bibinfo  {journal} {Annual Review of Condensed Matter
  Physics}\ }\textbf {\bibinfo {volume} {10}},\ \bibinfo {pages} {295--313}
  (\bibinfo {year} {2019})}\BibitemShut {NoStop}%
\bibitem [{\citenamefont {Motrunich}(2005)}]{PhysRevB.72.045105}%
  \BibitemOpen
  \bibfield  {author} {\bibinfo {author} {\bibfnamefont {Olexei~I.}\
  \bibnamefont {Motrunich}},\ }\bibfield  {title} {\enquote {\bibinfo {title}
  {{Variational study of triangular lattice spin-1/2 model with ring exchanges
  and spin liquid state in ${\kappa}$-(ET)$_2$Cu$_{2}${(CN)}$_{3}$}},}\ }\href
  {\doibase 10.1103/PhysRevB.72.045105} {\bibfield  {journal} {\bibinfo
  {journal} {Phys. Rev. B}\ }\textbf {\bibinfo {volume} {72}},\ \bibinfo
  {pages} {045105} (\bibinfo {year} {2005})}\BibitemShut {NoStop}%
\bibitem [{\citenamefont {Lee}\ and\ \citenamefont
  {Lee}(2005)}]{PhysRevLett.95.036403}%
  \BibitemOpen
  \bibfield  {author} {\bibinfo {author} {\bibfnamefont {Sung-Sik}\
  \bibnamefont {Lee}}\ and\ \bibinfo {author} {\bibfnamefont {Patrick~A.}\
  \bibnamefont {Lee}},\ }\bibfield  {title} {\enquote {\bibinfo {title} {{U(1)
  Gauge Theory of the Hubbard Model: Spin Liquid States and Possible
  Application to
  ${\kappa}\mathrm{\text{{-}}}(\mathrm{BEDT}\mathrm{\text{{-}}}\mathrm{TTF}{)}_{2}{\mathrm{Cu}}_{2}(\mathrm{CN}{)}_{3}$}},}\
  }\href {\doibase 10.1103/PhysRevLett.95.036403} {\bibfield  {journal}
  {\bibinfo  {journal} {Phys. Rev. Lett.}\ }\textbf {\bibinfo {volume} {95}},\
  \bibinfo {pages} {036403} (\bibinfo {year} {2005})}\BibitemShut {NoStop}%
\bibitem [{\citenamefont {Rozen}\ \emph {et~al.}(2021)\citenamefont {Rozen},
  \citenamefont {Park}, \citenamefont {Zondiner}, \citenamefont {Cao},
  \citenamefont {Rodan-Legrain}, \citenamefont {Taniguchi}, \citenamefont
  {Watanabe}, \citenamefont {Oreg}, \citenamefont {Stern}, \citenamefont
  {Berg}, \citenamefont {Jarillo-Herrero},\ and\ \citenamefont
  {Ilani}}]{Rozen_2021}%
  \BibitemOpen
  \bibfield  {author} {\bibinfo {author} {\bibfnamefont {Asaf}\ \bibnamefont
  {Rozen}}, \bibinfo {author} {\bibfnamefont {Jeong~Min}\ \bibnamefont {Park}},
  \bibinfo {author} {\bibfnamefont {Uri}\ \bibnamefont {Zondiner}}, \bibinfo
  {author} {\bibfnamefont {Yuan}\ \bibnamefont {Cao}}, \bibinfo {author}
  {\bibfnamefont {Daniel}\ \bibnamefont {Rodan-Legrain}}, \bibinfo {author}
  {\bibfnamefont {Takashi}\ \bibnamefont {Taniguchi}}, \bibinfo {author}
  {\bibfnamefont {Kenji}\ \bibnamefont {Watanabe}}, \bibinfo {author}
  {\bibfnamefont {Yuval}\ \bibnamefont {Oreg}}, \bibinfo {author}
  {\bibfnamefont {Ady}\ \bibnamefont {Stern}}, \bibinfo {author} {\bibfnamefont
  {Erez}\ \bibnamefont {Berg}}, \bibinfo {author} {\bibfnamefont {Pablo}\
  \bibnamefont {Jarillo-Herrero}}, \ and\ \bibinfo {author} {\bibfnamefont
  {Shahal}\ \bibnamefont {Ilani}},\ }\bibfield  {title} {\enquote {\bibinfo
  {title} {{Entropic evidence for a Pomeranchuk effect in magic-angle
  graphene}},}\ }\href {\doibase 10.1038/s41586-021-03319-3} {\bibfield
  {journal} {\bibinfo  {journal} {Nature}\ }\textbf {\bibinfo {volume} {592}},\
  \bibinfo {pages} {214--219} (\bibinfo {year} {2021})}\BibitemShut {NoStop}%
\bibitem [{\citenamefont {Saito}\ \emph {et~al.}(2021)\citenamefont {Saito},
  \citenamefont {Yang}, \citenamefont {Ge}, \citenamefont {Liu}, \citenamefont
  {Taniguchi}, \citenamefont {Watanabe}, \citenamefont {Li}, \citenamefont
  {Berg},\ and\ \citenamefont {Young}}]{Saito_2021}%
  \BibitemOpen
  \bibfield  {author} {\bibinfo {author} {\bibfnamefont {Yu}~\bibnamefont
  {Saito}}, \bibinfo {author} {\bibfnamefont {Fangyuan}\ \bibnamefont {Yang}},
  \bibinfo {author} {\bibfnamefont {Jingyuan}\ \bibnamefont {Ge}}, \bibinfo
  {author} {\bibfnamefont {Xiaoxue}\ \bibnamefont {Liu}}, \bibinfo {author}
  {\bibfnamefont {Takashi}\ \bibnamefont {Taniguchi}}, \bibinfo {author}
  {\bibfnamefont {Kenji}\ \bibnamefont {Watanabe}}, \bibinfo {author}
  {\bibfnamefont {J.~I.~A.}\ \bibnamefont {Li}}, \bibinfo {author}
  {\bibfnamefont {Erez}\ \bibnamefont {Berg}}, \ and\ \bibinfo {author}
  {\bibfnamefont {Andrea~F.}\ \bibnamefont {Young}},\ }\bibfield  {title}
  {\enquote {\bibinfo {title} {{Isospin Pomeranchuk effect in twisted bilayer
  graphene}},}\ }\href {\doibase 10.1038/s41586-021-03409-2} {\bibfield
  {journal} {\bibinfo  {journal} {Nature}\ }\textbf {\bibinfo {volume} {592}},\
  \bibinfo {pages} {220--224} (\bibinfo {year} {2021})}\BibitemShut {NoStop}%
\bibitem [{\citenamefont {Hazzard}\ \emph {et~al.}(2012)\citenamefont
  {Hazzard}, \citenamefont {Gurarie}, \citenamefont {Hermele},\ and\
  \citenamefont {Rey}}]{PhysRevA.85.041604}%
  \BibitemOpen
  \bibfield  {author} {\bibinfo {author} {\bibfnamefont {Kaden R.~A.}\
  \bibnamefont {Hazzard}}, \bibinfo {author} {\bibfnamefont {Victor}\
  \bibnamefont {Gurarie}}, \bibinfo {author} {\bibfnamefont {Michael}\
  \bibnamefont {Hermele}}, \ and\ \bibinfo {author} {\bibfnamefont {Ana~Maria}\
  \bibnamefont {Rey}},\ }\bibfield  {title} {\enquote {\bibinfo {title}
  {{High-temperature properties of fermionic alkaline-earth-metal atoms in
  optical lattices}},}\ }\href {\doibase 10.1103/PhysRevA.85.041604} {\bibfield
   {journal} {\bibinfo  {journal} {Phys. Rev. A}\ }\textbf {\bibinfo {volume}
  {85}},\ \bibinfo {pages} {041604} (\bibinfo {year} {2012})}\BibitemShut
  {NoStop}%
\bibitem [{\citenamefont {Assaad}(2005)}]{ASSAAD2005}%
  \BibitemOpen
  \bibfield  {author} {\bibinfo {author} {\bibfnamefont {F.~F.}\ \bibnamefont
  {Assaad}},\ }\bibfield  {title} {\enquote {\bibinfo {title} {{Phase diagram
  of the half-filled two-dimensional SU(N) Hubbard-Heisenberg model: A quantum
  Monte Carlo study }},}\ }\href {\doibase 10.1103/PhysRevB.71.075103}
  {\bibfield  {journal} {\bibinfo  {journal} {Phys. Rev. B}\ }\textbf {\bibinfo
  {volume} {71}},\ \bibinfo {pages} {75103} (\bibinfo {year}
  {2005})}\BibitemShut {NoStop}%
\bibitem [{\citenamefont {Wang}\ \emph {et~al.}(2019)\citenamefont {Wang},
  \citenamefont {Wang},\ and\ \citenamefont {Wu}}]{PhysRevB.100.115155}%
  \BibitemOpen
  \bibfield  {author} {\bibinfo {author} {\bibfnamefont {Da}~\bibnamefont
  {Wang}}, \bibinfo {author} {\bibfnamefont {Lei}\ \bibnamefont {Wang}}, \ and\
  \bibinfo {author} {\bibfnamefont {Congjun}\ \bibnamefont {Wu}},\ }\bibfield
  {title} {\enquote {\bibinfo {title} {{Slater and Mott insulating states in
  the SU(6) Hubbard model}},}\ }\href {\doibase 10.1103/PhysRevB.100.115155}
  {\bibfield  {journal} {\bibinfo  {journal} {Phys. Rev. B}\ }\textbf {\bibinfo
  {volume} {100}},\ \bibinfo {pages} {115155} (\bibinfo {year}
  {2019})}\BibitemShut {NoStop}%
\bibitem [{\citenamefont {Huang}\ \emph {et~al.}(2023)\citenamefont {Huang},
  \citenamefont {Chen}, \citenamefont {Huang}, \citenamefont {Setty},
  \citenamefont {Gao}, \citenamefont {Shi}, \citenamefont {Liu}, \citenamefont
  {Zhang}, \citenamefont {Yilmaz}, \citenamefont {Vescovo}, \citenamefont
  {Hashimoto}, \citenamefont {Lu}, \citenamefont {Yakobson}, \citenamefont
  {Dai}, \citenamefont {Chu}, \citenamefont {Si},\ and\ \citenamefont
  {Yi}}]{huang2023nonfermi}%
  \BibitemOpen
  \bibfield  {author} {\bibinfo {author} {\bibfnamefont {Jianwei}\ \bibnamefont
  {Huang}}, \bibinfo {author} {\bibfnamefont {Lei}\ \bibnamefont {Chen}},
  \bibinfo {author} {\bibfnamefont {Yuefei}\ \bibnamefont {Huang}}, \bibinfo
  {author} {\bibfnamefont {Chandan}\ \bibnamefont {Setty}}, \bibinfo {author}
  {\bibfnamefont {Bin}\ \bibnamefont {Gao}}, \bibinfo {author} {\bibfnamefont
  {Yue}\ \bibnamefont {Shi}}, \bibinfo {author} {\bibfnamefont {Zhaoyu}\
  \bibnamefont {Liu}}, \bibinfo {author} {\bibfnamefont {Yichen}\ \bibnamefont
  {Zhang}}, \bibinfo {author} {\bibfnamefont {Turgut}\ \bibnamefont {Yilmaz}},
  \bibinfo {author} {\bibfnamefont {Elio}\ \bibnamefont {Vescovo}}, \bibinfo
  {author} {\bibfnamefont {Makoto}\ \bibnamefont {Hashimoto}}, \bibinfo
  {author} {\bibfnamefont {Donghui}\ \bibnamefont {Lu}}, \bibinfo {author}
  {\bibfnamefont {Boris~I.}\ \bibnamefont {Yakobson}}, \bibinfo {author}
  {\bibfnamefont {Pengcheng}\ \bibnamefont {Dai}}, \bibinfo {author}
  {\bibfnamefont {Jiun-Haw}\ \bibnamefont {Chu}}, \bibinfo {author}
  {\bibfnamefont {Qimiao}\ \bibnamefont {Si}}, \ and\ \bibinfo {author}
  {\bibfnamefont {Ming}\ \bibnamefont {Yi}},\ }\href@noop {} {\enquote
  {\bibinfo {title} {Non-fermi liquid behavior in a correlated flatband
  pyrochlore lattice},}\ } (\bibinfo {year} {2023}),\ \Eprint
  {http://arxiv.org/abs/2311.01269} {arXiv:2311.01269 [cond-mat.str-el]}
  \BibitemShut {NoStop}%
\bibitem [{\citenamefont {Penc}\ \emph {et~al.}(2004)\citenamefont {Penc},
  \citenamefont {Shannon},\ and\ \citenamefont
  {Shiba}}]{PhysRevLett.93.197203}%
  \BibitemOpen
  \bibfield  {author} {\bibinfo {author} {\bibfnamefont {Karlo}\ \bibnamefont
  {Penc}}, \bibinfo {author} {\bibfnamefont {Nic}\ \bibnamefont {Shannon}}, \
  and\ \bibinfo {author} {\bibfnamefont {Hiroyuki}\ \bibnamefont {Shiba}},\
  }\bibfield  {title} {\enquote {\bibinfo {title} {Half-magnetization plateau
  stabilized by structural distortion in the antiferromagnetic heisenberg model
  on a pyrochlore lattice},}\ }\href {\doibase 10.1103/PhysRevLett.93.197203}
  {\bibfield  {journal} {\bibinfo  {journal} {Phys. Rev. Lett.}\ }\textbf
  {\bibinfo {volume} {93}},\ \bibinfo {pages} {197203} (\bibinfo {year}
  {2004})}\BibitemShut {NoStop}%
\bibitem [{\citenamefont {Knizhnik}\ and\ \citenamefont
  {Zamolodchikov}(1984)}]{KNIZHNIK198483}%
  \BibitemOpen
  \bibfield  {author} {\bibinfo {author} {\bibfnamefont {V.G.}\ \bibnamefont
  {Knizhnik}}\ and\ \bibinfo {author} {\bibfnamefont {A.B.}\ \bibnamefont
  {Zamolodchikov}},\ }\bibfield  {title} {\enquote {\bibinfo {title} {{Current
  algebra and Wess-Zumino model in two dimensions}},}\ }\href {\doibase
  https://doi.org/10.1016/0550-3213(84)90374-2} {\bibfield  {journal} {\bibinfo
   {journal} {Nuclear Physics B}\ }\textbf {\bibinfo {volume} {247}},\ \bibinfo
  {pages} {83--103} (\bibinfo {year} {1984})}\BibitemShut {NoStop}%
\end{thebibliography}%

\end{document}